\documentclass[12pt]{article}
\usepackage{graphicx}
\usepackage{epsfig}

\def\be{\begin{eqnarray}}
\def\ee{\end{eqnarray}}

\newcommand{\ice}[1]{\relax}

\newcommand{\sll}{/\kern-4pt l}
\newcommand{\slp}{p\kern-5pt/}
\newcommand{\slq}{q\kern-5.5pt/}

\newcommand{\dDp}{\frac{d^Dp}{(2\pi)^D}}
\newcommand{\Disc}{{\rm Disc\,}}

\newcommand{\Li}{{\rm Li}}
\newcommand{\Cl}{{\rm Cl}}

\newcommand{\eps}{\varepsilon}
\newcommand{\pfrac}[2]{\left(\frac{#1}{#2}\right)}
\newcommand{\simle}{\ \hbox{\raise2pt\rlap{$<$}%
 \lower3pt\rlap{$\sim$}\phantom{$<$}}\ }
\newcommand{\dalembertian}{\hbox{$\,\vbox{\hrule\hbox{\vrule%
  \vbox{\kern7pt}\kern7pt\vrule}\hrule}\,$}}
\begin{document}
\begin{flushright}
MZ-TH/05-08\\
hep-ph/0506286\\
June 2005\\
\end{flushright}

\begin{center}
{\Large\bf On the evaluation of a certain class of Feynman diagrams}\\[12pt]
{\Large\bf in $x$-space: Sunrise-type topologies at any loop order}

\vspace{1truecm}

{\large \bf S.~Groote$^{1,2}$, J.G.~K\"orner$^2$ and
  A.A.~Pivovarov$^{2,3}$}\\[.4truecm]
$^1$Tartu \"Ulikooli Teoreetilise F\"u\"usika Instituut,\\
  T\"ahe 4, EE-51010 Tartu, Estonia\\[.3truecm]
$^2$Institut f\"ur Physik der Johannes-Gutenberg-Universit\"at,\\
  Staudinger Weg 7, D-55099 Mainz, Germany\\[.3truecm]
$^3$Institute for Nuclear Research of the\\
  Russian Academy of Sciences, Moscow 117312, Russia
\end{center}

\begin{abstract}
We review recently developed new powerful techniques to compute a class of
Feynman diagrams at any loop order, known as sunrise-type diagrams. These
sunrise-type topologies have many important applications in many different
fields of physics and we believe it to be timely to discuss their evaluation
from a unified point of view. The method is based on the analysis of the
diagrams directly in configuration space which, in the case of the
sunrise-type diagrams and diagrams related to them, leads to enormous
simplifications as compared to the traditional evaluation of loops in momentum
space. We present explicit formulae for their analytical evaluation for
arbitrary mass configurations and arbitrary dimensions at any loop order. We
discuss several limiting cases of their kinematical regimes which are e.g.\
relevant for applications in HQET and NRQCD. We  completely solve the problem
of renormalization using simple formulae for the counterterms within
dimensional regularization. An important application is the computation of
the multi-particle phase space in $D$-dimensional space-time which we discuss.
We present some examples of their numerical evaluation in the general case of
$D$-dimensional space-time as well as in integer dimensions $D=D_0$ for
different values of dimensions including the most important practical cases
$D_0=2,3,4$. Substantial simplifications occur for odd integer space-time
dimensions where the final results can be expressed in closed form through
elementary functions. We discuss the use of recurrence relations naturally 
emerging in configuration space for the calculation of special series of
integrals of the sunrise topology. We finally report on results for the
computation of an extension of the basic sunrise topology, namely the
spectacle topology and the topology with an irreducible loop addition.
\end{abstract}

PACS {??.??}

\newpage

\tableofcontents
\newpage

\section{Introduction and motivation}
In the present--day analysis of various physical processes the complexity 
of the problem often makes it impossible to find a complete mathematical
description of the experimentally measured or observed quantities.
In order to proceed one usually simplifies the mathematical model which 
describes the bulk of the phenomena. The more delicate effects are accounted
for through an expansion in terms of a small parameter which has to be
defined such that one retains full quantitative control in the analysis.
The art of a physicist consists in splitting the whole analysis into a main 
part and into small corrections to the main part for which one develops a
perturbative expansion. One may quote one of the greatest physicists of the 
last century, L.D. Landau, who used to say that physics starts when a 
small parameter of the problem has been found. The usefulness and power of
such an approach has been proven over and over for many years. An example that
comes into mind are the first accurate calculations analyzing high precision
data from celestial mechanics in terms of Newton's law of gravitation. When
observational data on the motion of the planets became more accurate it was no
longer sufficient to take into account only their gravitational interaction
with the sun but it was necessary also to include the small corrections
arising from the mutual gravitational interactions between the planets
themselves. This was done through perturbation theory since the full equations
for the planetary movements with mutual interaction fully taken into account
are too difficult to treat analytically even within standard classical
mechanics. A remarkable fact is that these corrections are very small but
still have to be taken into account since the data are extremely precise. This
is well illustrated by a historical example taken from astronomy which was the
most precise scientific discipline in the past: the planets of the solar
system close to Earth were directly observed while an eighth planet -- Neptune
-- was first introduced in order to explain slight perturbations in the orbit
of Uranus. The orbit of Neptune was theoretically calculated with high
precision and the planet was directly observed later at the predicted place.

Let us now turn to high energy physics. The main interest of fundamental 
research in high energy physics is to find new physical phenomena and to place
them into a consistent logical picture of the natural world which mankind
inhabits. It is generally believed that the structure of fundamental
interactions (except gravity) is qualitatively understood at the energies 
available at present accelerators. Basically there are two trends for
the search of new physics. One is aiming at a direct discovery of new physical
phenomena and fundamental constituents of matter (elementary particles) by
moving to higher and higher energies. This requires ever more powerful
accelerating facilities while the precision of the experimental detection
and theoretical calculations can remain at a moderate level. The other is an 
indirect search for new physics at low energies (mainly through radiative 
corrections) which is based on very precise experimental data at moderate
energies and a high accuracy of the theoretical calculations~\cite{Langacker,
kuhn0,Hollik:1998hv,higgsLEP,Hollik:2000ap}. A very recent example of the
second category is the ongoing discussion about the possible appearance of
deviations from the Standard Model (SM) of strong and electroweak interaction
in new data on the muon anomalous magnetic moment~\cite{gminus2,
Hayakawa:2001bb, Pivovarov:2001mw,Groote:2001vu}. This two-fold road to
scientific discovery is well known and has been pursued for ages as 
illustrated by the above example from celestial mechanics.

The muon anomalous magnetic moment has been computed in Quantum
Electrodynamics (QED) -- the most precise physical theory of the present time.
QED is a highly accurate theory of electromagnetic particle interactions where
predictions for physical observables are calculated within perturbation 
theory in the form of a series in the expansion parameter of the theory, the
fine structure constant $\alpha$ with a numerical value of
$\alpha^{-1}=137.036$ at zero momentum. The fact that $\alpha$ is rather small
makes the expansions well convergent. The muon anomalous magnetic moment has
been computed through the expansion in the small ratio of lepton masses and
the expansion in the coupling constant up to the order $\alpha^4$. The most
recent result reads~\cite{Schwinger:1948iu,Kinoshita:1990wp,Kinoshita:2004wi} 
\begin{equation}\label{mamm}
\alpha_\mu({\rm QED})=A_1+A_2(m_\mu/m_e)+A_2(m_\mu/m_\tau)
  +A_3(m_\mu/m_e,m_\mu/m_\tau)
\end{equation}
where for instance~\cite{Hughes:1999fp}
\begin{eqnarray}
A_1&=&0.5\pfrac\alpha\pi-0.328478965\ldots\pfrac\alpha\pi^2
  +1.181241456\ldots\pfrac\alpha\pi^3\nonumber\\&&
  -1.5098(384)\pfrac\alpha\pi^4+4.393(27)\cdot 10^{-12}.\nonumber
\end{eqnarray}
The theoretical expressions for the coefficients of this series involve
complicated multidimensional integrals the evaluation of which is the main
difficulty in the calculation of a series expansion for the muon anomalous
magnetic moment. A part of the coefficients in Eq.~(\ref{mamm}) have been 
calculated analytically whereas the numbers in the higher order terms have to
be evaluated numerically~\cite{Hughes:1999fp,Mohr:2000ie}.

With the advent of powerful computers a number of problems can now be
approached through direct numerical calculations. The trajectories of
satellites, for instance, are computed through direct calculations with
computers. There are also important applications in Quantum Field Theory (QFT)
especially, in the theory of strong interaction formulated on the lattice.
However, at present the achieved accuracy is not yet sufficient leaving plenty 
of room for analytical computations. In addition, calculations in perturbation
theory are often formulated in such a way that a numerical approach is not
efficient or is unreliable due to the accumulation of rounding errors. The
situation is quite analogous to the problem of long-time weather forecasts in
meteorology where the direct numerical approaches are extremely unstable from
the computational point of view. Rounding errors of the numerical evaluation
of a large number of Feynman diagrams are such that the final accuracy is low
because of huge cancellations between separate contributions of separate
diagrams. Although an analytic evaluation is definitely preferable it cannot 
always be done due to technical difficulties in calculating the relevant
Feynman integrals.

In Quantum Field Theory only a very limited number of problems have a full
exact solution. Perturbation theory is a general tool for investigating
the realistic physical situations. This is especially true in the theory of
strong interactions and in the Standard Model in general. The classical
examples are the analysis of $e^+e^-$ annihilation into hadrons and the
analysis of tau decays where perturbation theory is used in its most advanced
form for the calculation up to very high orders of the strong coupling
constant~\cite{Chetyrkin:1979bj,Gorishnii:1990vf}. In the latter case such an
analysis has lead to the most accurate extraction of strong interaction
parameters at low energies~\cite{Braaten:1991qm,Dorokhov:2004vf,Pivovarov:wp}.
The attempt to extrapolate the calculations to all orders for some limited
subsets of diagrams or contributions are also quite useful~\cite{Coquereaux:vp,
Zakharov:1992bx,renPi,Beneke:1998ui}. The coefficients of the expansions are
given by complicated integrals which have a useful graphical representation in
terms of Feynman diagrams. Graphical techniques are very efficient and
convenient in the bookkeeping of the various loop contributions and are in
wide use. In higher orders these graphs have many loops which are associated
with the loop integrals. One speaks of multi-loop calculations for the
evaluation of the coefficients of the perturbative expansion in QFT. Due to
big advances in symbolic computing the most advanced symbolic programs can
nowadays generate the terms of the perturbative expansion without actually
drawing any graphs. Nevertheless, the terminology has survived. Nowadays the
methods of QFT (and its terminology) is starting to penetrate into fields
other than high energy physics such as polymer physics, hydro- and
aerodynamics and weather forecasts which are much closer to real life.
Therefore, the calculation of the complicated integrals associated with
Feynman diagrams in a perturbative expansion is important in high energy
physics as well as in many other areas of physics.

While the problem of calculating multi-loop integrals will not be solved in a
general form in the foreseeable future, there is a continuing interest in the 
analysis of particular diagrams in the efforts to enlarge the number of known
analytical results (as a review, see e.g.~\cite{Grozin:2002nb}). Especially
promising in this regard is the analysis of subclasses of diagrams with
simpler topological structures but with an arbitrary number of loops. The most
promising in this respect is the simple topology of the so-called generalized
sunrise diagrams (sometimes also called sunset, water melon, banana, or
basketball diagrams). They will be referred to as the sunrise-type diagrams in
this report. They can be computed for an arbitrary number of loops which makes
them an ideal tool to obtain clues to the structure of more complicated
multi-loop diagrams~\cite{Mendels:wc,Mendels:qe,Berends:1993ee}.

Recently there has been a renewal of interest in the calculation of diagrams 
of the sunrise-type topology for different numbers of loops involving
different loop masses and/or external momenta of the diagram (see
e.g.~\cite{Mastrolia:2002gt,Mastrolia:2002tv,Czyz:2002re,Schroder:2002re}).
The classical (or genuine) two-loop sunrise diagram with different values of
the internal masses has recently been studied in some detail using the
traditional momentum space approach (see e.g.~\cite{Post:1997dk,Post:1996gg,
Rajantie:1996cw,Berends:1997vk,Gasser:1998qt} and references therein). In the
present report we review the configuration space technique (also called
$x$-space technique) for the calculation of sunrise-type diagrams. The
configuration space technique is best suited for the topology of the diagram.
It considerably reduces the complexity of the calculation and allows for new
qualitative and quantitative results for this important particular class of
Feynman integrals~\cite{Mendels:wc,Groote:1998ic,Groote:1998wy,Groote:1999cx,
Groote:2000kz,Delbourgo:2003zi}. Configuration space techniques can be used to
verify known results obtained within other techniques both analytically and
numerically~\cite{Caffo:2002wm,Caffo:2002ch,Ligterink:1999mu,Fleischer:1999mp} 
and to investigate some general features of Feynman diagram
calculation~\cite{Passarino:2001wv,Czyz:2002re,Groote:1999cx,Groote:1999cn,
Laporta:2001dd,Suzuki:2000wj,Davydychev:2000na,Aste:2003wi}. We list features
of the $x$-space technique in turn with the aim to show how efficient the
method is.

The diagrams with sunrise-type topology form a subset of the general set of
diagrams with a given number of loops. They appear in various specific physics
applications:
\begin{itemize}
\item multi-loop calculations in general~\cite{Bauberger:1994by,Kazakov:1986mu}
\item the analysis of static properties of baryons using sum rule techniques in
  QCD~\cite{Ovchinnikov:1991mu,Pivovarov:1991nk,Groote:1999zp,Ioffe:kw,
  Groote:2000py,Grozin:1992td,Furnstahl:1995nd,Jin:1997pb,Kras,Ovchinnikov:zx}
\item the properties of glueballs extracted from the perturbative analysis of
  two-point correlators~\cite{Groote:2001vr,Kataev:1982gr}
\item the multi-particle phase space for multi-body decays of a
  single-particle phase space~\cite{Bashir:2001ad,Bardin:1999ak,Bardin:2002gs}
\item lattice QCD calculations~\cite{Wetzorke:2002mx,Doi:2004xe}
\item mixing of neutral mesons~\cite{Narison:1994zt}
\item Chiral perturbation theory (ChPT) and effective theories for Goldstone
  modes in higher orders of the momentum expansion~\cite{Post:1997dk,
  Gasser:1998qt,Weinberg:1978kz}
\item analysis of exotic states in QFT: multi-quark states in QCD or
  pentaquarks using sum rule techniques~\cite{Larin:1986yt,Sakai:1999qm,
  Jaffe:2003sg,Zhu:2003ba,Balitsky:ps,Chetyrkin:2000tj,Jaffe:2004wv,
  Sugiyama:2003zk,Ishii:2004qe}
\item general questions of QFT: sum rules in two-dimensional
  QED~\cite{Schwinger:1951nm,Pivovarov:1985zw} and properties of baryons in
  the large $N_c$ limit of QCD~\cite{Witten:1979kh,tHooftNc}
\item effective potentials for symmetry 
  breaking~\cite{effpot,Jackiw:1980kv,Jackiw:cv,Chung:1999xm,Chung:1998mz}
\item finite temperature calculations~\cite{Rajantie:1996cw,Kajantie:2003ax,
  Hatsuda:1991du,Gross:1980br,Nishikawa:2003js,Andersen:2000zn,Appelquist:vg,
  Doi:2004jp}
\item applications in nuclear physics~\cite{Yang:2003bv,VanHees:2001pf,
  Mehta:2003mi,Kulagin:kb,Meissner:2004yy,Bijnens:2005mi}
\item phase space integrals for particles in jets where the momentum along
  the direction of the jet is fixed~\cite{Mirkes}
\item applications in solid state physics~\cite{Platter:2002yr}
\end{itemize}

The principal aim of this report is to assemble all the necessary tools needed
for the computation of multi-loop Feynman diagrams with sunrise-type topology.
We discuss the whole known spectrum of different methods to analyze these
integrals. Amomg these are concise analytical evaluation techniques, expansion
in small parameters such as masses and momenta (or inverse masses and momenta),
expansions in special kinematical regions as the threshold regime in the
Minkowskian domain, integral representations for integrals including their 
discontinuity across physical cuts in terms of analytic functions of the
external momentum, and, finally, efficient, fast, simple and stable numerical
procedures. In the Euclidean domain the numerical procedures derived from our
representation are efficient and reliable, i.e.\ stable against error
accumulation. In order to acquaint the reader with the tools given given
in the report we start with basic notations and relations.

We present a comprehensive report on configuration space techniques including
many technical details such that an interested reader can use this report as a
practical guide for practical calculations. We have included a great deal of
mathematical material such that the report is self-contained to a large 
degree. In most cases it is not necessary to consult mathematical handbooks in
order to understand the calculations. We also give a rather large sample of
worked-out examples. The calculational methods used are rather well-suited for
further development and can easily be tailored to the further specific needs
of the potential user. Using the results of this paper one can create one's
own software for an efficient evaluation of the many quantities of interest
that can be extracted from the analysis of sunrise-type diagrams.

The report is organized as follows. In Sec.~2 we briefly summarize some
general notions and their application to sunrise-type diagrams. In Sec.~2.1 we
introduce the configuration space representation of sunrise-type diagrams and
fix our notation. In Sec.~2.2 we specify what is meant by computing a
sunrise-type diagram. In Sec.~2.3 we discuss the ultraviolet (UV) divergence
structure of sunrise-type diagrams and present recipes to regularize the UV
divergences either by subtraction or by dimensional regularization. In
Sec.~2.4 we comment on the spectral density of sunrise-type diagrams and its
connection to phase space. Sec.~3 is devoted to some explicit configuration
space calculations involving both analytical and numerical methods the results
of which are compared to previously known results where other calculational
techniques have been used. One of the examples is the direct computation of
the spectral density of sunrise-type diagrams without taking recourse to
Fourier transforms. The results are important for multi-body phase space
calculations. In Sec.~4 we discuss methods to find asymptotic expansions in
different kinematical regimes of mass and/or momentum configurations in the
Euclidean domain. Sec.~5 contains a generalization to non-standard propagators
and non-scalar cases. In Sec.~6 we generalize the configuration space
technique to other topologies. Sec.~7 contains our conclusions. Some of the
lengthier formulas are relegated to the appendices, together with useful
mathematical material about Bessel functions and Gegenbauer polynomials and
a short treatise about cuts and discontinuities as they are occuring in the
main text.

\section{Basic notions and relations for sunrise-type diagrams}

\begin{figure}[t]\begin{center}
\epsfig{figure=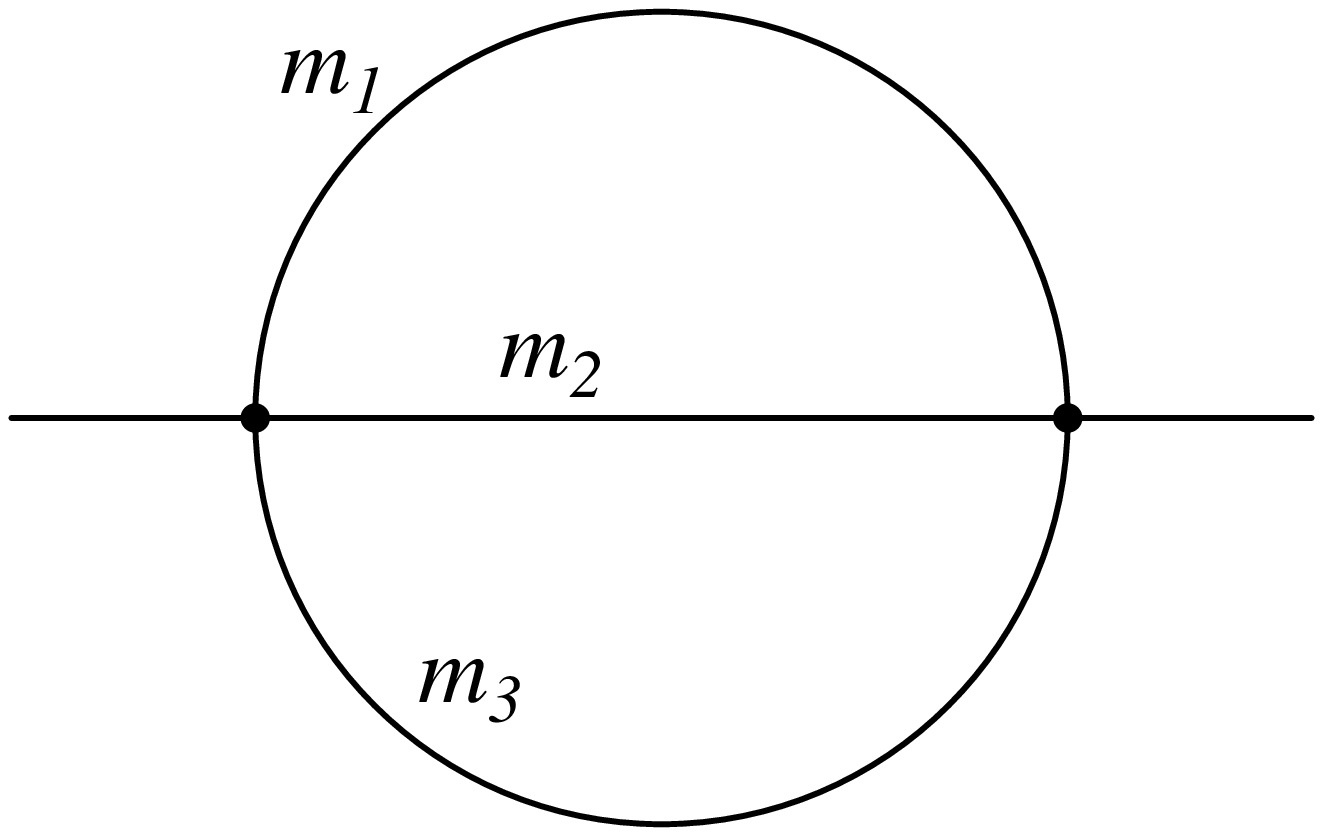, scale=0.3}\lower0pt\hbox{(a)}\qquad
\epsfig{figure=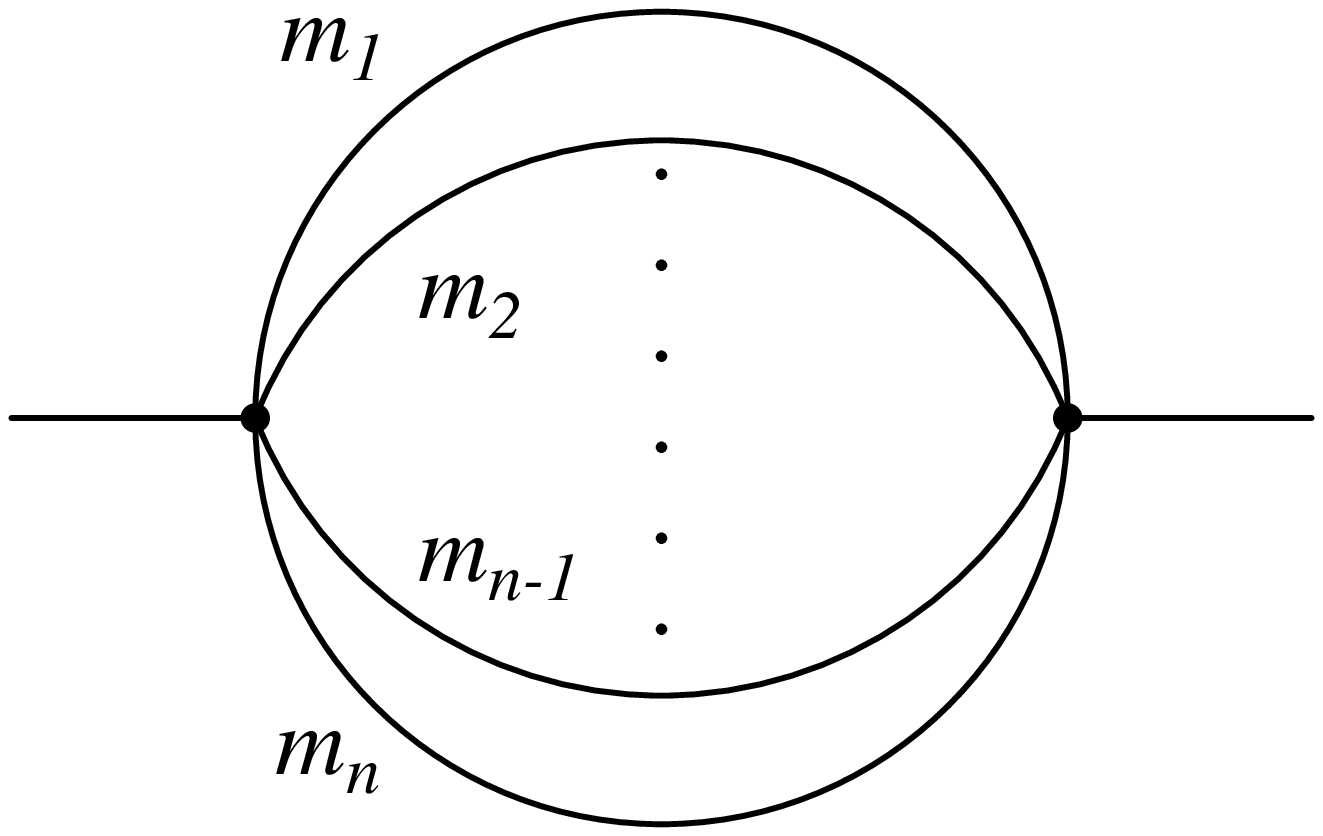, scale=0.3}\lower0pt\hbox{(b)}
\caption{\label{sundi}Genuine sunrise (a) and general topology
of the class of sunrise-type diagrams (b)}
\end{center}\end{figure}

Sunrise-type diagrams are graphic representations of the $n$-loop two-point
correlation functions in QFT with $(n+1)$ internal propagators connecting the
initial and final vertex. The well-studied genuine sunrise diagram shown in
Fig.~\ref{sundi}(a) is the leading order perturbative correction to the lowest
order propagator in $\phi^4$-theory, i.e.\ it is a two-point two-loop diagram
with three internal lines. This diagram emerges as a correction to the Higgs
boson propagation in the Standard Model. It also naturally appears in some
effective theory approaches to critical phenomena and studies of phase
transitions in QFT. The corresponding leading order perturbative correction in
$\phi^3$-theory is a one-loop diagram which can be considered as a
oversimplified case of the prior example.
  
The two-loop case is a standard starting point for the calculation of
radiative corrections in QFT. It emerges in a huge variety of physical
applications. Fortunately it can be analytically computed in any desired
kinematical regime for any values of the relevant parameters. In this respect
it does not present a real challenge as far as multi-loop calculations are
concerned. We shall, nevertheless, discuss the genuine sunrise diagram in some
detail in order to illustrate the efficiency of our configuration space
methods. We shall compare our results with known exact analytical results
obtained with other techniques. This provides a mutual check on the
correctness of the results. A straightforward generalization of this topology
to the multi-loop case is a correction to the free propagator in
$\phi^{n+2}$-theory that contains $n$ loops and $(n+1)$ internal lines (see
Fig.~\ref{sundi}(b)).

A subclass of the general sunrise-type diagram shown in Fig.~\ref{sundi}(b) is
the case when the two external momenta vanish. This subclass is referred to as
the subclass of vacuum bubble diagrams with sunrise-type topology which will
be referred to as vacuum bubbles. They have a simpler structure than the true
sunrise-type diagrams since the number of mass scales is reduced by one for
the vacuum bubbles. We shall frequently return to the discussion of vacuum
bubbles in the main text.

\subsection{Definitions and notation}

A $n$-loop sunrise-type diagram is a two-point correlation function with
$(n+1)$ propagators connecting points $x$ and $y$ and as such is explicitly
given by a product of simple propagators built from the basic two-point
correlation function $D(x,y,m)$,
\begin{equation}
\Pi(x,y)=\prod_{i=1}^{n+1}D(x,y,m_i)
\end{equation}
(and/or their derivatives if necessary). The quantity $D(x,y,m)$ may be the
propagator of a free massive particle with mass $m$ or a more general
two-point correlation function. The sunrise-type diagrams are the leading
order contribution to a two-point correlation function of two local currents
$j_{n+1}(x)$ of the form 
\begin{equation}
j_{n+1}(x)={\cal D}_{\mu_1}\phi_1\cdots {\cal D}_{\mu_{n+1}}\phi_{n+1}  
\end{equation}
where the fields $\phi_i$ have masses $m_i$ and where ${\cal D}_\mu$ is a 
derivative with multi-index $\mu\!=\!\{\mu_1,\ldots,\mu_k\}\!$ standing for 
${\cal D}_\mu=\partial^k/\partial x_{\mu_1}\ldots\partial x_{\mu_k}$. The 
sunrise-type diagrams are contained in the leading order expression for the 
polarization function
\begin{equation}
\Pi(x,y)=\langle Tj_{n+1}(x)j_{n'+1}(y)\rangle 
\end{equation}
where the brackets mean quantum mechanical averaging over the ground state
which is explicitly given by a product of propagators and/or their derivatives,
\begin{equation}
\Pi(x,y)={\cal D}_{\mu_1\nu_1}(x,y,m_1)\cdots
  {\cal D}_{\mu_{n+1}\nu_{n+1}}(x,y,m_{n+1}).
\end{equation}
As the standard case we will consider a translation invariant situation in
which the propagator depends on the difference of the arguments only,
$D(x,y,m)=D(x-y,m)$. An exception to this important standard case would be the 
existence of an arbitrary external field. However, in the present report we 
will not further discuss this possibility. In the standard case one of the 
vertices can conveniently be placed at the origin, say $y=0$. The propagator
is then a function of $x$ only, $D(x,m)$. The basic expression for the
sunrise-type diagram in configuration space reads
\begin{equation}\label{xsun}
\Pi(x)=\prod_{i=1}^{n+1}D(x,m_i).
\end{equation}
$D(x,m)$ represents a free propagator of a massive particle with mass $m$ in
$D$-dimensional (Euclidean) space-time. It is given by
\begin{equation}\label{xprop}
D(x,m)=\frac1{(2\pi)^D}\int\frac{e^{i(p\cdot x)}d^Dp}{p^2+m^2}
  =\frac{(mx)^\lambda K_\lambda(mx)}{(2\pi)^{\lambda+1}x^{2\lambda}}
\end{equation}
where we have used $D=2\lambda+2$. $K_\lambda(z)$ is the modified Bessel
function of the second kind (sometimes also known as McDonald function, see
Appendix~\ref{sec_bessel}) defined by Eq.~(\ref{defK}).

The massless propagator can be obtained from Eq.~(\ref{xprop}) by taking the
limit $m\to 0$ (or more precisely the limit $mx\to0$ at fixed $x$ in terms of
the dimensionless quantity $mx$). It reads
\begin{equation}\label{xprop0}
D(x,0)=\frac1{(2\pi)^D}\int\frac{e^{i(p\cdot x)}d^Dp}{p^2}
  =\frac{\Gamma(\lambda)}{4\pi^{\lambda+1}x^{2\lambda}}.
\end{equation}
and where $\Gamma(\lambda)$ is Euler's Gamma function.

In some physics applications one may have more general basic two-point
functions (``modified  propagators'') for massive particles. An example is the
calculation of the three-point function necessary for the determination of a
particle form factor at zero momentum transfer when the external momentum of
the current vanishes. This momentum configuration reduces the diagram to a
sunrise-type diagram with more complicated basic two-point lines (see
e.g.~\cite{krasulinForm}). This phenomenon frequently occurs also in
calculations of various quantities in Chiral Perturbation Theory
(ChPT)~\cite{Gasser:2002am}. Formally, these two-point functions cannot be
considered as propagators of a physical particle but emerge as effective basic
two-point functions. The simplest modification of the basic propagator which
frequently appears in applications is the occurrence of higher powers of the
standard propagator. In momentum space it has the form
\begin{equation}
\tilde D^{(\mu)}(p,m)=\frac1{(p^2+m^2)^{\mu+1}}.
\end{equation}
For integer $\mu$ these cases appear if one e.g.\ considers mass derivatives
of the propagator as they are needed in certain parameter expansions. For
instance, one can consider 
\begin{equation}
\tilde D'(p,m)=-\frac{\partial}{\partial m^2}\frac1{(p^2+m^2)}
\end{equation}
In the same way a derivative of the momentum itself will also increase the
power of the propagator. For instance, the second order scalar derivative 
$ \dalembertian = \partial_\mu \partial^\mu $ of the standard propagator will
result in
\begin{equation}
\partial_\mu\partial^\mu\frac1{p^2+m^2}
=\frac{-2}{(p^2+m^2)^2}+\frac{8p^2}{(p^2+m^2)^3}.
\end{equation}
Higher powers of propagators occur in many applications. Therefore, any
calculational technique should be well suited to treat such higher powers. It
turns out that the configuration space method very naturally accommodates
higher order derivatives.

In configuration space higher powers of the standard propagator with mass $m$
are explicitly given by (cf.\ Eq.~(\ref{xprop}))
\begin{equation}\label{primeD}
\tilde D^{(\mu)}(x,m)=
\frac1{(2\pi)^D}\int\frac{e^{i(p\cdot x)}d^Dp}{(p^2+m^2)^{\mu+1}}
  =\frac1{(2\pi)^{\lambda+1}2^\mu\Gamma(\mu+1)}{\left(\frac mx
  \right)^{\lambda-\mu}K_{\lambda-\mu}(mx)}.
\end{equation}
the additional power of denominator factors (or, equivalently, the number of
derivatives in masses) is denoted in the corresponding Feynman diagram by dots
placed on the line (though even the power can be non-integer). Note that the
modified propagator has the same functional form as the usual propagator. The
main difference is the change of the value of the index of the Bessel function
$K_{\lambda}(mx)\to K_{\lambda-\mu}(mx)$. This fact unifies and simplifies the
analysis of sunrise-type diagrams that contain more complicated basic
propagators (analytical expressions that correspond to one line of the
diagrams) as they emerge in some physical applications or are used for the
calculation of other diagrams through recurrence relations
techniques~\cite{Avdeev:1995eu}.

\subsection{Momentum versus configuration space representation:\\
  Fourier transform in the evaluation of sunrise--type integrals}

The calculation of loop diagrams in perturbative Quantum Field Theory is
associated with the evaluation of complicated multi-dimensional integrals,
see e.g.~\cite{Itzykson:1980rh}. It is obvious that the expression for the
sunrise-type diagram in configuration space given by Eq.~(\ref{xsun})
contains no integration at all even if it represents a multi-loop diagram. In
this respect it is a kind of tree-level diagram that usually appears in the
lowest orders of perturbative QFT. However, ``to calculate'' a sunrise diagram
means in most cases to find certain integrals of Eq.~(\ref{xsun}) over $x$
with some weight functions depending on the quantity in question. The simplest
situation corresponds to a computation of the values of the vacuum bubbles,
i.e.\ expressions for the diagrams without external momenta (or with vanishing
external momenta, see e.g.~\cite{Broadhurst:1998rz}). In case of the 
sunrise-type topology the value of the vacuum bubble is given by integrals
with a simple weight function,
\begin{equation}\label{bubble}
\tilde\Pi(0)=\int\Pi(x)d^Dx=\int D(x,m_1)\cdots D(x,m_{n+1})d^Dx.
\end{equation}
This integral provides the simplest example of calculating a sunrise-type
diagram. We like to emphasize that this calculation is almost always a must as
it allows one to determine the counterterms related to UV divergences of
sunrise-type diagrams that always occur in realistic physical models. Due to
rotation invariance the integral is one-dimensional as the angular integration
can be explicitly done. Thus
\begin{eqnarray}
\tilde\Pi(0)&=&\int\Pi(x)d^Dx=\int D(x,m_1)\cdots D(x,m_{n+1})d^Dx\nonumber\\
  &=&\frac{2\pi^{\lambda+1}}{\Gamma(\lambda+1)}\int D(x,m_1)\cdots
  D(x,m_{n+1})x^{2\lambda+1}dx.
\end{eqnarray}
The possibility to compute all vacuum bubbles in terms of one-dimensional
integrals for any number of loops is a big advantage of the  configuration
space technique. This guarantees that the complete renormalization program can
be explicitly accomplished in terms of one-dimensional integrals. It should be
clear that this is a huge simplification.

The most familiar example for the evaluation of a sunrise-type diagram is the
procedure of finding its value in momentum space, or put differently, the
evaluation of its Fourier transform. It corresponds to non-bubble diagrams
with two external lines as the integrals then requires the computation of the
Fourier transform of $\Pi(x)$,
\begin{equation}
\tilde\Pi(p)=\int\Pi(x)e^{i(p\cdot x)}d^Dx.
\end{equation}
The required integrals are basically scalar integrals. This makes the angular
integration in Eq.~(\ref{bubble}) simple in $D$-dimensional space-time. The 
needed formula is the integration over the angular variables $d^D\hat x$
which, in explicit form, reads
\begin{equation}\label{angint}
\int d^D\hat xe^{i(p\cdot x)}=2\pi^{\lambda+1}
  \left(\frac{px}2\right)^{-\lambda}J_\lambda(px)
\end{equation}
where $p=|p|$, $x=|x|$, and $J_\lambda(z)$ is the Bessel function of the first
kind. The integration $\int d^D\hat x$ over the rotationally invariant measure
$d^D\hat x$ on the unit sphere in $D$-dimensional (Euclidean) space-time means
integration over angles only. The generalization of Eq.~(\ref{angint}) to more
complicated integrands with an additional tensor structure
$x^{\mu_1}\cdots x^{\mu_k}$ is straightforward and leads to different orders
$J_{\lambda+l}(z)$ of the Bessel function for different irreducible tensors
(denoted by the number $l$) after angular averaging (integration over angles).
The corresponding order of the Bessel function can be inferred from the
expansion of the plane wave function $e^{i(p\cdot x)}$ which appears as a
weight function upon computing the Fourier transform in a series of Gegenbauer
polynomials $C^\lambda_l(w)$ (see Appendix~\ref{sec_gegenbauer}). The
Gegenbauer polynomials are orthogonal on a $D$-dimensional unit sphere, and
the expansion of the plane wave in terms of Gegenbauer polynomials
reads~\cite{Chetyrkin:pr}
\begin{equation}\label{gegen}
e^{i(p\cdot x)}=\Gamma(\lambda)\left(\frac{px}2\right)^{-\lambda}
  \sum_{l=0}^{\infty} i^l (\lambda+l)J_{\lambda+l}(px)
  C^\lambda_l\left(\frac{(p\cdot x)}{px}\right).
\end{equation}
Besides the definition of the Gegenbauer polynomials by a characteristic
polynomial as given in Eq.~(\ref{gegengen}), the expansion in
Eq.~(\ref{gegen}) can also serve as a definition of these polynomials.

The representation in Eq.~(\ref{gegen}) allows one to single out an
irreducible tensorial structure from the angular integration given by the
Fourier integral in Eq.~(\ref{bubble}). Integration techniques involving
Gegenbauer polynomials for the computation of multi-loop massless diagrams 
are described in detail in~\cite{Chetyrkin:pr} where many useful relations can
be found (see also~\cite{Terrano:1980af} where the calculation in momentum
space was considered). One arrives at a representation of the Fourier
transform of a sunrise-type diagram that is given in terms of the
one-dimensional integral
\begin{eqnarray}\label{psun}
\tilde\Pi(p)&=&2\pi^{\lambda+1}\int_0^\infty\left(\frac{px}2\right)^{-\lambda}
  J_\lambda(px)\Pi(x)x^{2\lambda+1}dx\nonumber\\
  &=&2\pi^{\lambda+1}\int_0^\infty\left(\frac{px}2\right)^{-\lambda}
  J_\lambda(px)D(x,m_1)\cdots D(x,m_{n+1})x^{2\lambda+1}dx
\end{eqnarray}
which is simple to analyze. The representation given by Eq.~(\ref{psun}) is
quite universal.

\subsection{Regularization and renormalization}

While the configuration space expression for the sunrise-type diagram contains
no integration it does not always represent an integrable function of $x$ in
configuration space for general values of the space-time dimension $D$. The
expression given by Eq.~(\ref{xsun}) can have non-integrable singularities at
small $x$ for a sufficiently large number of propagators if the space-time 
dimension is $D>2$.

Upon multiplication of many propagators one may get a function which is 
singular at small $x$. Each propagator by itself can be considered as a
distribution. However, distributions do not form an algebra and their
multiplication is not well-defined~\cite{Bogoliubov:1957gp}. In this respect
the polarization function $\Pi(x)$ in Eq.~(\ref{xsun}) is not completely
defined as a proper distribution. In attempting to integrate such a function
over the whole $x$-space we encounter infinities in the form of ultraviolet
(UV) divergences~\cite{Bogoliubov}. Therefore, the computation of the Fourier
transform of $\Pi(x)$ requires regularization (for instance, dimensional
regularization) and subtraction. Configuration space provides a nice
environment for this.

Note that all these statements depend on the dimension of space-time. The
vicinity of $D=2$ is special since the leading singularity of the basic
propagator is only logarithmic and the multiplication of propagators does not
lead to non-integrable singularities. Therefore, the UV properties of
two-dimensional models that are often considered as simplified models in QFT
or that emerge in real physical applications as a result of some given
approximations~\cite{Lipatov:1995pn,Lipatov:1998wd,Lipatov:1998mb}. They are
much softer than those of realistic four-dimensional models (see
e.g.~\cite{Krasnikov:1991hz}).

In general, however, one has to consider an arbitrary space-time dimension
$D$. The modern way of dealing with singularities in QFT in particular in
multi-loop calculations is mostly based on dimensional 
regularization~\cite{'tHooft:1972fi,'tHooft:1973us,'tHooft:1973mm} (as a
review see~\cite{Chetyrkin:1996ia}). Dimensional regularization is
characterized by considering the space-time dimension $D$ to be an arbitrary
parameter, for instance $D=D_0-2\eps$ where $D_0$ is an integer
dimension.\footnote{The usual choice of the integer space-time dimension $D_0$
for real physical applications is $D_0=4$, but other space-time dimensions
$D_0=2$, $D_0=3$, $D_0=5$ or $D_0=6$ are being used in other applications.
$D_0=3$ is natural for near threshold expansions or expansions in the
Minkowskian domain in general.} In case of non-integer space-time dimension
$D$ the integration is well-defined, while the limit $D\to D_0$ is singular,
consisting usually of poles in $D-D_0$. These pole parts are extremely simple
to handle in configuration space. Indeed, in configuration space the
regularization is given by adding the $\delta$-function $\delta(x)$ and its
derivatives to $\Pi(x)$. As we will see in the following, this procedure
corresponds to adding a polynomial in momentum space (local counterterms
according to the Bogoliubov-Parasiuk-Hepp-Zimmermann (BPHZ)
theorem~\cite{Bogoliubov:1957gp,Hepp:1966eg,Lowenstein:1974uk,Anikin:1973ra}).
The finite parts can be found numerically for any number of loops by doing a
single one-dimensional integration. Therefore, the renormalization in
$x$-space becomes
\begin{equation}\label{renormC}
\Pi^R(x)=\Pi(x)+C_0\delta(x)+C_1\dalembertian\delta(x)+\ldots
+C_r\dalembertian^r\delta(x)
\end{equation}
where $\dalembertian=\partial_\mu\partial^\mu$. The coefficients $C_i$ are
functions of the regularization parameter $\eps$ which are singular in the
limit $\eps\to 0$. The extraction of the coefficients $C_i$ is explained in
the subsequent sections for some special cases. The number $r$ of counterterms
is determined by dimensional arguments such as power counting and can be
easily related to the number of propagators and the space-time dimension.
The actual number of poles depends on the mass configuration. 

\subsection{The spectrum as discontinuity across the physical cut}

From the physics point of view one important aspect of our analysis of
sunrise-type diagrams is the construction of the spectral decomposition of the
diagrams. For the two-point correlation function we determine the
discontinuity across the physical cut in the complex plane of the squared
momentum, $p^2=-m^2\pm i0$ which is referred to as the spectral density
\[\rho(s)=\frac1{2\pi i}\Disc\tilde\Pi(p)\Big|_{p^2=-s}\]
(cf.\ Appendix~\ref{sec_disc}). Note that the spectral density of a
sunrise-type diagram is finite for any number of loops. It turns out that the
configuration space technique allows one to compute the spectral density in a
very efficient manner. The analytic structure of the correlator $\Pi(x)$ (or
the spectral density of the corresponding polarization operator) can be
determined directly in configuration space without having to compute its
Fourier transform first. The technique for the direct construction of the
spectral density of sunrise-type diagrams introduced in~\cite{Groote:1998ic}
is based on an integral transform in configuration space which in turn is
given by the inversion of the relevant dispersion relation.

The dispersion representation (or spectral decomposition) of the polarization
function in configuration space has the form
\begin{equation}\label{disprel}
\Pi(x)=\int_0^\infty\rho(s)D(x,\sqrt{s})ds=\int_0^\infty\rho(m^2)D(x,m)dm^2
\end{equation}
where $\sqrt s=m$. This representation was used for sum rule applications
in~\cite{Pivovarov:ij,Chetyrkin:yr} where the spectral density for the two-loop
sunrise diagram was given in two-dimensional space-time~\cite{Pivovarov:jm}.
The representation in momentum space is more familiar and is referred to as
the K\"all\'en-Lehmann representation~\cite{Lehmann} of the two-point
correlation function. In the Euclidean domain it is given by
\[
\tilde\Pi(p)=\int_0^\infty\frac{\rho(s)ds}{s+p^2}.
\]
This expression can of course be obtained by taking the Fourier transform of
both sides of Eq.~(\ref{disprel}). 

With the explicit form of the propagator in configuration space given by
Eq.~(\ref{xprop}), the representation in Eq.~(\ref{disprel}) turns out to be a
particular example of the Hankel transform, namely the
$K$-transform~\cite{Meijer,Erdelyi}. Up to inessential factors of $x$ and $m$,
Eq.~(\ref{disprel}) reduces to the generic form of the $K$-transform for a
conjugate pair of functions $f$ and $g$,
\begin{equation}
g(x)=\int_0^\infty f(y)K_\nu(xy)\sqrt{xy}\,dy.
\end{equation}
The inverse of this transform is known to be given by
\begin{equation}
f(y)=\frac1{\pi i}\int_{c-i\infty}^{c+i\infty}g(x)I_\nu(xy)\sqrt{xy}\,dx
\end{equation}
where $I_\nu(x)$ is a modified Bessel function of the first kind and the
integration runs along a vertical contour in the complex plane to the right
of the right-most singularity of the function $g(x)$~\cite{Erdelyi}. In
order to obtain a representation for the spectral density $\rho(m^2)$ of a
sunrise-type diagram in general $D$-dimensional space-time one needs to
apply the inverse $K$-transform to the particular case given by
Eq.~(\ref{disprel}). One has
\begin{equation}\label{invK}
m^\lambda\rho(m^2)=\frac{(2\pi)^\lambda}{i}
  \int_{c-i\infty}^{c+i\infty}\Pi(x)x^{\lambda+1}I_\lambda(mx)dx.
\end{equation}
From Eq.~(\ref{invK}) we obtain an explicit analytical representation for the
spectral density $\rho(s)$ as a contour integral of the polarization function
which reads
\begin{equation}\label{exactrho}
\rho(s)=\frac{(2\pi)^\lambda}{is^{\lambda/2}}\int_{c-i\infty}^{c+i\infty}
  I_\lambda(x\sqrt s)\Pi(x)x^{\lambda+1}dx
\end{equation}
where $I_\lambda(z)$ is a modified Bessel function of the first kind and the
integration runs along a vertical contour in the complex plane to the right
of the right-most singularity of $\Pi(x)$.

The inverse transform given by Eq.~(\ref{exactrho}) completely solves the
problem of determining the spectral density $\rho(s)$ of the general class of
sunrise-type diagrams by reducing it to the computation of a one-dimensional
integral along the contour in a complex plane. This is valid for the general
class of sunrise-type diagrams with any number of internal lines and different
masses. For $(n+1)$ internal lines with $(n+1)$ equal masses $m$ 
the spectral
density reads 
\begin{equation}
\rho(s)=\frac{m^{\lambda(n+1)}}{i(2\pi)^{n\lambda+n+1}s^{\lambda/2}}
  \int_{c-i\infty}^{c+i\infty}I_\lambda(x\sqrt s)
  \left(K_\lambda(mx)\right)^{n+1}x^{1-n\lambda}dx.
\end{equation}
Because the contour can bypass the area of small values of $x$, the integral
is finite, as it was stated before. The spectral density in turn can be used
to restore the finite part of the correlation function in the momentum space
representation. The path traced here is is an alternative to the calculation
of the Fourier transform for sunrise-type diagrams.

In the standard, or momentum representation, the polarization function
$\tilde\Pi(p)$ is calculated from a $n$-loop diagram with $n$ $D$-dimensional
integrations over the entangled loop momenta. It is clear that the momentum 
space evaluation becomes very difficult when the number of internal lines 
becomes large.

\begin{figure}[ht]\begin{center}
\epsfig{figure=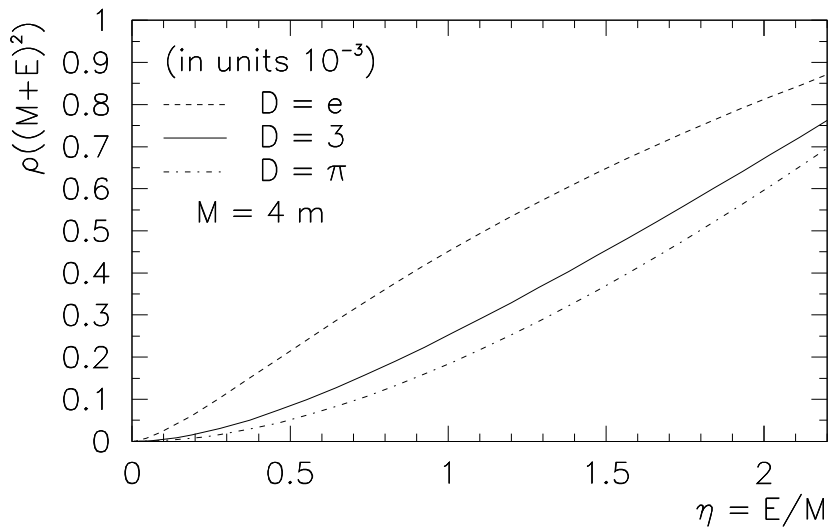,scale=0.7}
\caption{\label{fudas234}The spectral density for the four-line sunrise-type
diagram with equal masses for $D=e=2.718\ldots$, $D=3$, and $D=\pi=3.14\ldots$
space-time dimensions.}
\end{center}\end{figure}

In order to demonstrate the applicability of our method described in more
detail in~\cite{Groote:1998ic,Groote:1998wy,Groote:1999cx}, we show in
Fig.~\ref{fudas234} the result for the spectral density for the four-line
sunrise-type diagram with $D=e=2.718\ldots$, $D=3$, and $D=\pi=3.14\ldots$. We
have chosen these exotic values for $D$ to demonstrate the power of
configuration space techniques. To the best of our knowledge there is no 
other method which allows one to compute the spectral density with such 
completeness and ease, i.e.\ through a one-dimensional integral. The technique
was first suggested in~\cite{Groote:1998ic} and has been studied in detail in
Refs.~\cite{Bashir:2001ad,Delbourgo:2003zi}.

Note, finally, that the spectral density of a sunrise-type diagram is a
representation of multi-particle phase space with the number of particles
equal to the number of lines of the diagram. Our formula for the spectral
density is a striking example of how the configuration-space technique can be
used to determine the phase space of physical processes. Efficient techniques
for the calculation of the multi-particle phase space are important for
Monte-Carlo simulations and cross section calculations for multi-particle
final states as they will occur at the upcoming Large Hadron Collider (LHC)
under construction at CERN. 

\section{``Tool box'' for a practical analysis}

In this section we give examples and practical prescriptions of how to proceed
with the calculation of sunrise-type diagrams containing only a single Bessel
function and powers of $x$. The massless case can be solved analytically in 
closed form for any number $(n+1)$ of internal lines since it contains only a
single Bessel function and powers of $x$. The case for one massive line and
$n$ massless lines (with two Bessel functions) is completely solvable
analytically as well. The first nontrivial situation emerges with two massive
lines. In this case a complete analytical solution is unknown in most of the
cases, the exception being the case of equal masses that can be integrated
analytical through hypergeometric functions. However, when one separates the
diagram into a singular or pole part and a finite part, both parts can be
calculated either analytically (for the pole part) or numerically (for the
finite part) in very general situations. This will be demonstrated in this
section.

\subsection{Extraction of poles in dimensional regularization}

Renormalization counterterms can be constructed analytically for general mass
configurations and any number of loops. As mentioned earlier, in dimensional
regularization the UV divergences become manifest as poles in $\eps\sim D-D_0$.
In configuration space the UV divergences are related to short distances. 
In order to analyze the structure of singularities at small $x$, one has to 
expand the massive propagators at small $x$ (or, effectively, in the masses).
In order to obtain the requisite conterterms one then integrates over $x$
which is analytically feasible for $D\ne D_0$, i.e.\ $\eps\ne 0$.
Nevertheless, one should ensure convergence also at large $x$ which requires
an appropriate IR regularization. Only this guarantees that singularities that
emerge after integration are related to small $x$ divergences only (see
e.g.~\cite{Pivovarov:re}). This technical constraint can be met for instance
by keeping one (or two) of the massive propagators in the integrand
unexpanded. Since the massive propagator falls off exponentially at large $x$,
this procedure is sufficient to ensure convergence at large $x$.

The possibility of retaining one propagator unexpanded exists because the
integrals of one or a product of two Bessel functions are known
analytically~\cite{Watson,Prudnikov,Gradshteyn}. This nice feature provides a
tool for obtaining counterterms analytically. The necessary formulae read
\begin{eqnarray}
\int_0^\infty x^{\mu-1}K_\nu(mx)dx
  &=&2^{\mu-2}m^{-\mu}\Gamma\left(\frac{\mu+\nu}2\right)
  \Gamma\left(\frac{\mu-\nu}2\right)\\
\int_0^\infty x^{2\alpha-1}K_\mu(mx)K_\mu(m x)dx
  &=&\frac{2^{2\alpha-3}}{m^{2\alpha}\Gamma(2\alpha)}
  \Gamma(\alpha+\mu)\Gamma(\alpha)\Gamma(\alpha)\Gamma(\alpha-\mu).
\label{KKint}
\end{eqnarray}
In general, the finite parts of the relevant diagrams resulting from the
subtraction of counterterms can be calculated only numerically. Still there
are some simple examples where an analytic calculation is feasible. One of
these examples is the three-loop sunrise-type diagram with two massive 
and two massless lines at vanishing external momentum, the three-loop vacuum
bubble. There exist an analytical expression for the value of this particular
vacuum bubble in configuration space. It is given by
\begin{equation}\label{mm00bub}
\tilde\Pi(0)=\int D(x,m)^2D(x,0)^2d^Dx=\int
  \left(\frac{(mx)^\lambda K_\lambda(mx)}{(2\pi)^{\lambda+1}x^{2\lambda}}
  \right)^2\left(\frac{\Gamma(\lambda)}{4\pi^{\lambda+1}x^{2\lambda}}
  \right)^2 d^Dx.
\end{equation}
While the angular integration in $D$-dimensional space-time is trivial, the
problem of the residual radial integration is solved by using
Eq.~(\ref{KKint}). The result for the integral in Eq.~(\ref{mm00bub}) reads
\begin{equation}\label{mm00res}
\tilde\Pi(0)=\frac{(m^2)^{3\lambda-1}}{(4\pi)^{3(\lambda+1)}}\
\frac{\Gamma(\lambda)^2\Gamma(1-\lambda)\Gamma(1-2\lambda)^2
  \Gamma(1-3\lambda)}{\Gamma(\lambda+1)\Gamma(2-4\lambda)}.
\end{equation}
This result corresponds to the quantity $M_1$ in Ref.~\cite{Broadhurst:1991fi}
where it constitutes the simplest basis element for the computation of massive
three-loop diagrams in a general three-loop topology. Again, any number of
massless lines can be added.

The result given by Eq.~(\ref{mm00res}) is obtained in a concise form and
valid for any dimension $D$. Therefore its pole parts can be explicitly
singled out through a direct expansion into a Laurent series near an integer
value of the space-time dimension $D$. Taking $D=4-2\eps$ (i.e.\
$\lambda=1-\eps$) and expanding the result for small $\eps$ one readily finds
\[
\tilde\Pi(0)=\frac{(m^2)^{2-3\eps}}{3(4\pi)^{3(2-\eps)}}
  \left(\frac1{\eps^3}+\frac7{2\eps^2}+\frac{25+6\zeta(2)}{4\eps}
  -\frac{5-42\zeta(2)-56\zeta(3)}8+O(\eps)\right).
\]
Using this example one can take a closer look at the two-dimensional case. As
was stated before, the case $D_0=2$ is much softer and the UV divergences are
practically absent. Hence one could come to the conclusion that the expression
given by Eq.~(\ref{mm00res}) should be perfectly finite at $D=D_0=2$. In terms
of $D=2\lambda+2$ the limit corresponds to taking $\lambda\to 0$. Therefore,
one might be surprised to find a singularity $\Gamma(\lambda)^2$ in
Eq.~(\ref{mm00res}). However, this is not a UV singularity. Because massless
propagators are not defined in the two-dimensional world, the limit
$\lambda\to 0$ is ill-defined in this case. Indeed, looking at
Eq.~(\ref{xprop0}), the singularity of this sunrise-type diagram is exactly
the product of two singular massless propagators. Power counting for the
integral expression in Eq.~(\ref{xprop0}) shows that the singularity is caused
by the infrared (IR) region. As this is an IR problem, the afore mentioned
BPHZ theorem does not apply and the divergence cannot be treated with local
counterterms in the usual manner.

Returning to the more general case, a closed-form evaluation (as in
Eq.~(\ref{mm00res})) is not possible and an explicit subtraction is required
to separate singular and finite parts. In our case it is convenient to use
momentum subtraction which in fact is the oldest renormalization method.
Momentum subtraction consists of subtracting a polynomial at some fixed
momentum point. For massive diagrams the expansion point $p=0$ is safe (i.e.\
has no infrared (IR) singularity), and the prescription is realized by
expanding the function
\begin{equation}
\left(\frac{px}2\right)^{-\lambda}J_\lambda(px)
\end{equation}
(which is the kernel or weight function of the integral transformation in
Eq.~(\ref{psun})) in a Taylor series around $p=0$ in terms of a polynomial
series in $p^2$ (cf.\ Eq.~(\ref{serJfac}) in Appendix~\ref{sec_bessel}). The
subtraction at order $N$ is achieved by writing
\begin{equation}\label{subN}
\left[\left(\frac{px}2\right)^{-\lambda}J_\lambda(px)\right]_N
  =\left(\frac{px}2\right)^{-\lambda}J_\lambda(px)
  -\sum_{k=0}^N\frac{(-1)^k}{k!\Gamma(\lambda+k+1)}
  \left(\frac{px}2\right)^{2k}
\end{equation}
and by keeping $N$ terms in the expansion on the right hand side. After
substituting the expansion Eq.~(\ref{subN}) into Eq.~(\ref{psun}) one
obtains the momentum subtracted polarization function
\begin{equation}\label{psunsub}
\tilde\Pi_{\rm mom}(p)=\tilde\Pi(p)-\sum_{k=0}^N
  \frac{p^{2k}}{k!}\left(\frac{d}{d p^2}\right)^k\tilde\Pi(p)|_{p^2=0}
\end{equation}
which is finite if the number of subtractions $N$ is sufficiently high. The
function $\tilde\Pi(p)$ is divergent as well as any of the derivatives on the 
right hand side of Eq.~(\ref{psunsub}). The divergences require
regularization. However, the difference, i.e.\ the quantity
$\tilde\Pi_{\rm mom}(p)$ is finite and independent of any regularization used
to give a meaning to each individual term in Eq.~(\ref{psunsub}). Note that
the expansion in Eq.~(\ref{subN}) is a polynomial in $p^2$ in accordance with
the general structure of the $R$-operation~\cite{Bogoliubov}. The number $N$
of necessary subtractions is determined by the divergence index of the diagram
and can be found according to the standard power counting rules. The
subtraction at the origin $p=0$ is allowed if there is at least one massive
line in the diagram along with an arbitrary number of massless lines. If there
are no massive internal lines at all, the corresponding diagram can easily be
calculated analytically and the problem of subtraction is trivial. After
having performed the requisite subtraction, one can take the limit
$D\rightarrow D_0$ in Eq.~(\ref{psun}) where $D_0$ is an integer. The diagram
as a whole becomes finite after subtraction which reflects the topology of the
sunrise-type diagram: there is no divergent subdiagram in the sense of
Bogoliubov.

In selecting the momentum subtraction in order to regularize the singularities
we have selected a particular subtraction or renormalization scheme. The
coefficients $C_i$ in Eq.~(\ref{renormC}) depend on this choice. While the
highest coefficient related to the strongest singularity is unique, the other
coefficients are scheme dependent. The standard scheme used in the literature
is the $\overline{\rm MS}$ scheme. The transition from the momentum
subtraction scheme to the $\overline{\rm MS}$ scheme is achieved by the
familiar redefinition of the measure
\begin{equation}
d^Dx\to(4\pi\mu^2e^{-\gamma_E})^\eps d^Dx
\end{equation}
with $\mu$ being the renormalization scale within dimensional regularization
and $\gamma_E$ being Euler's constant $\gamma_E=0.577\dots$

\subsection{Singular and finite parts for particular parameter values of
 the genuine sunrise: warm-up example demonstrating the technique}

As an example of calculating the counterterms and evaluating the remaining
finite part of the diagram we present the special case of the genuine sunrise
diagram with three massive lines, i.e.\ the two-loop sunrise-type diagram. 
For special values of the external momenta a closed-form expression for the
genuine sunrise diagram is known within dimensional regularization using
momentum space techniques~\cite{Berends:1997vk}. We shall reproduce this
result in the following. The one-dimensional integral to be analyzed is given
by
\begin{equation}
\tilde\Pi(p)=2\pi^{\lambda+1}\int_0^\infty\left(\frac{px}2\right)^{-\lambda}
  J_\lambda(px)D(x,m_1)D(x,m_2)D(x,m_3)x^{2\lambda+1}dx.
\end{equation}
To determine the finite part we first use momentum subtraction and split
$\tilde\Pi(p)$ up into its finite and infinite (but dimensionally
regularized) part,
\begin{equation}
\tilde\Pi(p)=\tilde\Pi_{\rm mom}(p)+\tilde\Pi_{\rm sing}(p)
\end{equation}
where $\tilde\Pi_{\rm mom}(p)$ is a momentum subtracted polarization function
and $\tilde\Pi_{\rm sing}(p)$ is a counterterm in dimensional regularization.
Power counting shows that only two subtractions are necessary. The explicit
expression for the momentum subtracted polarization function reads
\begin{eqnarray}
\tilde\Pi_{\rm mom}(p)&=&2\pi^{\lambda+1}\int_0^\infty
  \left[\left(\frac{px}2\right)^{-\lambda}J_\lambda(px)\right]_1
  D(x,m_1)D(x,m_2) D(x,m_3)x^{2\lambda+1}dx\nonumber\\
  &=&2\pi^{\lambda+1}\int_0^\infty\left[\left(\frac{px}2\right)^{-\lambda}
  J_\lambda(px)-\frac1{\Gamma(\lambda+1)}
  +\frac{p^2x^2}4\frac1{\Gamma(\lambda+2)}\right]\nonumber\\&&\qquad
  \times D(x,m_1)D(x,m_2)D(x,m_3)x^{2\lambda+1}dx
\end{eqnarray}
while the pole part is given by a first order polynomial in $p^2$ of the form
$A+Bp^2$. The actual values of the coefficients $A$ and $B$ depend on the
regularization scheme that has been used. We use dimensional regularization
and obtain
\begin{eqnarray}
\label{gensing}
\tilde\Pi_{\rm sing}(p)&=&A+p^2B\nonumber\\
  &=&\frac{2\pi^{\lambda+1}}{\Gamma(\lambda+1)}\int_0^\infty D(x,m_1)D(x,m_2)
  D(x,m_3)x^{2\lambda+1}dx\nonumber\\&&
  -p^2\frac{2\pi^{\lambda+1}}{4\Gamma(\lambda+2)}\int_0^\infty
  x^2D(x,m_1)D(x,m_2) D(x,m_3)x^{2\lambda+1}dx.
\end{eqnarray}
With the help of the representation given in Eq.~(\ref{gensing}) we can apply
our strategy in a  straightforward manner. The coefficients $A$ and $B$ are
simple numbers independent of $p$. They contain divergent parts (regularized
within dimensional regularization) and need to be computed only once to
recover the function $\tilde\Pi(p)$ for any $p$. In the momentum subtracted
part one can forego the regularization (it is finite by the $R$-operation) and
perform the one-dimensional integration numerically for $D=4$ (i.e.\
$\lambda=1$). 

In the particular case of the genuine sunrise diagram the necessary integrals
are known analytically in closed form and can be found in integral tables (see
e.g.~\cite{Prudnikov}). However, using these tables may not always be
convenient and we therefore again present a simplified approach which allows
one to deal with even more complicated cases in a simpler manner. Let us
specify to the particular case $p=m_1+m_2-m_3$ (which corresponds to a
pseudothreshold) where an analytical answer for the integral
exists~\cite{Berends:1997vk}. For simplicity we choose $m_1=m_2=m_3/2=m$. Then
$p=0$ and $\tilde\Pi_{\rm mom}(0)=0$ (which is a regular function at this
Euclidean point). In this case the counterterm $Bp^2$ vanishes because the
quantity $B$ is finite at finite $\eps$. We thus only have to consider the
coefficient $A$. On the other hand our considerations are completely general
since for arbitrary $p$ one only requires more terms in the $p^2$-expansion.
For the special mass configuration considered here the result
of~\cite{Berends:1997vk} reads
\begin{equation}\label{gensunref}
\tilde\Pi^{\rm ref}(0)=\pi^{4-2\eps}\frac{m^{2-4\eps}
  \Gamma^2(1+\eps)}{(1-\eps)(1-2\eps)}
  \left[-\frac3{\eps^2}+\frac{8\ln 2}{\eps}
  -8\ln^2 2\right]+O(\eps).
\end{equation}
This result can be extracted from the first line of Eq.~(\ref{gensing}) using
the integral tables given in~\cite{Prudnikov}. However, in the general mass
case the necessary formulas are rather cumbersome. Even for the special mass
configuration considered here they are not so simple. We therefore discuss a
short cut which allows one to obtain results immediately without having to
resort to integral tables. What we really need is an expansion in $\eps$.
Basically we need the integral
\begin{equation}
\int_0^\infty D(x,m)D(x,m)D(x,2m)x^{2\lambda+1}dx
\end{equation}
which is of the general form
\begin{equation}\label{KKKint}
\int_0^\infty x^\rho K_\mu(mx)K_\mu(m x)K_\mu(2mx)dx 
\end{equation}
($\mu=\lambda$ and $\rho=1-\lambda$ in our case). For a product of two Bessel 
functions in the integrand (without the last one in the above equation) the
result of the integration is known analytically and is given by
Eq.~(\ref{KKint}). Let us reduce the problem at hand to Eq.~(\ref{KKint}) and
do our numerical evaluations with functions in four-dimensional space-time
where no regularization is necessary. To do so we subtract the leading
singularities at small $\xi$ from the last Bessel function in
Eq.~(\ref{KKKint}) using the series expansion near the origin given by
Eq.~(\ref{serKfac}). After decomposing the whole answer into finite and
singular parts according to $(2\pi)^{2D}A=F+S$ (where the total normalization
of~\cite{Berends:1997vk} has been adopted in the definition of $F$ and $S$) we
find for the singular part
\begin{eqnarray}\label{partS}
S&=&\frac{(2\pi)^D m^{2\lambda}}{\Gamma(\lambda+1)}
  \int_0^\infty x^{2(1-\lambda)-1} K_\lambda(mx)K_\lambda(m x)
  \frac{\Gamma(\lambda)}2\left[1+\frac{(mx)^2}{1-\lambda}
  -\frac{\Gamma(1-\lambda)}{\Gamma(1+\lambda)}(mx)^{2\lambda}\right]dx
  \nonumber\\
  &=&\frac{(2\pi)^D m^{2-4\eps}}{\Gamma(\lambda+1)}\int_0^\infty
  \xi^{2\eps-1}K_\lambda(\xi)K_\lambda(\xi)\frac{\Gamma(\lambda)}2
  \left[1+\frac{\xi^2}{1-\lambda}-\frac{\Gamma(1-\lambda)}{\Gamma(1+\lambda)}
  \xi^{2\lambda}\right]d\xi\nonumber\\
  &=&\pi^{4-2\eps}\frac{m^{2-4\eps}\Gamma^2(1+\eps)}
  {(1-\eps)(1-2 \eps)}\left[-\frac3{\eps^2}
  +\frac{8\ln 2}{\eps}+8(2-2\ln 2-\ln^2 2)\right]+O(\eps).
\end{eqnarray}
The pole contributions coincide with the result in Eq.~(\ref{gensunref}) while
the finite part is different. It is corrected by the finite expression
\begin{eqnarray}
F&=&\frac{(2\pi)^Dm^{2\lambda}}{\Gamma(\lambda+1)}
  \int_0^\infty x^{2(1-\lambda)-1}K_\lambda(mx)K_\lambda(m x)\nonumber\\&&
  \times\left\{(mx)^\lambda K_\lambda(2mx)
  -\frac{\Gamma(\lambda)}2\left[1+\frac{(mx)^2}{1-\lambda}
  -\frac{\Gamma(1-\lambda)}{\Gamma(1+\lambda)}(mx)^{2\lambda}\right]\right\}dx.
\end{eqnarray}
Because $F$ is finite (no strong singularity at small $x$) one can put
$\lambda=1$ to obtain
\begin{equation}\label{partF}
F=16\pi^4m^2\int_0^\infty\frac{dx}xK_1(x)K_1(x)
  \left\{xK_1(2x)-\frac12\left[1+x^2(-1+2\gamma_E+2\ln x)\right]\right\}.
\end{equation}
Doing the numerical integration in Eq.~(\ref{partF}) results in the expression 
\begin{equation}
F=16\pi^4m^2[-0.306853\ldots\ ].
\end{equation}
One can restore all $\eps$-dependence in the normalization factors as in
Eqs.~(\ref{gensunref}) and~(\ref{partS}) because $F$ is not singular in
$\eps$, so that this change of normalization is absorbed by the $O(\eps)$
symbol. One obtains
\begin{equation}\label{partFabs}
F=\pi^{4-2\eps}\frac{m^{2-4\eps}\Gamma^2(1+\eps)}{(1-\eps)(1-2\eps)}
  \left[-0.306853\ldots\ \right]+O(\eps).
\end{equation}
Adding $F$ from Eq.~(\ref{partFabs}) to $S$ from Eq.~(\ref{partS}) one obtains
a numerical result for the finite part of Eq.~(\ref{gensunref}). In order to
establish the full coincidence with the result of Eq.~(\ref{gensunref}), note
that 
\[
-0.306853\ldots\ =-(1-\ln 2)\times 1.0000\ldots.
\]
Of course we have used the analytical expression $(1-\ln 2)$ for the numerical
quantity $0.306853\ldots$ for illustrative reasons because we knew the final
answer.

We emphasize that there is nothing new in computing the polarization function
related to this diagram at any $p$. Some finite part appears from the momentum
subtracted polarization function. One then needs another counterterm for
non-zero $p^2$. The computation of these additional terms proceeds in analogy
to what has been done before. It is even simpler because the singularity at
small $x$ is weaker and only one subtraction of the Bessel function is
necessary. This computational technique works for any complex value $p^2$.

\subsection{Three-loop sunrise-type diagram for special momenta}

In a similar manner one can treat the three-loop sunrise-type diagram with two
equal masses $M$, one mass $m$ with values between zero and $M$ and a massless
line for outer momentum $p^2=-m^2$. This particular sunrise-type diagram was
analyzed by Mastrolia~\cite{Mastrolia:2002tv} whose results we confirm. The
starting integral expression reads
\begin{eqnarray}
\tilde\Pi(p)&=&2\pi^{\lambda+1}\int_0^\infty\pfrac{px}2^{-\lambda}
  J_\lambda(px)D(x,M)D(x,M)D(x,m)D(x,0)x^{2\lambda+1}dx\nonumber\\
  &=&\frac{M^{2\lambda}m^\lambda\Gamma(\lambda)}{2(2\pi)^{3(\lambda+1)}}
  \int_0^\infty\pfrac{px}2^{-\lambda}J_\lambda(px)K_\lambda(Mx)K_\lambda(Mx)
  K_\lambda(mx)x^{1-3\lambda}dx.\qquad
\end{eqnarray}
The integral has the same form as in the previous example. It is obvious that
we can proceed in the same way as before: first expand the integrand in $p$ 
and $m$ in order to find the counterterms, and second determine the finite
contribution for $p^2=-m^2$ numerically. However, in this case we have to
expand the Bessel function $J_\lambda(px)$ up to the second order in $p^2$.
For the singular part we obtain
\begin{eqnarray}
\tilde\Pi_{\rm sing}(p)&=&A+p^2B+\frac12p^4C\nonumber\\
  &=&\frac{2\pi^{\lambda+1}}{\Gamma(\lambda+1)}
  \int_0^\infty D(x,m_1)D(x,m_2)D(x,m_3)x^{2\lambda+1}dx\nonumber\\&&
  -p^2\frac{2\pi^{\lambda+1}}{4\Gamma(\lambda+2)}
  \int_0^\infty x^2D(x,m_1)D(x,m_2)D(x,m_3)x^{2\lambda+1}dx\nonumber\\&&
  +p^4\frac{2\pi^{\lambda+1}}{32\Gamma(\lambda+3)}
  \int_0^\infty x^4D(x,m_1)D(x,m_2)D(x,m_3)x^{2\lambda+1}dx
\end{eqnarray}
where
\begin{eqnarray}
A&=&\frac{2M^{4-6\eps}}{(4\pi)^{3(2-\eps)}\Gamma(2-2\eps)\Gamma(2-\eps)}
  \Bigg(\frac{1+2y^2}{6\eps^3}+\frac{2+8y^2-y^4-24y^2\ln y}{24\eps^2}
  \nonumber\\&&
  +\frac1\eps\left(-\frac3{16}(2+y^4)+\frac23(1+2y^2)\zeta(2)-y^2\ln y
  +\frac14y^4\ln y+y^2\ln^2y\right)\Bigg),\nonumber\\
B&=&\frac{2M^{2-6\eps}}{(4\pi)^{3(2-\eps)}\Gamma(2-2\eps)\Gamma(3-\eps)}
  \Bigg(\frac{2+y^2}{12\eps^2}+\frac{2+y^2-6y^2\ln y}{12\eps}\Bigg),\nonumber\\
C&=&\frac{2M^{-6\eps}}{(4\pi)^{3(2-\eps)}\Gamma(2-2\eps)\Gamma(4-\eps)}
  \Bigg(\frac1{6\eps}\Bigg)
\end{eqnarray}
and $y=m/M$.

\subsection{Infinite parts of sunrise-type diagrams with different masses
at any number of loops: general techniques}

If one deals with sunrise-type diagrams with different masses one has to be
careful to use a symmetric subtraction procedure. Otherwise one looses the
symmetry of the original diagram. Such an asymmetry would arise if one would
expand one of the Bessel functions in the integrand of the sunrise-type
diagram as was done in the previous example for the massive line with mass
$m$. If one wants to obtain a result which is symmetric under the exchange of
the masses $m_i$ as expected from the topology of the diagram and from the
initial form of the integral we have to subtract a counterterm which is
symmetric in these masses. In calculating the genuine sunrise diagram one can
follow this procedure by using a damping function such as $e^{-\mu^2x^2}$ to
suppress contributions for large values of $x$, corresponding to IR
singularities. This is done by introducing the factor
$1=e^{\mu^2x^2}e^{-\mu^2x^2}$ into the integrand. In this case all massive
propagators along with the factor $e^{\mu^2x^2}$ can be expanded in terms of
small $x$ values. The expansion of the propagators is straightforward and can
be obtained from Eqs.~(\ref{serJfac}) and~(\ref{serKfac}) in
Appendix~\ref{sec_bessel}. The final integration can then be done formally by
using the identity
\begin{equation}
\int_0^\infty x^{p-1}e^{-\mu^2x^2}dx=\frac12\mu^{-p}\Gamma(p/2)
\end{equation}
which is nothing but the definition of Euler's Gamma function.

As an illustration we compute the four-loop bubble diagram which was also
calculated by Laporta~\cite{Laporta:2002pg} whose results we confirm. In the
sample results below we extract the normalization factor
\begin{equation}\label{normfac}
{\cal N}_n=\pfrac{(4\pi)^{2-\eps}}{\Gamma(1+\eps)}^n
\end{equation}
for the $n$-loop (or $(n+1)$-line) sunrise-type diagram. Extraction of this
factor renders all pole parts to be rational numbers. The results of the
calculation are
\begin{eqnarray}\label{poleparts}
{\cal N}_1\tilde\Pi_1(p^2)&=&\frac1\eps+O(\eps^0),\nonumber\\
{\cal N}_2\tilde\Pi_2(p^2)&=&m^2\left\{-\frac3{2\eps^2}-\frac9{2\eps}
  \right\}-\frac{p^2}{4\eps}+O(\eps^0),\nonumber\\
{\cal N}_3\tilde\Pi_3(p^2=0)&=&m^4\left\{\frac2{\eps^3}+\frac{23}{3\eps^2}
  +\frac{35}{2\eps}\right\}+O(\eps^0),\nonumber\\
{\cal N}_3\tilde\Pi_3(p^2=-m^2)&=&m^4\left\{\frac2{\eps^3}+\frac{22}{3\eps^2}
  +\frac{577}{36\eps}\right\}+O(\eps^0),\nonumber\\
{\cal N}_4\tilde\Pi_4(p^2=0)&=&m^6\left\{-\frac5{2\eps^4}-\frac{35}{3\eps^3}
  -\frac{4565}{144\eps^2}-\frac{58345}{864\eps}\right\}+O(\eps^0),\nonumber\\
{\cal N}_4\tilde\Pi_4(p^2=-m^2)&=&m^6\left\{-\frac5{2\eps^4}-\frac{45}{4\eps^3}
  -\frac{4255}{144\eps^2}-\frac{106147}{1728\eps}\right\}+O(\eps^0),\nonumber\\
{\cal N}_5\tilde\Pi_5(p^2=0)&=&m^8\left\{\frac3{\eps^5}+\frac{33}{2\eps^4}
  +\frac{1247}{24\eps^3}+\frac{180967}{1440\eps^2}+\frac{898517}{3456\eps}
  \right\}+O(\eps^0),\nonumber\\
{\cal N}_5\tilde\Pi_5(p^2=-m^2)&=&m^8\left\{\frac3{\eps^5}+\frac{16}{\eps^4}
  +\frac{49}{\eps^3}+\frac{6967}{60\eps^2}+\frac{1706063}{7200\eps}
  \right\}+O(\eps^0),\nonumber\\
{\cal N}_6\tilde\Pi_6(p^2=0)&=&m^{10}\left\{-\frac7{2\eps^6}
  -\frac{133}{6\eps^5}-\frac{238}{3\eps^4}-\frac{77329}{360\eps^3}\right.
  \nonumber\\&&\qquad\left.-\frac{21221921}{43200\eps^2}
  -\frac{2596372387}{2592000\eps}\right\}+O(\eps^0).\qquad
\end{eqnarray}
The coefficient of the leading singularity in $\eps$ is independent of $p^2$. 
In Appendix~\ref{sec_sing} we list results for unequal mass configurations up
to four-loop order. When setting all masses equal the results of the general
mass case can be seen to agree with the above equal mass results.

\subsection{Finite part for the genuine sunrise with three different masses}

In configuration space the genuine sunrise diagram with three different masses 
$m_1$, $m_2$, and $m_3$ is given by
\begin{equation}
\Pi(x)=D(x,m_1)D(x,m_2)D(x,m_3).
\end{equation}
The Fourier transform of this polarization function in the Euclidean domain
reads
\begin{equation}
\tilde\Pi(p)=2\pi^{\lambda+1}\int_0^\infty\pfrac{px}2^{-\lambda}
  J_\lambda(px)D(x,m_1)D(x,m_2)D(x,m_3)x^{2\lambda+1}dx.
\end{equation}
In dimensional regularization the singular parts ($D=4-2\eps$, i.e.\
$\lambda=1-\eps$) are given by
\begin{equation}
\tilde\Pi_{\rm sing}(p)=\frac{\mu^{-4\eps}}{\pi^{4-2\eps}}
  \Bigg\{-\frac{m_1^2+m_2^2+m_3^2}{512\eps^2}
  +\frac1\eps\left(\sum_{i=1}^3\frac{m_i^2\ln(m_ie^{\gamma\prime}/2\mu)}{128}
  -\frac{p^2}{1024}\right)\Bigg\}
\end{equation}
where $\gamma'=\gamma_E/2-3/4$. The counterterms for the Fourier transform
are obtained as explained before by introducing $1=e^{\mu^2x^2}e^{-\mu^2x^2}$
and expanding the relevant Bessel functions together with $e^{\mu^2x^2}$. This
is the reason for the explicit dependence on the arbitrary parameter $\mu$.
In this way one readily finds the pole parts for any diagram with any mass
configuration, i.e.\ one finds the coefficients $C_i$ of the counterterms in
Eq.~(\ref{renormC}). Even if an external momentum goes through the diagram,
the pole parts can be still calculated easily since they are given by
integrals over the expanded integrand. The momentum subtracted finite part
is given by
\begin{equation}
\tilde\Pi_{\rm fin}(p)=2\pi^{\lambda+1}\int_0^\infty
  \left[\pfrac{px}2^{-\lambda}J_\lambda(px)\right]_ND(x,m_1)D(x,m_2)D(x,m_3)
  x^{2\lambda+1}e^{\mu^2x^2}e^{-\mu^2x^2}dx
\end{equation}
where $N$ is the order of the expansion needed to ensure integrability.
Naturally $N$ has the same order as the one used earlier in order to extract
the counterterms. It is obvious, though, that the expansion can terminate at
positive powers of $x$. Finally, one can set $\eps=0$ to calculate the finite
part numerically. As a technical remark note that most numerical integration
routines will run into problems if the upper limit extends to infinity.
However, already for values of the order $x\sim 1$ the integrand is negligibly
small, so that the integration can be cut off at this point. Also the region
around the origin might cause trouble for the convergence of the numerical
integration. In this case the integration interval can be subdivided with an
interval close to the origin, and the whole integrand in this subinterval can
be expanded in $x$ and integrated analytically. For specific values of $p^2$,
$m_1$, $m_2$, and $m_3$ we were able to reproduce results given in the
literature (see e.g.\ results given in Refs.~\cite{Caffo:2002ch,SonDo}).

\subsection{Calculation of the finite part: further examples\label{finipart}}

\begin{figure}\begin{center}
\epsfig{figure=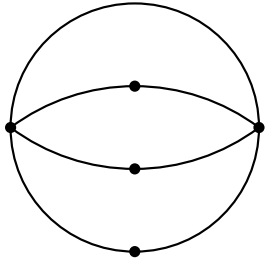, scale=0.5}
\caption{\label{diag2221}Three-loop massive sunrise-type bubble. A dot on a
line corresponds to one differentiation with respect to the propagator mass
squared}
\end{center}\end{figure}

The examples presented so far are well-known and have been obtained before by
using techniques different from the ones presented here. While we can
numerically compute any sunrise-type diagram with any arbitrary number of
internal massive lines it is not always possible to find corresponding
analytical expressions to compare with. Beyond two loops only a few examples
can be found in the literature. For instance, we have calculated the result
for the vacuum bubble diagram shown in Fig.~\ref{diag2221}
(see also Sec.~\ref{sec_diag2221}). For $p\to 0$ we obtain (cf.\
Eq.~(\ref{serJfac}))
\begin{equation}
\pfrac{px}2^{-\lambda}J_\lambda(px)\to\frac1{\Gamma(\lambda+1)}.
\end{equation}
One therefore has
\begin{equation}
\tilde\Pi(0)=\frac{2\pi^{\lambda+1}}{\Gamma(\lambda+1)}
  \int_0^\infty\left(D^{(1)}(x,m)\right)^3D(x,m)x^{2\lambda+1}dx
\end{equation}
where $D^{(1)}(x,m)$ is defined in Eq.~(\ref{primeD}). In order to compare
with the literature~\cite{Broadhurst:1991fi}, we multiply with a relative
factor ${\cal N}_3m^{2+6\eps}$ (cf.\ Eq.~(\ref{normfac})) to obtain
\begin{equation}
B_\eps={\cal N}_3m^{2+6\eps}\tilde\Pi(0)
  =\frac{2^{2-2\eps}}{(1-\eps)\Gamma(1+\eps)^3\Gamma(1-\eps)}
  \int_0^\infty K_{-\eps}^3(x)K_{1-\eps}(x)x^{2+2\eps}dx.
\end{equation}

The integral remains finite even for $\eps\to 0$ and, after finite
integration, we obtain the numerical result 
\begin{equation}
B_0=4\int_0^\infty K_0^3(x)K_1(x)x^2dx=2.1035995805\ldots
\end{equation}
A comparison with the results presented in Ref.~\cite{Broadhurst:1991fi} 
shows full numerical agreement. Since we know that the quantity $B_0$ contains
only a single transcendental number, namely $\zeta(3)$, where $\zeta(z)$ is
Riemann's $\zeta$-function~\cite{Broadhurst:1991fi,Ogreid:1999mf}, we can take
its numerical value $\zeta(3)=1.202056903\ldots$ to obtain the result
\[
B_0=\zeta(3)\cdot 1.7500000\ldots=\frac74\zeta(3)
\]
which is nothing but the analytical result given in 
Ref.~\cite{Broadhurst:1991fi}.

We conclude that the configuration space method is simple and allows one to
numerically compute diagrams with sunrise-type topology in a very efficient
manner. When the structure of the transcendentality of the result is known for
a particular diagram (as in the latest example where only $\zeta(3)$ is
present) or when it can be obtained from an educated guess looking at the
topology of the diagram \cite{kreimer97} our numerical technique can be used
to restore the rational coefficients of these transcendentalities which can be
thought of as elements of the basis for a class of diagrams.

\subsection[Higher order expansion in $\eps$: general techniques and the
  four-loop vacuum bubble]{Higher order expansion in $\eps$:\\
  general techniques and the four-loop vacuum bubble}

After having determined the singular parts of the Laurent series expansion
using the damping factor method what remains to be done is to calculate the 
coefficients of the positive powers of $\eps$ in the $\eps$-expansion
including the finite $\eps^0$ term. Higher order terms in the $\eps$-expansion
are needed if the sunrise-type diagram is inserted into a divergent diagram or
when one is using the integration-by-parts recurrence relation which can
generate inverse powers of $\eps$. In order to determine these higher order
terms one needs to resort to the Bessel function method. What is technically
needed is to develop a procedure for the $\eps$-expansion of Bessel functions.

The $O(\eps)$ term can be obtained analytically. Indeed, within dimensional
regularization the propagator in configuration space contains Bessel functions
with non-integer index depending on the regularization parameter $\eps$, as
they occurred for instance in Sec.~\ref{finipart} at an intermediate step. In
order to expand the Bessel function in the parameter $\eps$ entering its
index, we use Eq.~(\ref{serKind}) for the derivative of the modified Bessel
function of the second kind with respect to its index near integer values of
this index. For instance, we obtain
\begin{eqnarray}
K_{-\eps}(x)&=&K_0(x)+O(\eps^2),\nonumber\\[7pt]
K_{1-\eps}(x)&=&K_1(x)-\frac\eps xK_0(x)+O(\eps^2),\nonumber\\
K_{2-\eps}(x)&=&K_2(x)-\frac{2\eps}xK_1(x)-\frac{2\eps}{x^2}K_0(x)+O(\eps^2)
  \quad\ldots\qquad
\end{eqnarray}
For the more general cases one can use the integral representation
\begin{equation}
K_\nu(z)=\int_0^\infty\cosh(\nu t)e^{-z\cosh t}dt
\end{equation}
for the index expansion. However, one can only obtain numerical results in
this case. In expanding the integrand, we obtain
\begin{eqnarray}\label{Keps}
K_{-\eps}(z)&=&\sum_{n=0}^\infty\frac{\eps^{2n}}{(2n)!}f_{2n}(z),\nonumber\\
K_{1-\eps}(z)&=&\sum_{n=0}^\infty\frac{\eps^{2n}}{(2n)!}a_{2n}(z)
  -\sum_{n=0}^\infty\frac{\eps^{2n+1}}{(2n+1)!}b_{2n+1}(z)
\end{eqnarray}
where
\begin{eqnarray}
f_k(z)&=&\int_0^\infty t^ke^{-z\cosh t}dt,\nonumber\\
a_k(z)&=&\int_0^\infty t^k\cosh t\,e^{-z\cosh t}dt,\nonumber\\
b_k(z)&=&\int_0^\infty t^k\sinh t\,e^{-z\cosh t}dt. 
\end{eqnarray}
Due to the recurrence relations in the index $\nu$ for the family $K_\nu(z)$
(cf.\ Sec.~\ref{sec_reduct}) the two formulas in Eq.~(\ref{Keps}) suffice to
calculate the Bessel function $K_\nu(z)$ for any $\nu$. The two functions
$a_k(z)$ and $b_k(z)$ in Eq.~(\ref{Keps}) are in turn related to the functions
$f_k(z)$,
\begin{equation}
a_k(z)=-\frac d{dz}f_k(z),\qquad b_k(z)=\frac kzf_{k-1}(z).
\end{equation}
Using these relations, we obtain
\begin{equation}
K_{1-\eps}(z)=-\sum_{n=0}^\infty\frac{\eps^{2n}}{(2n)!}
  \left(\frac{d}{dz}+\frac\eps z\right)f_{2n}(z).
\end{equation}
 
\begin{figure}[t]\begin{center}
\epsfig{figure=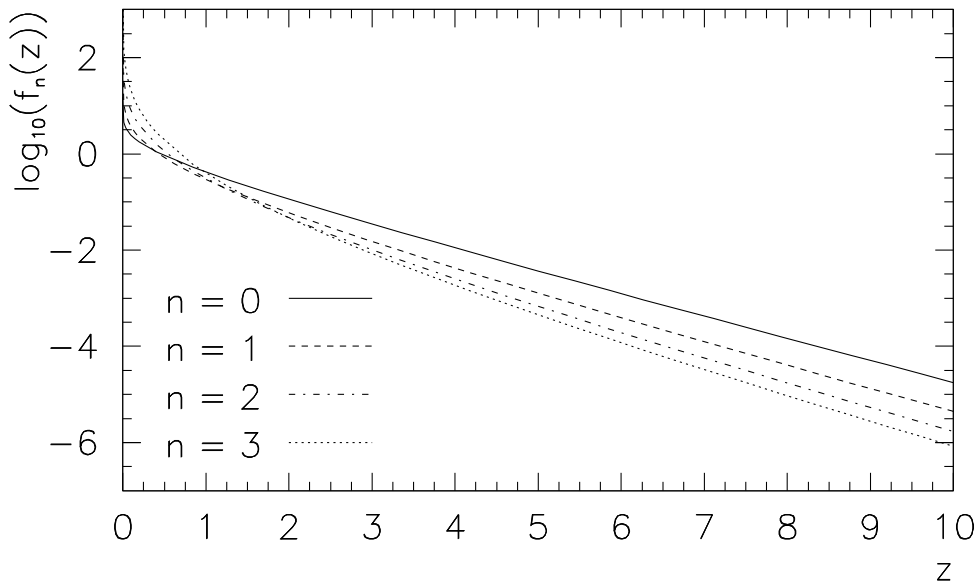, scale=0.8}
\caption{\label{fig03}Comparison of functions $f_n(z)$ for different values of
  $n$, plotted on a logarithmic scale}
\end{center}\end{figure}

The family of functions $f_k(z)$ is rather close to the original set of Bessel
functions $K_\nu(z)$ (cf.\ Fig.~\ref{fig03}) and can easily be studied both
analytically and numerically. The limits at $z\to 0$ and $z\to\infty$ are
known analytically and are simple. They allow for an efficient interpolation
for intermediate values of the argument.
The functions $f_k(z)$ satisfy the differential equation
\begin{equation}
\left(\frac{d^2}{dz^2}+\frac1z\frac{d}{dz}-1\right)f_k(z)
  =\frac{k(k-1)}{z^2}f_{k-2}(z)
\end{equation}
related to the Bessel differential equation. The small $z$ behaviour
\begin{equation}
f_k(z)=\frac1{k+1}\ln^{k+1}\pfrac1z
\left(1+O\left(\frac1{\ln(z)}\right)\right)
\end{equation}
can be found by using yet another representation for the function $f_k(z)$,
\begin{equation}
f_k(z)=\int_1^\infty\frac{e^{-zu}}{\sqrt{u^2-1}}\ln^k(u+\sqrt{u^2-1})du
\end{equation}
or directly from the behaviour of the function $K_\nu(x)$ at small $x$,
\begin{equation}
K_{-\eps}(x)=\frac1\eps\sinh\left(\eps\ln(1/x)\right).
\end{equation}

While in four-dimensional space-time the massive propagator contains the
Bessel function $K_{1-\eps}(z)$, the basic function in $D=2$ dimensional
space-time is the Bessel function $K_{-\eps}(z)$ since the propagator reads
\begin{equation}
D(x,m)|_{D=2}=\frac1{2\pi}K_0(mx).
\end{equation}
As an example for the numerical evaluation of the $\eps$-expansion using
Bessel functions we consider a toy model integral related to the one-loop case
in two dimensions. We select this example because the integral is finite and
analytically known, so that we can compare our numerical calculation with the
exact answer. Using Eq.~(\ref{KKint}), we obtain
\begin{equation}
2\int K_{-\eps}(x)K_{-\eps}(x)x\,dx=\Gamma(1+\eps)\Gamma(1-\eps).
\end{equation}
The expansion in $\eps$ is given by
\begin{equation}
\Gamma(1-\eps)\Gamma(1+\eps)=\frac{\pi\eps}{\sin(\pi\eps)}
  =1+\frac{\pi^2\eps^2}6+\frac{7\pi^4\eps^4}{360}+O(\eps^6)
\end{equation}
On the other hand, we can use the expansion
\begin{equation}
K_{-\eps}(x)=f_0(x)+\frac{\eps^2}2f_2(x)+\frac{\eps^4}{24}f_4(x)+O(\eps^6)
\end{equation}
to rewrite the integral in the form 
\begin{eqnarray}
2\int_0^\infty K_{-\eps}(x)K_{-\eps}(x)x\,dx&=&2\int_0^\infty f_0(x)^2x\,dx
  +2\eps^2\int_0^\infty f_0(x)f_2(x)x\,dx\nonumber\\&&
  +\frac{\eps^4}6\int_0^\infty f_0(x)f_4(x)x\,dx
  +\frac{\eps^4}2\int_0^\infty f_2(x)^2x\,dx+O(\eps^6).\qquad
\end{eqnarray}
Using the explicit expressions for the functions $f_k$ we have checked by 
numerical integration that the identities
\begin{eqnarray}
2\int_0^\infty f_0(x)^2x\,dx&=&1,\nonumber\\
2\int_0^\infty f_0(x)f_2(x)x\,dx&=&\frac{\pi^2}6,\nonumber\\
\frac16\int_0^\infty(f_0(x)f_4(x)+3f_2(x)^2)x\,dx&=&\frac{7\pi^4}{360}
\end{eqnarray}
are valid numerically with very high degree of accuracy. We have implemented
our algorithm for the $\eps$-expansion of sunrise-type diagrams as a simple
code in Wolfram's Mathematica system for symbolic manipulations and checked
its work-ability and efficiency.

When analyzing the $\eps$-expansion one realizes that integrals over products 
of modified Bessel functions of the second kind result in integrals over
products of functions $f_k(z)$. But because the analytical behaviour of the
functions $f_k(z)$ is quite similar to the original Bessel functions, the
numerical integration is again easy to perform. For example, in the case of
the four-loop bubble diagram, the integral
\begin{equation}
\tilde\Pi_4(p^2=0)=\int D(x,m)^5d^Dx
\end{equation}
can be calculated by subtracting the expansion (cf. Appendix~\ref{sec_smallx})
\begin{eqnarray}
\Delta(x)&=&\frac1{4\pi^{\lambda+1}x^{2\lambda}}\Bigg\{
\Gamma(\lambda)+\left(\pfrac x2^2\frac{\Gamma(\lambda)}{1-\lambda}
  -\pfrac x2^{2\lambda}\frac{\Gamma(1-\lambda)}\lambda\right)
  \nonumber\\&&\qquad\qquad\qquad
  +\pfrac x2^2\left(\pfrac x2^2\frac{\Gamma(\lambda)}{2(1\lambda)(2-\lambda)}
  -\frac x2^{2\lambda}\frac{\Gamma(1-\lambda)}{\lambda(\lambda+1)}\right)
  \Bigg\}
\end{eqnarray}
from each of the propagators except for one which we keep as IR regulator.
We obtain
\begin{eqnarray}\label{V4main}
\tilde\Pi_4(0)&=&\int D(x,m)\Big(D(x,m)-\Delta(x)+\Delta(x)\Big)^4
  d^Dx\nonumber\\
  &=&\int D(x,m)\Big(\left(D(x,m)-\Delta(x)\right)^4
  +4\left(D(x,m)-\Delta(x)\right)^3\Delta(x)\nonumber\\[3pt]&&
  +6\left(\Delta(x,m)-\Delta(x)\right)^2\Delta(x)^2
  +4\left(D(x,m)-\Delta(x)\right)\Delta(x)^3+\Delta(x)^4\Big)d^Dx.\qquad
\end{eqnarray}
The last two terms in Eq.~(\ref{V4main}) can be integrated analytically and
can be expanded to an arbitrary order in $\eps$. They contain the singular
contributions, i.e.\ all poles in $\eps$ and are expressible through Euler's
$\Gamma$-functions. As expected, the pole part coincides with the expression
in Eq.~(\ref{poleparts}). Because the analytical expression (as given up to
order $\eps^3$ in Appendix~\ref{sec_fourloop}) is rather lengthy, we present
only its numerical evaluation,
\begin{eqnarray}
{\cal N}_4\tilde\Pi_4^{\rm ana}(0)&=&m^6\Big(-2.5\eps^{-4}
  -11.6666667\eps^{-3}-31.701389\eps^{-2}-67.528935\eps^{-1}\nonumber\\&&\qquad
  -15871.965743-142923.10240\eps-701868.64762\eps^2\nonumber\\&&\qquad\qquad
  -2486982.5547\eps^3+O(\eps^4)\Big).\qquad
\end{eqnarray}
The remaining first three terms in Eq.~(\ref{V4main}) can be integrated
numerically for $D=4$ using the expansion of the Bessel functions in terms of
the functions $a_k(z)$ and $b_k(z)$~\cite{Groote:2004qq}. No regularization is
necessary since these contributions are regular at small $x$. The analytical
expression for the functions to be integrated is again rather long. The zeroth
order $\eps$-coefficient is found in Appendix~\ref{sec_numint} (see also the
discussion of the integration procedure given in Appendix~\ref{sec_numint}).
However, as shown in Fig.~\ref{numplot5}, the integrands themselves show a very
smooth behaviour which renders the numerical integration rather simple. We
obtain
\begin{equation}
{\cal N}_4\tilde\Pi_4^{\rm num}(0)=m^6\Big(15731.745122+142349.56687\eps
  +699112.42072\eps^2+2468742.6339\eps^3+O(\eps^4)\Big).
\end{equation}
The sum of both parts gives
\begin{eqnarray}
{\cal N}_4\tilde\Pi_4(0)&=&m^6\Big(-2.5\eps^{-4}-11.6666667\eps^{-3}
  -31.701389\eps^{-2}-67.528935\eps^{-1}\nonumber\\&&\qquad
  -140.220621-573.53553\eps-2756.22690\eps^2-18239.9208\eps^3+O(\eps^4)
  \Big)\qquad
\end{eqnarray}
which confirms the known result~\cite{Laporta:2002pg}. We have not included 
as many digits as are given in Ref.~\cite{Laporta:2002pg}.

\begin{figure}[t]\begin{center}
\epsfig{figure=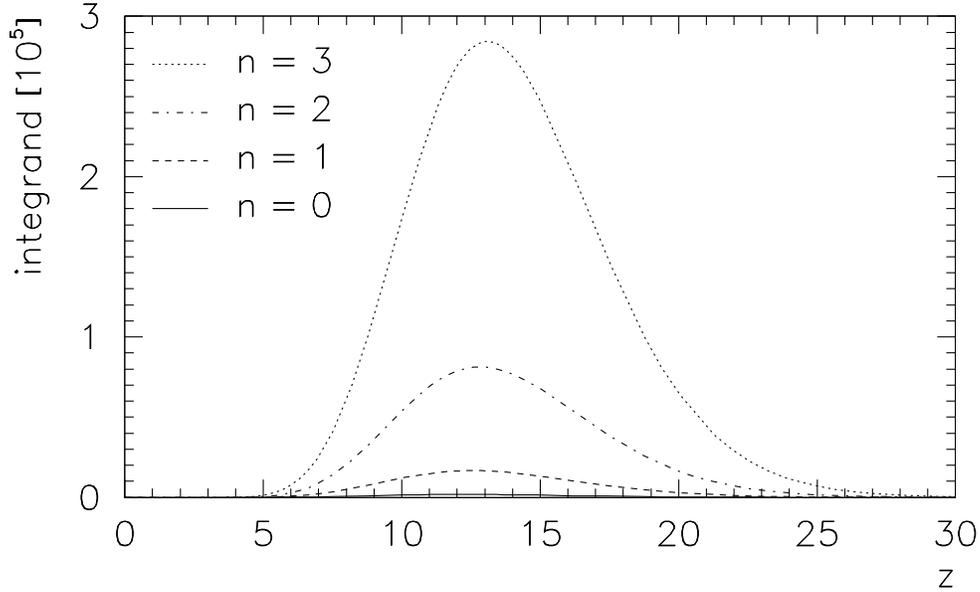, scale=0.8}
\caption{\label{numplot5}Integrands for numerical integration in case of the
  four-loop bubble at different orders $O(\eps^n)$ in $\eps$}
\end{center}\end{figure}

It is not difficult to extend the analysis to higher orders in $\eps$ or to a
larger number of significant digits in the coefficients of the
$\eps$-expansion. However, since the technique is rather straightforward and
simple we do not consider it worthwhile to extend the calculations into these
directions. If the need arises, the potential user can tailor and optimize his
or her programming code to obtain any desired accuracy and/or order of the
$\eps$-expansion. In our evaluation we have used standard tools provided by
Wolfram's Mathematica system for symbolic manipulations which allows one to 
reliably control the accuracy of numerical calculations. Even at this early 
stage of improvement it is obvious that our algorithm is extremely simple and 
efficient.

\begin{figure}[t]\begin{center}
\epsfig{figure=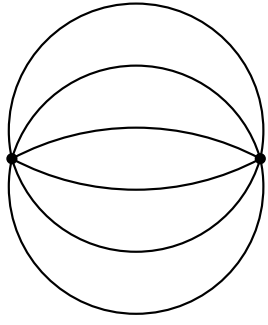, scale=0.5}
\caption{\label{fig5loop}A five loop vacuum bubble}
\end{center}\end{figure}

\subsection{$\eps$-expansion of the five-loop vacuum bubble}

In this paragraph we present results for the next-order sunrise-type diagram
with $p^2=0$, the five-loop vacuum bubble (see Fig.~\ref{fig5loop}). We have
chosen to extend our calculation to the five-loop case since there exist no
results on the five-loop bubble in the literature. The integral representation
of the five-loop bubble with equal masses $m$ is given by 
\begin{equation}
\tilde\Pi_5(p^2=0)=\int D(x,m)^6d^Dx.
\end{equation}
Evaluating numerically the analytical part one obtains
\begin{eqnarray}
{\cal N}_5\tilde\Pi_5^{\rm ana}(0)&=&m^8\Big(3\eps^{-5}+16.5\eps^{-4}
  +51.95833\eps^{-3}+125.6715\eps^{-2}+259.9876\eps^{-1}\nonumber\\&&\qquad
  -1360392.5934-16888723.177\eps-111392297.46\eps^2\nonumber\\&&\qquad\qquad
  -518606741.1\eps^3+O(\eps^4)\Big)\qquad
\end{eqnarray}
while the numerical integration of the nonsingular part gives
\begin{eqnarray}
{\cal N}_5\tilde\Pi_5^{\rm num}(0)
  &=&m^8\Big(1360739.9485+16886269.683\eps\nonumber\\&&\qquad
  +111360751.91\eps^2+518295438.0\eps^3+O(\eps^4)\Big).
\end{eqnarray}
The sum of both contributions is given by
\begin{eqnarray}
{\cal N}_5\tilde\Pi_5(0)&=&m^8\Big(3\eps^{-5}+16.5\eps^{-4}+51.95833\eps^{-3}
  +125.6715\eps^{-2}+259.9876\eps^{-1}\nonumber\\&&\qquad
  +347.3551-2453.494\eps-31545.55\eps^2-311303.1\eps^3+O(\eps^4)\Big)\qquad.
\end{eqnarray}
One observes huge cancellation effects between the terms obtained by the
analytical calculation and the numerical integration. Apparently the
subtraction procedure chosen here is non-optimal. As mentioned before, the
subtraction procedure should really be optimized for any given problem in
order to avoid a necessity to retain high numerical precision at intermediate
steps of the calculation. Nevertheless, our non-optimized simple subtraction
procedure already works quite reliably with available standard computational
tools.

\subsection{The spectral density, the discontinuity across the physical cut 
and the evaluation of the multi-particle phase space}

\begin{figure}[t]\begin{center}
\epsfig{figure=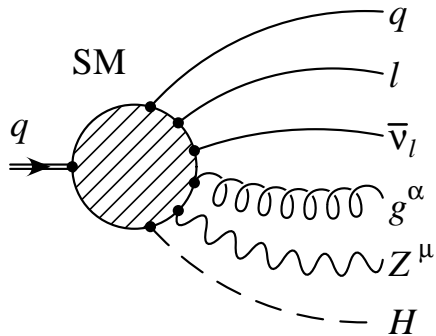, scale=0.5}
\caption{\label{mediag3} Example of a multi-body decay of a quark in the
Standard Model}
\end{center}\end{figure}

Finally, we discuss the spectral density which can also be interpreted as the
phase space volume of a multi-body decay of a single particle. An example of
such a multi-body decay is shown in Fig.~\ref{mediag3}. At a given value of
$s$ the value of the spectral density gives the phase space volume of the
system of particles corresponding to the lines of the sunrise diagram. The
phase space volume for $(n+1)$ particles with masses $m_1,\ldots, m_{n+1}$ is
computed according to~\cite{Eidelman:2004wy}
\[
\Phi_{n+1}(p;p_1,\ldots,p_{n+1})=\int \frac{d^3p_1}{(2\pi)^3E_1}\ldots
  \frac{d^3p_{n+1}}{(2\pi)^32E_{n+1}}\delta^{(4)}(p_1+\ldots+p_{n+1}-p)
\]
where $E_i=\sqrt{{\vec p}_i^2+m_i^2}$. The optical theorem relates the
imaginary part of a Feynman amplitude to on-shell propagation of its internal
lines and therefore to the product of an outer state with its conjungate via
so-called cutting rules~\cite{Cutkosky:1960sp} (for a recent review see also
Ref.~\cite{Zhou:2004gm}). Therefore the discontinuity across a physical cut of 
the sunrise-type diagram with $(n+1)$ internal lines carrying $(n+1)$ masses
$m_1,\ldots, m_{n+1}$ leads to the $(n+1)$ particle phase space in a natural
way. We obtain
\[
\Phi_{n+1}(p;p_1,\ldots,p_{n+1})=2\pi\rho(s)|_{p^2=s}
\]
which means that, up to a constant factor, the phase space is given by the
spectral density.

\newpage

\section{Asymptotic analysis of sunrise-type diagrams}

In multi-loop calculations it is very useful to use expansions in different
regimes of its kinematical variables and masses, corresponding each to the
smallness of one of the parameters involved in the problem. After defining an
appropriate scale, the separation of scales can considerably simplify the
description in the given region of energy or momentum transfer for a
particular physical process. Such an analysis is needed when, starting from
an underlying full theory, one wants to formulate an effective theory for a 
given limited energy scale. Taking for instance Quantum Chromodynamics (QCD)
as the underlying theory, the best-known of such effective theories is the
Heavy Quark Effective Theory (HQET) which analyzes the behaviour of a heavy
quark near its mass shell which leads to a simplified description of physical
processes subject to this approximation~\cite{Isgur:1989vq}. The
effective theory is directly constructed through an expansion of the
expression originally written in QCD. Its analysis, therefore, relies on the
expansion of Feynman diagrams in a special regime. Another important example
is the expansion near the production threshold, leading to an effective theory
termed nonrelativistic QCD (NRQCD). NRQCD is a more complicated case of an
effective theory since one introduces new variables in NRQCD which differ from
those defined in QCD. The most involved example is Chiral Perturbation Theory
(ChPT) where the variables are Goldstone modes which are not directly related
to QCD (and where it is still difficult to construct a direct correspondence).
Nevertheless, the common thread of all these effective theories is that they
use  perturbative expansions in terms of diagrams in given specific
kinematical regimes.

When sunrise-type diagrams enter these analyses, they usually provide the
leading order contribution or serve as laboratory for checking more elaborate
techniques. In this connection the expansion of sunrise-type diagrams in
configuration space happens to be very efficient for many different regimes. 
Having the closed form expression in Eq.~(\ref{psun}) for the correlation
function $\tilde\Pi(p)$ at hand, different asymptotic regimes in the
parameters (masses, external momentum) can easily be analyzed by calculating
the corresponding expansions. The first three subsections in this section will
deal with examples in the Euclidean domain. Starting with the fourth
subsection, we will consider some limiting cases in the Minkowskian domain
which is the domain of physical states. The calculation of asymptotic cases is
more involved in this case. The calculations are performed for the spectrum
given by Eq.~(\ref{exactrho}) which contains the physical content of the
problem.

\subsection{Large momentum expansion: close to the massless limit}

Loop integrations considerably simplify for massless internal lines. This
holds true for the momentum space representation as well as for the
configuration space representation of the loops. In some cases particle mass
corrections to the massless approximation can become numerically important.
For instance, in QCD the analysis of hadronic processes involving the strange
quark require some special care since the mass of the $s$-quark is not small
compared to the masses of the lightest quarks. On the other hand, the
$s$-quark mass is still much smaller than the QCD scale $\Lambda_{\rm QCD}$.
The expansion in the ratio of the $s$-quark mass to the relevant scale of the
considered process is therefore very effective. This expansion is especially
fruitful for the analysis of tau lepton decays~\cite{Chetyrkin:1998ej,
Korner:2000ef,Gorbunov:2004wy}.

Mass corrections to the large $p^2$ behaviour in the Euclidean domain (i.e.\
expansions in $m_i^2/p^2$) are obtained by expanding the massive propagators
under the integration sign in Eq.~(\ref{psun}) in terms of the masses $m_i$.
The final integration is performed by using the identity
\begin{equation}\label{largemom}
\int_0^\infty x^\mu J_\lambda(px)dx
=2^\mu p^{-\mu-1}\frac{\Gamma\left(
  (\lambda+\mu+1)/2\right)}{\Gamma\left((\lambda-\mu+1)/2\right)}.
\end{equation}
Note that all these manipulations are straightforward and can be easily
implemented in a system of symbolic computations. Some care is necessary,
however, when poles of the $\Gamma$-function are encountered which reflect the
presence of artificial infrared singularities. The framework for dealing with
such problems is well-known (see e.g.\ Refs.~\cite{Chetyrkin:nn,
Tkachov:1999nk}).

\begin{figure}\begin{center}
\epsfig{figure=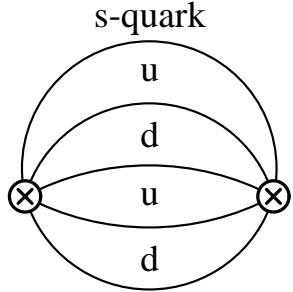, scale=0.5}
\caption{\label{pentaScorr}Leading order contribution to the two-point
correlator in the sum rule analysis of light pentaquark states}
\end{center}\end{figure}

A further physics application concerns the study of pentaquark properties
within the QCD sum rule approach. In Fig.~\ref{pentaScorr} we show the leading
order contribution to the relevant two-point correlator of the interpolating
pentaquark currents which contains one massive $s$-quark line. One therefore
has to determine the mass corrections to the two-point correlator. Another
application of the proposed techniques concerns the analysis of certain scalar
mesons which can be viewed as four-quark states involving the $s$-quark. For
instance, the scalar meson $a_0$ can be considered to be a $KK$ molecule built
from two $K$-mesons~\cite{Braun:1988kv}. Yet another example is the analysis
of $K^0-\bar K^0$ mixing in the two-point function approach~\cite{Pich:1985ab}.
Note, however, that the approach based on the three-point function analysis
appears to be more direct~\cite{Chetyrkin:1986vj}.

\subsection{Small momentum expansion: close to the vacuum bubble}

The basic expression in Eq.~(\ref{psun}) is also well suited for finding the
power series expansion in $p^2$ because the values of the polarization
function and its derivatives at $p^2=0$ can easily be obtained. The
convenience of a $p^2$-expansion is demonstrated by making use of the basic
identity for differentiating the Bessel function (cf.\ Eq.~(\ref{diffJ})),
\begin{equation}\label{diffJp}
\frac{d^k}{d(p^2)^k}\left(\frac{px}2\right)^{-\lambda}J_\lambda(px)
  =\left(-\frac{x^2}4\right)^k\left(\frac{px}2\right)^{-\lambda-k}
  J_{\lambda+k}(px).
\end{equation}
Differentiation, therefore, results in an expression which has the same
functional structure as the original function. This is convenient for
numerical computations. Note that sufficiently high order derivatives become
UV-finite. Operationally, this is obvious since the subtraction polynomial
vanishes after sufficiently high derivatives are taken. However, this can also
be seen explicitly from Eq.~(\ref{diffJp}) where high powers of $x^2$ suppress
the singularity of the product of propagators at small $x$. The key integral
is given by
\begin{equation}
\int_0^\infty x^{\mu-1}K_\nu(Mx)dx
  =2^{\mu-2}M^{-\mu}\Gamma\left(\frac{\mu+\nu}2\right)
  \Gamma\left(\frac{\mu-\nu}2\right).
\end{equation}
The main physical applications are given by calculations within ChPT and HQET.
The technique is especially useful for the analysis of form factors. Another
application is given by expansions of heavy quark correlators through the
conformal mapping in order to find an approximate spectral density from the
calculation of moments~\cite{Chetyrkin:1995ii,Chetyrkin:1997mb,Fleischer:dc}.
Finally, applications can be found in the analysis of some specific rules for
massive exotic states by calculating the moments of the spectral density.

\subsection{Dominating mass expansion: near the static limit}

Consider a hadron with one large mass which is of the same order as the
external momentum and all other masses are small. An example is a heavy meson
consisting of the $c$-quark and light quark(s). In the Euclidean domain the
polarization function is not difficult to compute in this limit because of the
simplicity of the configuration space representation and the high speed of
convergence of the ensuing numerical procedures. One can do even better since
this special limit can be done analytically. When expanding the propagators in
the limit of small masses one encounters powers of $x$ and $\ln(mx)$ (or,
within the framework of dimensional regularization, non-integer powers of $x$).
The remaining functions are the weight function (Bessel function of the first
kind) and the propagator of the heavy particle with the large mass $M$ which
is given by the modified Bessel function of the second kind. The general
structure of the terms in the series expansion that contribute to
$\tilde\Pi(p)$ is given by
\begin{equation}\label{JKint}
2\pi^{\lambda+1}\int_0^\infty\left(\frac{px}2\right)^{-\lambda}
  J_\lambda(px)K_\nu(Mx)x^{2\rho}dx
\end{equation}
The integrations in Eq.~(\ref{JKint}) can be done in closed form by using the
basic integral representation
\begin{eqnarray}
\lefteqn{\int_0^\infty x^\mu J_\lambda(px)K_\nu(Mx)dx}\nonumber\\
  &=&\frac{p^\lambda\Gamma((\lambda+\mu+\nu+1)/2)
  \Gamma((\lambda+\mu-\nu+1)/2)}{2^{1-\mu}M^{\lambda+\mu+1}
  \Gamma(\lambda+1)}\nonumber\\&&\qquad\times\
  _2F_1\left((\lambda+\mu+\nu+1)/2,
  (\lambda+\mu-\nu+1)/2;\lambda+1;-p^2/M^2\right)
\end{eqnarray}
where $_2F_1(a,b;c;z)$ is the hypergeometric function~\cite{Gradshteyn}. The
corresponding integrals contain integer powers of logarithms. They can be 
computed by differentiation with respect to $\mu$. Note that the maximal power
of the logarithm is determined by the number of light propagators and does not
increase with the order of the expansion: any light propagator contains only
one power of the logarithm, as can be seen from the expansion of the Bessel
function $K_\nu(\xi)$ at small $\xi$~\cite{Gradshteyn}.

Physical applications in this regime are mainly sum rules for hadrons
including a heavy flavour~\cite{Bagan:1994dy}. These are the sum rules for the
$\Lambda_c$ or $\Lambda_b$ baryons where the lowest order contribution to the
sum rules are given by the genuine two-loop sunrise-type diagram. In the case
of the $D$ and $B$ mesons only the one-loop (degenerate sunrise) diagram 
contributes.

\subsection{Threshold expansion: close to nonrelativistic physics}

The threshold region is very important for the description of heavy quarkonia
states. In the threshold region the necessary calculations considerably
simplify if one uses an appropriate effective theory. This effective theory is
designed to remove the degrees of freedom that can be treated perturbatively
from the outset and considers only the dynamics of the essential modes
relevant for the considered energy scales. For heavy quarkonia near the
threshold the effective theory is constructed on the basis of the
non-relativistic approximation for the strong interaction and is called
nonrelativistic QCD (NRQCD). Due to the use of NRQCD there has been
a great advance in describing the heavy flavour production near threshold
during the last years~\cite{Kuhn:1998uy,Penin:1998zh,Melnikov:1998pr,
Beneke:1997jm,Czarnecki:1997vz,Hoang:2000yr,Hoang:1999zc,Nagano:1999nw,
Pivovarov:1999is,Smirnov:1997gx}.

The threshold region of a sunrise-type diagram is determined by the condition
$q^2+M^2\simeq 0$ where $q$ is the Euclidean momentum and $M=\sum_im_i$ is the
threshold value for the spectral density. We introduce the Minkowskian
momentum $p$ defined by $p^2=-q^2$ which allows for an analytic continuation 
to the physical cut. Operationally this analytic continuation can be performed
by replacing $q\rightarrow ip$. To analyze the region near threshold we use
the parameter $\Delta=M-p$ which can take complex values. The parameter
$\Delta$ is more convenient in the Euclidean domain while the parameter
$E=-\Delta=p-M$ is the actual energy counted from threshold and is used in
phenomenological applications. The analytic continuation of the Fourier
transform to the Minkowskian domain has the form
\begin{equation}
\tilde\Pi(p)=2\pi^{\lambda+1}\int_0^\infty\pfrac{ipx}2^{-\lambda}
  J_\lambda(ipx)\Pi(x)x^{2\lambda+1}dx.
\end{equation}
For the threshold expansion we have to analyze the large $x$ behaviour of the
integrand, corresponding to the region which saturates the integral in the
limit $p\rightarrow M$ or, equivalently, $E\rightarrow 0$. It is convenient to
perform the analysis in a basis where the integrand has a simple large $x$
behaviour. The most important part of the integrand is the Bessel function
$J_\lambda(ipx)$ which, however, contains both rising and falling branches at
large $x$. This resembles the situation with the elementary trigonometric 
functions $\sin(z)$ and $\cos(z)$ to which the Bessel function $J_\lambda(z)$ 
is rather close in a certain sense. Indeed, the function $\cos(z)$ and
$\sin(z)$ are linear combinations of exponentials, namely
\begin{equation}
\cos(z)=\frac12\left(e^{iz}+e^{-iz}\right)
\end{equation}
and has also both rising and falling branches at large pure imaginary
arguments: the exponentials show simple asymptotic behaviour $e^{\pm z}$ at
$z=\pm i\infty$. The analogous statement is true for $J_\lambda(z)$ which can
be written as a sum of two Hankel functions,
\begin{equation}\label{hansum}
J_\lambda(z)=\frac12(H_\lambda^+(z)+H_\lambda^-(z))
\end{equation}
where $H_\lambda^\pm(z)=J_\lambda(z)\pm iY_\lambda(z)$. The Hankel functions
$H_\lambda^\pm(z)$ show a simple asymptotic behaviour at infinity,
\begin{equation}
H_\lambda^\pm(iz)\sim z^{-1/2}e^{\pm z}
\end{equation}
(cf.\ Eq.~(\ref{asyH})). Accordingly we split up $\tilde\Pi(p)$ into
$\tilde\Pi(p)=\tilde\Pi^+(p)+\tilde\Pi^-(p)$ with
\begin{equation}
\label{pm}
\tilde\Pi^\pm(p)=\pi^{\lambda+1}\int_0^\infty\pfrac{ipx}2^{-\lambda}
  H_\lambda^\pm(ipx)\Pi(x)x^{2\lambda+1}dx.
\end{equation}
The two parts $\tilde\Pi^\pm(p)$ of the polarization function $\tilde\Pi(p)$
have a completely different behaviour near threshold which allows one to
analyze them independently. This observation makes the subsequent analysis
straightforward.

\begin{figure}\begin{center}
\epsfig{figure=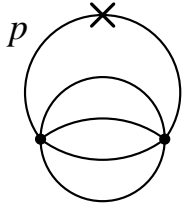, scale=0.5}
\caption{\label{thdiag1}Representation of $\tilde\Pi^+(p)$ as a standard
vacuum bubble with an additional line. The cross on the line denotes an
arbitrary number of derivatives of that line.}
\end{center}\end{figure}

We first consider the contribution of the $\tilde\Pi^+(p)$ part which reduces
to a regular sunrise-type diagram. Indeed, using the relation given by
Eq.~(\ref{KtoH}),
\begin{equation}
K_\lambda(z)=\frac{\pi i}2e^{i\lambda\pi/2}H_\lambda^+(iz)
\end{equation}
between Bessel functions of different kinds one can replace the Hankel
function $H_\lambda^+(ipx)$ with the Bessel function $K_\lambda(px)$. Since
the propagator of a massive particle (massive line in the diagram) is given by
the Bessel function up to a power in $x$, this substitution shows that the
weight function behaves like a propagator of an additional line with the
``mass'' $p$. The explicit expression is given by
\begin{equation}
\tilde\Pi^+(p)=\frac{(-2\pi i)^{2\lambda+1}}{(p^2)^\lambda}
  \int_0^\infty\Pi_+(x)x^{2\lambda+1}dx.
\end{equation}
The function $\Pi_+(x)=\Pi(x)D(x,p)$ is the polarization function of a new 
effective diagram which is equal to the initial polarization function
multiplied by a propagator with $p$ as mass parameter. We thus end up with a
vacuum bubble of the sunrise-type type with one additional line compared to
the initial diagram (see Fig.~\ref{thdiag1}). All derivatives of
$\tilde\Pi^+(p)\equiv\tilde\Pi^+(M-\Delta)$ with respect to $\Delta$ are
represented as vacuum bubbles with one additional line carrying rising
indices. Such diagrams can be efficiently calculated within the recurrence
relation technique developed in~\cite{Groote:1998ic,Groote:1998wy,
Groote:1999cx}. They possess no singularities at the production threshold
$p=M$. This can be seen by looking at the expansion for large $x$. The
behaviour at large $x$ is given by the asymptotic form of the functions for 
which one obtains
\begin{equation}
H^+(ipx)=\sqrt{\frac{2}{i\pi px}}e^{-px}(1+O(x^{-1})),\qquad
  K(mx)=\sqrt{\frac{\pi}{2 m x}}e^{-mx}(1+O(x^{-1}))
\end{equation}
(cf.\ Eqs.~(\ref{asyK}) and~(\ref{asyH})). The large $x$ range of the integral
(above a reasonably large cutoff parameter $\Lambda$) has the general form
\begin{equation}\label{pilap}
\tilde\Pi^+_\Lambda(M-\Delta)\sim\int_\Lambda^\infty x^{-a}e^{-(2M-\Delta)x}dx
\end{equation}
where
\begin{equation}\label{defa}
a=(n-1)(\lambda+1/2).
\end{equation}
The right hand side of Eq.~(\ref{pilap}) is an analytic function in $\Delta$
in the vicinity of $\Delta=0$. It exhibits no cut or other singularities near
the threshold and therefore does not contribute to the spectral density.

In contrast to the previous case, the integrand of the second part
$\tilde\Pi^-(p)$ contains $H^-(ipx)$ which behaves like a rising exponential
function at large $x$,
\begin{equation}
H^-(ipx)\sim x^{-1/2}e^{px}
\end{equation}
(cf.\ Eq.~(\ref{asyH}). Therefore, the integral is represented by
\begin{equation}\label{pilam}
\tilde\Pi^-_\Lambda(M-\Delta)\sim\int_\Lambda^\infty x^{-a}e^{-\Delta x}dx.
\end{equation}
The function $\tilde\Pi^-(M-\Delta)$ is non-analytic near $\Delta=0$ because
for $\Delta<0$ the integrand in Eq.~(\ref{pilam}) grows in the large $x$ region
and the integral diverges at the upper limit. Therefore the function which is
determined by this integral is singular at $\Delta<0$ ($E>0$) and requires
an interpretation for these values of the argument $\Delta$. The function is 
analytic in the complex $\Delta$-plane with a cut along the negative axis.
This cut corresponds to the physical positive energy cut. The discontinuity
across the cut gives rise to the non-vanishing spectral density of the diagram
(cf.\ Appendix~\ref{sec_disc}).

The integral for $\tilde\Pi^-(p)$ in Eq.~(\ref{pm}) cannot be done
analytically. In order to obtain an expansion for the spectral density near
the threshold in analytical form we make use of the asymptotic series
expansion for the function $\Pi(x)$ which crucially simplifies the integrands
but still preserves the singular structure of the integral in terms of the
variable $\Delta$. The asymptotic series expansion of the main part of each
propagator, i.e.\ of the modified Bessel function of the second kind, to the
order $N$ is given by Eq.~(\ref{asyK}). The asymptotic expansion of the
function $\Pi(x)$ therefore consists of an exponential factor $e^{-Mx}$ and
an inverse power series in $x$ up to the order $\tilde N$, where $\tilde N$
is closely related to $N$. It is this asymptotic expansion that determines the
singularity structure of the integral. We write the whole integral in the form
of a sum of two terms,
\begin{eqnarray}\label{pidias}
\tilde\Pi^-(p)&=&\pi^{\lambda+1}\int\pfrac{ipx}2^{-\lambda}H_\lambda^-(ipx)
  \left(\Pi(x)-\Pi^{as}_N(x)\right)x^{2\lambda+1+2\eps}dx\\&&
  +\pi^{\lambda+1}\int\pfrac{ipx}2^{-\lambda}H_\lambda^-(ipx)
  \Pi^{as}_N(x)x^{2\lambda+1+2\eps}dx=\tilde\Pi^{di}(p)+\tilde\Pi^{as}(p).
\nonumber
\end{eqnarray}
The integrand of the first term $\tilde\Pi^{di}(p)$ behaves as
$1/x^{\tilde N}$ at large $x$ while the integrand of the second term
accumulates all lower powers of the large $x$ expansion. Note that only the
large $x$ behaviour is essential for the near threshold expansion of the
spectral density. This fact has been taken into account in Eqs.~(\ref{pilap})
and~(\ref{pilam}) where we introduced a cutoff $\Lambda$. However, from the
practical point of view the calculation of the regularized integrals with an
explicit cutoff is inconvenient. The final result of the calculation -- the
spectral density of the diagram -- is independent of the cutoff, but the
integration is technically complicated if the cutoff is introduced. However,
in extending the integration over the whole region of the variable $x$ without
using the cutoff one immediately encounters divergences at small $x$ because
the asymptotic expansion is invalid in the region near the origin, so one is
not allowed to continue it to this region. The standard way to cope with such
a situation is to introduce dimensional regularization. It allows one to
deal with divergent expressions at intermediate stages of the calculation and
is technically simple because it does not introduce any cutoff and therefore
does not modify the integration region very much. Note that dimensional
regularization does not necessarily regularize all divergences in this case
(in contrast to the standard case of ultraviolet divergences) but it
nevertheless suffices for our purposes. We shall use a parameter $\eps$ to
regularize the divergences at small $x$.

The first part $\tilde\Pi^{di}(p)$ in Eq.~(\ref{pidias}) containing a
difference of the polarization function and its asymptotic expansion since the
integrand gives no contributions to the spectral density up to a given order
of the expansion in $\Delta$. This is so because the subtracted asymptotic
series to order $N$ cancels the inverse power behaviour of the integrand to
this degree $N$. The integrand decreases sufficiently fast for large values of
$x$ and the integral converges even at $\Delta=0$. Therefore this term is
inessential for the evaluation of the expansion of the spectral density up to
given order.

The expansion of the spectral density at small $E$ is determined only by the
integral $\tilde\Pi^{as}(p)$ in Eq.~(\ref{pidias}). This integral is still
rather complicated to compute but we can go a step further in its analytical
evaluation. Indeed, since the singular behaviour of $\tilde\Pi^{as}(p)$ is
determined by the behaviour at large $x$, we can replace the first factor,
i.e.\ the Hankel function, in the large $x$ region by its asymptotic expansion
up to some order $N$. We use Eq.~(\ref{asyH}) to obtain a representation
\begin{equation}\label{pidas}
\tilde\Pi^{das}(p)=\pi^{\lambda+1}\int\pfrac{ipx}2^{-\lambda}
  H_{\lambda,N}^{-as}(ipx)\Pi^{as}_N(x)x^{2\lambda+1+2\eps}dx.
\end{equation}
The index ``{\it das}'' stands for ``double asymptotic'' and indicates that
the integrand in Eq.~(\ref{pidas}) consists of a product of two asymptotic
expansions: one for the polarization function $\Pi(x)$ and another for the
Hankel function $H_{\lambda}(x)$ as weight (or kernel). Both asymptotic
expansions are straightforward and can be obtained from standard handbooks on
Bessel functions (cf.\ Appendix~\ref{sec_bessel}). We therefore arrive at our
final result: the integration necessary for evaluating the near threshold
expansion of the sunrise-type diagrams reduces to integrals of Euler's Gamma
function type, i.e.\ integrals containing exponentials and powers. Indeed, the
result of the expansion in Eq.~(\ref{pidas}) is an exponential function
$e^{-\Delta x}$ times a power series in $1/x$, namely
\begin{equation}\label{serx}
x^{-a+2\eps}e^{-\Delta x}\sum_{j=0}^{N-1}\frac{A_j}{x^j}
\end{equation}
where $a$ has already been defined in Eq.~(\ref{defa}) and the coefficients
$A_j$ are simple functions of the momentum $p$ and the masses $m_i$. The
expression in Eq.~(\ref{serx}) can be integrated analytically using
\begin{equation}
\int_0^\infty x^{-a+2\eps}e^{-\Delta x}dx=\Gamma(1-a+2\eps)\Delta^{a-1-2\eps}.
\end{equation}
The result is
\begin{equation}\label{serint}
\tilde\Pi^{das}(M-\Delta)
  =\sum_{j=0}^{N-1}A_j\Gamma(1-a-j+2\eps)\Delta^{a+j-1-2\eps}.
\end{equation}
This expression is our final representation for the part of the polarization
function of a sunrise-type diagram necessary for the calculation of the
spectral density near the production threshold (see Appendix~\ref{sec_disc}).
The spectral density is a function of $E=-\Delta$ and will be denoted by
$\tilde\rho(E)=\rho((M+E)^2)$ in the following. Starting from the main result
in Eq.~(\ref{serint}), we discuss the general structure in detail. In the case
where $a$ takes integer values, these coefficients result in
$1/\eps$-divergences for small values of $\eps$. The powers of $\Delta$ in
Eq.~(\ref{serint}) have to be expanded to first order in $\eps$ and give
\begin{equation}
\frac1{2\eps}\Delta^{2\eps}=\frac1{2\eps}+\ln\Delta+O(\eps).
\end{equation}
Because of
\begin{equation}
{\rm Disc\,}\ln(\Delta)\equiv\ln(-E-i0)-\ln(-E+i0)=-2\pi i\theta(E),
\end{equation}
$\tilde\Pi^{das}(M-\Delta)$ in Eq.~(\ref{serint}) contributes to the spectral
density. For half-integer values of $a$ the power of $\Delta$ itself has a
cut even for $\eps=0$. The discontinuity is then given by
\begin{equation}
{\rm Disc\,}\sqrt{\Delta}=-2i\sqrt{E}\,\theta(E).
\end{equation}
Our method to construct a threshold expansion thus simply reduces to the
analytical calculation of the integral in Eq.~(\ref{pidas}) which can be done
for arbitrary dimension and an arbitrary number of lines with different
masses. In the next paragraphs we use our technique to work out some specific
examples which demonstrate both the simplicity and efficiency of our method.

\begin{figure}[t]\begin{center}
\epsfig{figure=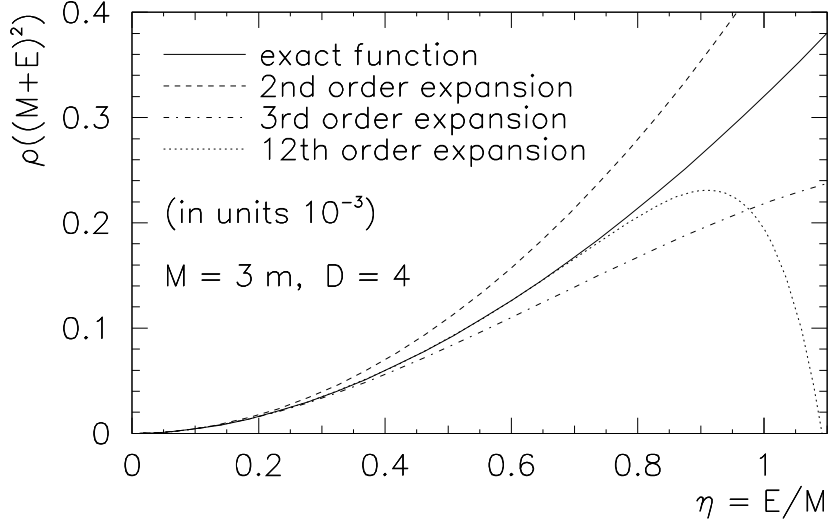, scale=0.7}
\caption{\label{fudas430}Various results for the spectral density for $n=3$
  equal masses in $D=4$ space-time dimensions as function of the threshold
  parameter $E/M$. Shown are the exact solution obtained by using
  Eq.~(\ref{exactrho4}) (solid curve) and threshold expansions for different
  orders taken from Eq.~(\ref{pidas430}) (dashed to dotted curves).}
\end{center}\end{figure}

\subsection{Equal mass genuine sunrise diagram at threshold}

The polarization function corresponding to the genuine sunrise diagram with 
three propagators with equal masses $m$ in $D=4$ space-time dimensions is
given by
\begin{equation}
\Pi(x)=\frac{m^3 K_1(mx)^3}{(2\pi)^6x^3}.
\end{equation}
The exact spectral density is given by the integral representation in
Eq.~(\ref{exactrho}) which for this particular case reads
\begin{equation}\label{exactrho4}
\rho(s)=\frac{2\pi}{i\sqrt s}\int_{c-i\infty}^{c+i\infty}
  I_1(x\sqrt s)\Pi(x)x^2dx.
\end{equation}
In order to obtain a threshold expansion of the spectral density in
Eq.~(\ref{exactrho4}) we use Eq.~(\ref{serint}) to calculate the expansion of
the appropriate part of the polarization function. To illustrate the procedure
we derive the explicit functional behaviour of the integrand in
Eq.~(\ref{pidas}) which is given by an asymptotic expansion at large $x$,
\begin{eqnarray}\label{ex}
\lefteqn{\pi^2\pfrac{ipx}2^{-1}H_{1,N}^{as}(px)\Pi_N^{as}(x)
  x^{3+2\eps}\ =\ \frac{m^{3/2}e^{(p-3m)x}}{(4\pi)^3p^{3/2}}x^{-3+2\eps}\
  \times}\nonumber\\&&\qquad\times\ \left\{1+\frac9{8mx}-\frac3{8px}
  +\frac9{128m^2x^2}-\frac{27}{64mpx^2}-\frac{15}{128p^2x^2}+O(x^{-3})\right\}.
\end{eqnarray}
From Eq.~(\ref{ex}) we can easily read off the coefficients $A_j$ that enter
the expansion in Eq.~(\ref{serx}). The spectral density is obtained by
performing a term-by-term integration of the series in Eq.~(\ref{ex}) and
by evaluating the discontinuity across the cut along the positive energy axis
$E>0$. The result reads
\begin{eqnarray}
\label{pidas430}
\lefteqn{\tilde\rho(E)\ =\ \frac{E^2}{384\pi^3\sqrt 3}
  \Bigg\{1-\frac12\eta+\frac7{16}\eta^2-\frac38\eta^3
  +\frac{39}{128}\eta^4-\frac{57}{256}\eta^5}\qquad\\&&
  +\frac{129}{1024}\eta^6-\frac3{256}\eta^7
  -\frac{4047}{32768}\eta^8+\frac{18603}{65536}\eta^9
  -\frac{248829}{524288}\eta^{10}+O(\eta^{11})\Bigg\}\nonumber
\end{eqnarray}
where the notation $\eta=E/M$, $M=3m$ is used. The simplicity of the
derivation is striking. By no cost it can be generalized to any number of
lines, arbitrary masses, and any space-time dimension. The genuine equal mass
sunrise has been chosen for definiteness only. It also allows us to compare
our results with results available in the literature. Eq.~(\ref{pidas430})
reproduces the expansion coefficients $\tilde a_j$ obtained in
Ref.~\cite{Davydychev:1999ic} (the fourth column in Table~1 of
Ref.~\cite{Davydychev:1999ic}) by a direct integration in momentum space
within the technique of region separation~\cite{Smirnov:1999bz}. In
Fig.~\ref{fudas430} the exact solution is shown together with expansions to
various orders.

\subsection{Three loops and a route to any number of loops at threshold}

The sunrise-type diagrams with four or more propagators cannot be easily done
by using the momentum space technique because it requires the multi-loop
integration of entangled momenta. As emphasized before the configuration 
space technique allows one to immediately generalize our previous results to
any number of lines (or loops) with no additional effort. Consider first the
three-loop case of sunrise-type diagrams. The polarization function of the
equal mass sunrise-type diagram with four propagators in $D=4$ space-time is
given by
\begin{equation}
\Pi(x)=\frac{m^4 K_1(mx)^4}{(2\pi)^8x^4}.
\end{equation}
The exact spectral density of this diagram can be obtained from
Eq.~(\ref{exactrho4}) while the near threshold expansion can be found using
Eq.~(\ref{serint}). The expansion of the spectral density near threshold reads
\begin{eqnarray}\label{pidas440}
\lefteqn{\tilde\rho(E)\ =\ \frac{E^{7/2}M^{1/2}}{26880\pi^5\sqrt2}
\Bigg\{1-\frac14\eta+\frac{81}{352}\eta^2-\frac{2811}{18304}\eta^3
  +\frac{17581}{292864}\eta^4}\\&&\kern-16pt
  +\frac{1085791}{19914752}\eta^5-\frac{597243189}{3027042304}\eta^6
  +\frac{4581732455}{12108169216}\eta^7
  -\frac{496039631453}{810146594816}\eta^8+O(\eta^9)\Bigg\}\nonumber
\end{eqnarray}
where $\eta=E/M$ and $M=4m$ is the threshold value. One sees the difference
with the previous three-line case. In Eq.~(\ref{pidas440}) the cut corresponds
to a square root branch cut while in the three-line case one has a logarithmic 
cut. One can easily figure out the reason for this by looking at the asymptotic
structure of the integrand. For an even number of lines (i.e.\ odd number of
loops) one has a square root branch cut, while for an odd number of lines (even
number of loops) one has a logarithmic branch cut. This statement generalizes 
from $D=4$ to any even space-time dimension. In the general case the structure
of the cut depends on the dimensionality of the space-time. The general
formula for the $n$-loop case reads
\begin{equation}\label{genthr}
\tilde\rho(E)\sim E^{(\lambda+1/2)n-1}(1+O(E)).
\end{equation}
For $D=4$ space-time dimension (i.e.\ $\lambda=1$) we can verify the result of
Ref.~\cite{Groote:1998ic,Groote:1998wy,Groote:1999cx},
\begin{equation}
\tilde\rho(E)\sim E^{(3n-2)/2}(1+O(E)).
\end{equation}

Returning to Eq.~(\ref{pidas440}) one has numerically
\begin{eqnarray}\label{sernum}
\lefteqn{\tilde\rho(E)\ =\ 8.5962\cdot 10^{-5}E^{7/2}M^{1/2}
   \Big\{1.000-0.250\eta+0.230\eta^2}\\&&
  -0.154\eta^3+0.060\eta^4+0.055\eta^5-0.197\eta^6+0.378\eta^7-0.612\eta^8
  +O(\eta^9)\Big\}\nonumber
\end{eqnarray}
where we have written down the coefficients up to three decimal places. It is
difficult to say anything definite about the convergence of this series. By
construction it is an asymptotic series. However, we stress that the practical
(or explicit) convergence can always be checked by comparing series expansions
like the one shown in Eq.~(\ref{sernum}) with the exact spectral density given
in Eq.~(\ref{exactrho4}) by numerical integration.

We conclude that the spectral density of the sunrise-type diagram can be
efficiently calculated within the configuration space technique. It does not
matter whether one is aiming for the exact result or an expansion the
configuration space technique which can readily deliver the desired result.
The exact formula in Eq.~(\ref{exactrho4}) as well as the threshold expansion
obtained from it can be used to calculate the spectral density for an
arbitrarily large number of internal lines. We stress that the case of
different masses does not lead to any complications within the configuration
space technique: the exact formula in Eq.~(\ref{exactrho}) and/or the near
threshold expansion work equally well for any configuration of masses. We do
not present plots for general cases of different masses because they are not
very illuminating as one can only see the common threshold. However, there is
an interesting kinematic regime for the different mass case which is important
for certain applications which, to the best of our knowledge, has not been
treated in the literature before. An analytical solution for the expansion of
the spectral density in this regime is given in the next subsection.

\subsection{New features of the threshold expansion: resummation of small mass
  effects for strongly asymmetric mass configurations}

The main question in constructing expansions is their practical usefulness and
the region of validity. The threshold expansion for equal or almost equal
masses breaks down for instance for $E\approx M=\sum m_i$. However, if the
masses are not equal, the region of the breakdown of the expansion is
determined by the mass with the smallest numerical value. The simplest example
where one can see this phenomenon is the analytical expression for the
spectral density of the simple loop (degenerate sunrise-type diagram) with two
different masses $m_1$ and $m_2$. In $D=4$ space-time dimensions (see e.g.\ 
Ref.~\cite{Groote:1998ic,Groote:1998wy,Groote:1999cx}) one has
\begin{equation}\label{rho42m}
\tilde\rho(E)=\frac{\sqrt{E(E+2m_1)(E+2m_2)(E+2M)}}{(4\pi(M+E))^2}
\end{equation}
where $M=m_1+m_2$. The threshold expansion is obtained by expanding the right
hand side of Eq.~(\ref{rho42m}) in $E$ for small values of $E$. If $m_2$ is
much smaller than $m_1$, the expansion breaks down at $E\approx 2m_2$. The
break-down of the series expansion can also be observed in more general cases.
If one of the masses (which we call $m_0$) is much smaller than the other
masses, the threshold expansion is only valid in a very limited region
$E\simle 2m_0$.

To generalize the expansion and extend it to the region of $E\sim M$ one has
to treat the smallest mass exactly. In this case one can use a method which
we call the resummation of small mass effects~\cite{Groote:2000kz}. This
method will be explained in the following. We start with the representation
\begin{equation}\label{pipas}
\tilde\Pi^{pas}(p)=\pi^{\lambda+1}\int\pfrac{ipx}2^{-\lambda}
  H_{\lambda,N}^{-as}(ipx)\Pi^{as}_{m_0}(x)x^{2\lambda+1+2\eps}dx
\end{equation}
which is the part of the polarization function contributing to the spectral
density. The integrand in Eq.~(\ref{pipas}) has the form
\begin{equation}
\Pi^{as}_{m_0}(x)=\Pi^{as}_{n-1}(x)D(m_0,x)
\end{equation}
where the asymptotic expansions are substituted for all the propagators except
for the one with the small mass $m_0$. This is indicated by the index
``{\it pas}'' in Eq.~(\ref{pipas}) which stands for ``partial asymptotic''.
The main technical observation leading to the generalization of the expansion
method is that $\tilde\Pi^{pas}(p)$ is still analytically computable in a
closed form. Indeed, the genuine integral to compute has the form
\begin{eqnarray}\label{intm0}
\lefteqn{\int_0^\infty x^{\mu-1}e^{-\tilde\alpha x}K_\nu(\beta x)dx
  \ =}\nonumber\\
  &=&\frac{\sqrt\pi(2\beta)^\nu}{(2\tilde\alpha)^{\mu+\nu}}
  \frac{\Gamma(\mu+\nu)\Gamma(\mu-\nu)}{\Gamma(\mu+1/2)}\
  {_2F_1}\left(\frac{\mu+\nu}2,\frac{\mu+\nu+1}2;\mu+\frac12;
  1-\frac{\beta^2}{\tilde\alpha^2}\right)
\end{eqnarray}
where $\tilde\alpha=\Delta-m_0$ and $\beta=m_0$. The integral
$\tilde\Pi^{pas}(p)$ in Eq.~(\ref{pipas}) is thus expressible in terms of
hypergeometric functions~\cite{AbramowitzStegun}. For the construction of the
spectral density one has to find the discontinuity of the right hand side of
Eq.~(\ref{intm0}). There are several ways to do this. We proceed by applying
the discontinuity operation to the integrand of the integral representation of
the hypergeometric function. The resulting integrals are calculated again in
terms of hypergeometric functions. Indeed,
\begin{eqnarray}\label{genuine}
\lefteqn{\frac1{2\pi i}\Disc\int_0^\infty x^{\mu-1}e^{\alpha x}
  K_\nu(\beta x)dx\ =}\nonumber\\
  &=&\frac{2^\mu(\alpha^2-\beta^2)^{1/2-\mu}}{\alpha^{1/2-\nu}\beta^\nu}
  \frac{\Gamma(3/2)}{\Gamma(3/2-\mu)}\ {_2F_1}\left(\frac{1-\mu-\nu}2,
  \frac{2-\mu-\nu}2;\frac32-\mu;1-\frac{\beta^2}{\alpha^2}\right)
\end{eqnarray}
where $\alpha=E+m_0$. The final expression in Eq.~(\ref{genuine}) completely
solves the problem of the generalization of the near threshold expansion
technique. For integer values of $\mu$ there are no singular Gamma functions
(with negative integer argument). Therefore we abolish the regularization
and set $\eps=0$ when using this expression. We have thus found a direct
transition from the polarization function (expressed through the integral) to
the spectral density in terms of one hypergeometric function for each genuine
integral. There is no need to use the recurrence relations available for
hypergeometric functions.

As an example we present the (over)simplified case of the two-line diagram
with masses $m$ and $m_0\ll m$ in four space-time dimensions. We cite this
example because the expansion of the spectral density and its generalized
expansion can be readily compared analytically with the exact result in
Eq.~(\ref{rho42m}). Both expansions, the pure and the resummed expansion,
will be compared with the exact anaytical result.

\begin{figure}[t]\begin{center}
\epsfig{figure=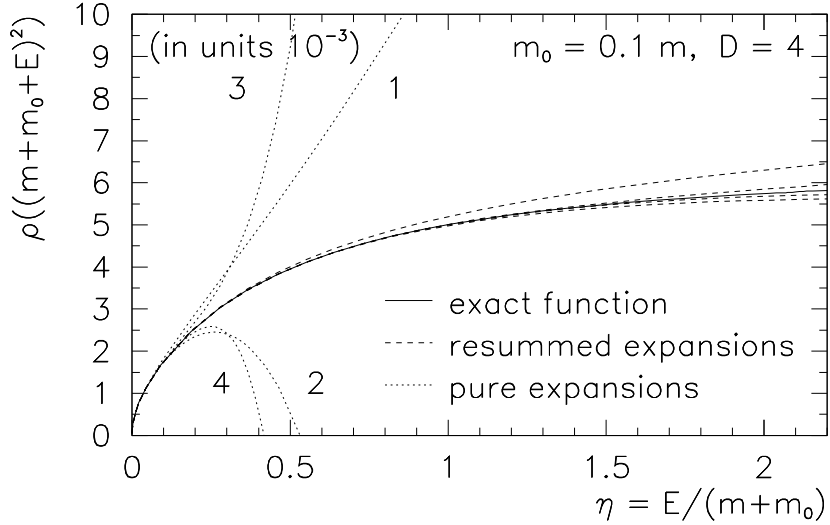, scale=0.7}
\caption{\label{fudas42t}Various solutions for the one-loop spectral density 
for two different masses $m$ and $m_0\ll m$ in $D=4$ space-time dimensions. 
Shown are the exact solution which is obtained by using Eq.~(\ref{exactrho4})
(solid curve), the pure threshold expansions using Eq.~(\ref{pidas42t})
(dotted curves), and the solutions for the resummation of small mass effects
such as in Eq.~(\ref{pipas42t}) (dashed curves). For both case the expansions
are shown from the first up to the fourth order in the asymptotic expansion.
For the pure threshold expansion the order is indicated explicitly.}
\end{center}\end{figure}

The pure expansion of the spectral density near threshold (the second order
asymptotic expansion should suffice to show the general features in a short
and concise form) is given by
\begin{eqnarray}\label{pidas42t}
\tilde\rho^{das}(E)&=&\frac{\sqrt{2m_0mE}}{8\pi^2M^{3/2}}
  \Bigg\{1+\left(\frac1m+\frac1{m_0}-\frac7M\right)\frac{E}4\nonumber\\&&
  -\left(\frac1{m_0^2}+\frac1{m^2}+\frac{12}{m_0m}-\frac{79}{M^2}\right)
  \frac{E^2}{32}+O(E^3)\Bigg\}
\end{eqnarray}
where $M=m+m_0$. As mentioned above, this series breaks down for $E>2m_0$
(see Eq.~(\ref{rho42m})). The analytical expression for the spectral density
of the polarization function in Eq.~(\ref{pipas}) within the generalized
asymptotic expansion based on Eq.~(\ref{genuine}) is given by
\begin{eqnarray}\label{pipas42t}
\lefteqn{\tilde\rho^{pas}(E)\ =\ \frac{\sqrt{mE(E+2m_0)}}{8\pi^2(E+M)^{3/2}}
  \Bigg\{{_2F_1}\left(0,\frac12;\frac32;1-\frac{m_0^2}{(E+m_0)^2}\right)}
  \nonumber\\&&+\frac{E(E+2m_0)}{8m(E+M)}\
  {_2F_1}\left(\frac12,1;\frac52;1-\frac{m_0^2}{(E+m_0)^2}\right)\\&&
  -\frac{E^2(E+2m_0)^2}{128m^2(E+M)^2}\left(1+\frac{16m(E+M)}{5(E+m_0)^2}
  \right)\
  {_2F_1}\left(1,\frac32;\frac72;1-\frac{m_0^2}{(E+m_0)^2}\right)+\ldots
  \Bigg\}.\nonumber
\end{eqnarray}
We have set the regularization parameter $\eps=0$ because the spectral density
is finite. With $\eps=0$ the resulting expressions for the hypergeometric
functions in Eq.~(\ref{genuine}) simplify. The first term in the curly
brackets of Eq.~(\ref{pipas42t}) is obviously equal to $1$ in this limit
because the first parameter of the hypergeometric function vanishes for
$\eps=0$. However, we keep Eq.~(\ref{pipas42t}) in its given form to show the
structure of the contributions. The generalized threshold expansion has the
form
\begin{eqnarray}\label{genStr}
\tilde\rho^{pas}(E)=g_0(E,m_0)+Eg_1(E,m_0)+E^2g_2(E,m_0)+\ldots
\end{eqnarray}
where the functions $g_j(E,m_0)$ represent effects of the resummation of small
mass effects and are not polynomials in the threshold parameter $E$. In the
simple one-loop case the hypergeometric functions reduce to elementary
functions. For instance,
\begin{eqnarray}
\label{resumSm}
\lefteqn{{_2F_1}\left(\frac12,1;\frac52;1-\frac{m_0^2}{(E+m_0)^2}\right)\ =}\\
  &=&\frac{3(E+m_0)}{2E(E+2m_0)}\left(E+m_0-\frac{m_0^2}{2\sqrt{E(E+2m_0)}}
  \ln\pfrac{E+m_0+\sqrt{E(E+2m_0)}}{E+m_0-\sqrt{E(E+2m_0)}}\right).\nonumber
\end{eqnarray}
Higher order contributions are given by hypergeometric functions with larger
numerical values of the parameters. They can be simplified by using Gaussian
recurrence relations for hypergeometric functions.\footnote{Note that
Eq.~(\ref{resumSm}) does not lead to the exact function in Eq.~(\ref{rho42m})
because terms of order $E^N$ are missing which originate from the difference
part $\tilde\Pi^{di}(p)$ of the correlator. The difference part simply
corrects the behaviour of the coefficient functions by the small mass
contributions.}

In Fig.~\ref{fudas42t} we compare the exact result of Eq.~(\ref{exactrho4})
with the pure expansion in Eq.~(\ref{pidas42t}) and the resummed expansion in
Eq.~(\ref{pipas42t}). While the pure expansion breaks down already at the order
$E\sim m_0$, the convergence of the expansion in Eq.~(\ref{pipas42t}) breaks
down only at $E\sim M=m+m_0$. The resummation leads to an essential
improvement of the convergence in comparison with the pure threshold expansion.
Further examples and their discussion can be found in
Ref.~\cite{Groote:2000kz}.

\subsection{Large $s$ expansion: Expansion of the spectrum at large energies}

In the Minkowskian domain where $s=p^2$ the situation is closely related to
the large $p$ expansion if all masses are small compared to the external
momentum. We start with a simple example which will be calculated in momentum
space as well as in configuration space. In the next paragraph we give more
involved examples.

As a first example for an expansion in $m^2/s$ we look at a case where the
spectral density is given by a finite series only. For the one-loop
sunrise-type diagram with one massive and one massless line we can use
Feynman parametrization and obtain in (Euclidean) momentum space
\begin{equation}
\tilde\Pi(p)=\frac{\Gamma(\eps)}{(4\pi)^{2-\eps}}
  \int_0^1(1-x)^{-\eps}\left(xp^2+m^2\right)^{-\eps}dx.
\end{equation}
In order to determine the spectral desity, the discontinuity can be calculated
already on the level of the integrand. A discontinuity appears if $xp^2+m^2<0$ 
which is satisfied for $p^2<-m^2$. We refer to Appendix~\ref{sec_disc} for
more details. If we parametrize $p^2=se^{i\varphi}$, we can reach the cut from
both sides $\varphi=-\pi$ and $\varphi=+\pi$ and can calculate the difference,
i.e.\ the discontinuity
\begin{equation}
\Disc(m^2-xs)^{-\eps}=\frac{2\pi i}{\Gamma(\eps)\Gamma(1-\eps)}
  (xs-m^2)^{-\eps}\theta(xs-m^2).
\end{equation}
For the spectral density we obtain
\begin{equation}
\rho(s)=\frac1{2\pi i}\Disc\tilde\Pi(p)\Big|_{p^2=-s}
  =\frac1{(4\pi)^{2-\eps}\Gamma(1-\eps)}\int_{m^2/s}^1(1-x)^{-\eps}
  \left(xs-m^2\right)^{-\eps}dx.
\end{equation}
Because the general factor is no longer singular, we can calculate the
integral for $\eps=0$ where the integrand is simply unity. Then we obtain
\begin{equation}\label{rho1loop}
\rho(s)=\frac1{(4\pi)^2}\int_{m^2/s}^1dx
  =\frac1{(4\pi)^2}\left(1-\frac{m^2}s\right).
\end{equation}
This result can also be obtained by using configuration space techniques. At
first sight one may think that this example is too simple to be worth to be
calculated using configuration space techniques. Later on we shall see that
for the multi-loop case the configuration space method is the only one which
leads to an analytical result.

In configuration space the spectral density of the one-loop diagram with one
massive and one massless line in four-dimensional spacetime is given by
Eq.~(\ref{exactrho4}). The integral contains the modified Bessel function of
the first kind, $I_1(z)$. Because $K_1(x)$ is a pure exponentially decreasing
function in its asymptotic expansion while $K_1(x)\pm i\pi I(x)$ increases
exponentially along the real axis if continued in the upper resp.\ lower
complex plane (note the ambiguity due to Stokes' phenomenon mentioned in
Appendix~\ref{sec_disc}), we can divide according to
\begin{eqnarray}
i\pi\int_\epsilon^{\epsilon+i\infty}I_1(z)f(z)dz
  &=&\int_\epsilon^{\epsilon+i\infty}(i\pi I_1(z)-K_1(z))f(z)dz
  +\int_\epsilon^{\epsilon+i\infty}K_1(z)dz,\\
i\pi\int_{\epsilon-i\infty}^\epsilon I_1(z)f(z)dz
  &=&\int_{\epsilon-i\infty}^\epsilon(i\pi I_1(z)+K_1(z))f(z)dz
  -\int_{\epsilon-i\infty}^\epsilon K_1(z)dz
\end{eqnarray}
and can lower the paths to the negative resp.\ positive real axis by using
Cauchy's theorem realizing that the quarter circle integrals in the
corresponding quadrant vanish (see Ref.~\cite{Groote:1998wy}). The origin and
the negative real axis is circumvented by two half-circle paths $C_-$ (lower
half plane) and $C_+$ (upper half plane) in positive direction. The result
found in Ref.~\cite{Groote:1998wy} reads
\begin{eqnarray}\label{pathint}
\lefteqn{i\pi\int_{c-i\infty}^{c+i\infty}I_1(x\sqrt s)\Pi(x)x^2dx
  \ =\ \int_\epsilon^\infty
  K_1(r\sqrt s)\left(2\Pi(r)-\Pi(e^{i\pi}r)-\Pi(e^{-i\pi}r)\right)r^2dr}\\&&
  +\int_{C_-}(i\pi I_1(z\sqrt s)+K_1(z\sqrt s))\Pi(z)z^2dz
  +\int_{C_+}(i\pi I_1(z\sqrt s)-K_1(z\sqrt s))\Pi(z)z^2dz.\nonumber
\end{eqnarray}
For our simple example we have to take the correlator function
$\Pi(x)=D(x,m)D(x,0)$. While the massless propagator is invariant under the
multiplication of $e^{\pm i\pi}=\pm 1$ of its argument, i.e.\
$D(-x,0)=D(x,0)$, for the massive propagator we obtain
\begin{equation}
D(e^{\pm i\pi}r,m)=\frac{(mr)}{(2\pi)^2r^2}\left(K_1(mr)\pm i\pi I_1(mr)\right)
\end{equation}
where $K_\lambda(e^{\pm i\pi}x)=e^{\mp i\pi\lambda}K_\lambda(x)\mp i\pi
I_\lambda(x)$ is used. We obtain $2\Pi(r)-\Pi(e^{i\pi}r)-\Pi(e^{-i\pi}r)=0$
which results in the vanishing of the first part in Eq.~(\ref{pathint}). We
are therefore left with the two semicircle integrals which we can combine in a
different manner
\begin{equation}
\rho(s)=\frac{-2\pi i}{\sqrt s}\int_{C_-+C_+}I_1(z\sqrt s)\Pi(z)z^2dz
  -\frac2{\sqrt s}\int_{C_--C_+}K_1(z\sqrt s)\Pi(z)z^2dz.
\end{equation}
In expanding the first integrand for small values of $r=|z|$ we obtain
\begin{equation}
I_1(z\sqrt s)\Pi(z)z^2=\frac{\sqrt s}{2(2\pi)^4z}+O(z),
\end{equation}
the integral results in
\begin{equation}
\int_{C_-+C_+}I_1(z\sqrt s)\Pi(z)z^2dz
  =\frac{\sqrt s}{2(2\pi)^4}\Big[\ln r+i\varphi\Big]_{-\pi}^\pi
  =\frac{i\sqrt s}{16\pi^3}.
\end{equation}
For the second integrand the expansion results in many more terms,
\begin{eqnarray}
K_1(z\sqrt s)\Pi(z)z^2&=&\frac1{(2\pi)^4z^3\sqrt s}
  +\frac1{4(2\pi)^4z\sqrt s}(m^2\ln(m^2)+s\ln s)\nonumber\\&&
  +\frac{m^2+s}{2(2\pi)^4z\sqrt s}\left(\ln\pfrac z2+\gamma_E-\frac12\right).
\end{eqnarray}
However one has
\begin{eqnarray}
\int_{C_--C_+}\frac{dz}{z^3}&=&\left[\frac{-1}{2z^2}\right]_{C_--C_+}
  \ =\ \left[\frac{-1}{2r^2e^{2i\varphi}}\right]_{\varphi=-\pi}^0
  -\left[\frac{-1}{2r^2e^{2i\varphi}}\right]_{\varphi=0}^\pi=0,\nonumber\\
\int_{C_--C_+}\frac{dz}z&=&\Big[\ln z\Big]_{C_--C_+}
  \ =\ \Big[\ln\epsilon+i\varphi\Big]_{\varphi=-\pi}^0
  -\Big[\ln\epsilon+i\varphi\Big]_{\varphi=0}^\pi\ =\ \pi-\pi=0.\qquad
\end{eqnarray}
The only non-vanishing contribution is given by
\begin{equation}
\int_{C_--C_+}\frac{dz}z\ln z=\frac12\Big[\ln^2z\Big]_{C_--C_+}=\pi^2.
\end{equation}
Therefore, we obtain
\begin{equation}
\int_{C_--C_+}K_1(z\sqrt s)\Pi(z)z^2dz=\frac{m^2+s}{32\pi^2\sqrt s}.
\end{equation}
The spectral density, finally, reads again
\begin{equation}
\rho(s)=\frac{-2\pi i}{\sqrt s}\ \frac{i\sqrt s}{16\pi^3}
  -\frac2{\sqrt s}\ \frac{m^2+s}{32\pi^2\sqrt s}
  =\frac1{16\pi^2}\left(1-\frac{m^2}s\right)
\end{equation}
as in Eq.~(\ref{rho1loop}).

\subsection{Large $s$ expansion for massive one- and multiloop diagrams}

A second example that we treat is the one-loop diagram with two different
masses $m_1$ and $m_2$. We start again with the calculation in momentum space
to obtain
\begin{equation}
\tilde\Pi(p)=\frac{\Gamma(2-D/2)}{(4\pi)^{D/2}}\int_0^1
  \left(x(1-x)p^2+(1-x)m_1^2+xm_2^2\right)^{D/2-2}dx.
\end{equation}
Again the discontinuity can be calculated already on the level of
integrands. Using Eq.~(\ref{discpow}) we obtain
\begin{eqnarray}
\lefteqn{\Disc\left(-x(1-x)s+(1-x)m_1^2+xm_2^2\right)^{-\eps}\ =}\\
  &=&\frac{2\pi i}{\Gamma(\eps)\Gamma(1-\eps)}
  \left(x(1-x)s-(1-x)m_1^2-xm_2^2\right)^{-\eps}
  \theta\left(x(1-x)s-(1-x)m_1^2-xm_2^2\right).\nonumber
\end{eqnarray}
In order to write down the spectral density we have to determine the limits
for $x$ given by the positiveness of the theta function. The zeros of the
argument of the theta function can be calculated to be
\begin{equation}
x_{1,2}=\frac12\left(1+\frac{m_1^2}s-\frac{m_2^2}s\right)
 \pm\frac12\sqrt{\lambda(1,m_1^2/s,m_2^2/s)}
\end{equation}
where
\begin{equation}
\lambda(x,y,z)=x^2+y^2+z^2-2xy-2xz-2yz
\end{equation}
is K\"all\'en's lambda function. In terms of these two zeros $x_1\le x_2$ we
can write the condition for the argument of the theta function as
\begin{equation}
(x-x_1)(x_2-x)s\ge 0\quad\Rightarrow\quad x_1\le x\le x_2.
\end{equation}
The two zeros take real values if
\begin{equation}
s^2\lambda(1,m_1^2/s,m_2^2/s)
  =\left(s-(m_1-m_2)^2\right)\left(s-(m_1+m_2)^2\right)\ge 0.
\end{equation}
This is the case for $s\ge(m_1+m_2)^2$ or $s\le(m_1-m_2)^2$. We do not consider
the part of the spectral density up to the pseudothreshold $(m_1-m_2)^2$
because in this case both $x_1$ and $x_2$ are negative. Starting from the
threshold $(m_1+m_2)^2$, both $x_1$ and $x_2$ take values between $0$ and $1$.
We obtain
\begin{eqnarray}
\rho(s)&=&\frac1{2\pi i}\Disc\tilde\Pi(-s)
  \ =\ \frac1{(4\pi)^{2-\eps}\Gamma(1-\eps)}\int_{x_1}^{x_2}
  s^{-\eps}(x-x_1)^{-\eps}(x_2-x)^{-\eps}dx\nonumber\\
  &=&\frac{s^{-\eps}\Gamma(1-\eps)}{(4\pi)^{2-\eps}\Gamma(2-2\eps)}
  \sqrt{\lambda(1,m_1^2/s,m_2^2/s)}
\end{eqnarray}
where we have used the substitution $x'=(x-x_1)/(x_2-x_1)$ and Euler's 
Beta function
\begin{equation}
B(\alpha,\beta)=\frac{\Gamma(\alpha)\Gamma(\beta)}{\Gamma(\alpha+\beta)}
  =\int_0^1x^{\alpha-1}(1-x)^{\beta-1}dx.
\end{equation}
For $D=4$, i.e.\ $\eps=0$, we obtain
\begin{equation}
\rho(s)=\frac1{(4\pi)^2}\sqrt{1-2\left(\frac{m_1^2}s+\frac{m_2^2}s\right)
  +\left(\frac{m_1^2}s-\frac{m_2^2}s\right)^2}.
\end{equation}
For the case $m_2=0$ we recover the spectral density of the previous example.
If we consider an expansion in terms of large values of $s$ compared to the 
threshold value $s=(m_1+m_2)^2$, i.e.\ if we expand in small values of $z$ for
$z=(m_1+m_2)^2/s$, we
obtain
\begin{equation}
\rho(s)=\frac1{(4\pi)^2}\left(1-\frac{m_1^2+m_2^2}s+O\pfrac1{s^2}\right).
\end{equation}
The same result shall now be calculated in a different way. Consider the
correlator function
\begin{equation}\label{PiDD}
\tilde\Pi(p)=2\pi^{\lambda+1}\int_0^\infty\pfrac{px}2^{-\lambda}J_\lambda(px)
D(x,m_1)D(x,m_2)x^{2\lambda+1}dx.
\end{equation}
We can expand the propagators $D(x,m_i)$ in $m_i/\sqrt s$, leading to pure
powers in $x$. Starting from Eq.~(\ref{largemom}), the integration of the
Bessel function of the first kind with (non-integer) powers of $x$ can be
cast into a more appropriate form
\begin{equation}
\int_0^\infty x^\mu\pfrac{px}2^{-\lambda}J_\lambda(px)x^{2\lambda+1}dx
  =\pfrac p2^{-2\lambda-2-\mu}\frac{\Gamma(\lambda+1+\mu/2)}{2\Gamma(-\mu/2)}.
\end{equation}
We can then immediately calculate the discontinuity and obtain
\begin{equation}
\frac1{2\pi i}\Disc\int_0^\infty x^\mu\pfrac{px}2^{-\lambda}
  J_\lambda(px)x^{2\lambda+1}dx
  =\frac{(s/4)^{-\lambda-1-\mu/2}}{2\Gamma(-\lambda-\mu/2)\Gamma(-\mu/2)}.
\end{equation}
Using the expansion of the Bessel functions $K_\lambda(z)$ at small values of
$z$, the expansion of the propagator for $\lambda$ close to $\lambda=1$ for
small mass $m$ (compared to $\sqrt s$) reads
\begin{eqnarray}
\lefteqn{D(x,m)\ =\ \frac{(mx)^\lambda K_\lambda(mx)}{(2\pi)^{\lambda+1}
  x^{2\lambda}}}\nonumber\\
  &=&\frac{2^{\lambda-1}\Gamma(\lambda)}{(2\pi)^{\lambda+1}x^{2\lambda}}
  \left(1-\frac1{1-\lambda}\pfrac{mx}2^2
  -\frac{\Gamma(1-\lambda)}{\Gamma(1+\lambda)}\pfrac{mx}2^{2\lambda}
  +O\left((mx)^4,(mx)^{2+2\lambda}\right)\right).\qquad
\end{eqnarray}
Eq.~(\ref{PiDD}) contains two of these propagators. We now expand up to the
power $m^2$ and keep $\lambda$ close to $1$. We obtain a zeroth order
contribution, two contributions of the order $m^2$, and two contributions of
the order $m^{2(1+\lambda)}$. We consider these contributions in turn.
\begin{itemize}
\item zeroth order ($\mu=-4\lambda$):
\begin{equation}
\frac{2\pi^{\lambda+1}(s/4)^{\lambda-1}}{2\Gamma(2\lambda)\Gamma(\lambda)}
  \pfrac{2^{\lambda-1}\Gamma(\lambda)}{(2\pi)^{\lambda+1}}^2\ \rightarrow\
  \frac{2\pi^2}{2\Gamma(2)\Gamma(1)}\pfrac{\Gamma(1)}{(2\pi)^2}^2
  \ =\ \frac1{16\pi^2}
\end{equation}
\item second order ($\mu=2-4\lambda$, $i=1,2$):
\begin{eqnarray}
\lefteqn{\frac{2\pi^{\lambda+1}(s/4)^{\lambda-2}}{2\Gamma(2\lambda-1)
  \Gamma(\lambda-1)}\pfrac{2^{\lambda-1}\Gamma(\lambda)}{(2\pi)^{\lambda+1}}^2
  \frac{(m_i^2/4)}{1-\lambda}\qquad
  \Big(\Gamma(\lambda)=(\lambda-1)\Gamma(\lambda-1)\Big)}\\
  &=&\frac{-2\pi^{\lambda+1}(s/4)^{\lambda-2}}{2\Gamma(2\lambda-1)
  \Gamma(\lambda)}\pfrac{2^{\lambda-1}\Gamma(\lambda)}{(2\pi)^{\lambda+1}}^2
  \frac{m_i^2}4\ \rightarrow\ \frac{-2\pi^2(s/4)^{-1}}{2\Gamma(1)
  \Gamma(1)}\pfrac{\Gamma(1)}{(2\pi)^{\lambda+1}}^2\frac{m_i^2}4
  =\frac{-m_i^2}{16\pi^2s}\nonumber
\end{eqnarray}
\item order $m^{2(1+\lambda)}$ (also called ``second order primed'',
  $\mu=-2\lambda$, $i=1,2$):
\begin{equation}
-\frac{(s/4)^{-1}}{\Gamma(-\lambda-\mu/2)\Gamma(-\mu/2)}
  \pfrac{2^{\lambda-1}\Gamma(\lambda)}{(2\pi)^{\lambda+1}}^2
  \frac{2\Gamma(1-\lambda)}{\Gamma(1+\lambda)}\pfrac{m_i^2}4^\lambda.
\end{equation}
\end{itemize}
Since the third contribution vanishes for $\mu\to-2\lambda$, only the first
two contributions have to be taken into account. One ends up with
\begin{equation}
\rho(s)=\frac1{16\pi^2}\left(1-\frac{m_1^2}s-\frac{m_2^2}s+O\pfrac1{s^2}\right)
\end{equation}
in agreement with the result obtained earlier. Moreover, we can predict the
result for a $n$-loop sunrise-type diagram with different masses
$m_1,\ldots m_{n+1}$ in four space-time dimensions where $\mu=-2(n+1)\lambda$
(zeroth order contribution), $\mu=2-2(n+1)\lambda$ (second order
contribution), and $\mu=-2n\lambda$ (second order primed contribution which
does not vanish for $n>1$),
\begin{eqnarray}
\rho(s)&=&2\pi^{\lambda+1}
  \pfrac{2^{\lambda-1}\Gamma(\lambda)}{(2\pi)^{\lambda+1}}^{n+1}
  \frac{(s/4)^{n\lambda-1}}{2\Gamma((n+1)\lambda)\Gamma(n\lambda)}
  \Bigg(1\,+\\&&
  +\frac{((n+1)\lambda-1)(n\lambda-1)}{1-\lambda}\sum_{i=1}^{n+1}
  \frac{m_i^2}s-\frac{\Gamma((n+1)\lambda)
  \Gamma(1-\lambda)}{\Gamma((n-1)\lambda)\Gamma(1+\lambda)}
  \sum_{i=1}^{n+1}\pfrac{m_i^2}s^\lambda+O\pfrac1{s^2}\Bigg).\nonumber
\end{eqnarray}
The second and third term have to be expanded in $\eps=1-\lambda$ in order
that the singularity in $\eps$ cancels. After a few simplifications using
expecially
\begin{equation}
\psi(n+1)=\psi(n)+\frac1n
\end{equation}
for the polygamma function, we finally obtain for $\lambda=1$
\begin{equation}
\rho(s)=\frac{s^{n-1}}{(4\pi)^{2n}n!(n-1)!}
  \left(1+n\sum_{i=1}^{n+1}\left((n-1)\left(\ln\pfrac{m_i^2}s
  +2(\psi(n)+\gamma_E)\right)-n\right)\frac{m_i^2}s+O\pfrac1{s^2}\right).
\end{equation}
This result is also valid for $n=1$ and again confirms our previous results.
Note that $(\psi(n)+\gamma_E)$ is a rational number because Euler's constant
cancels out. Indeed, we have
\begin{equation}
\psi(n)+\gamma_E=\sum_{k=1}^{n-1}\frac1k.
\end{equation}
Note, however, that for $n>1$ logarithmic contributions $\ln(m_i^2/s)$ appear
in the spectral density.

\newpage

\section{Non-standard propagators and other exotic settings}

In this section we deal with modifications of the standard propagators in 
sunrise-type diagrams in any space-time dimension. First we consider
odd-dimensional space-time which in most of the cases allows one to calculate
the sunrise-type integrals in closed form. Returning to space-time dimensions
close to four, we then deal with larger powers of propagators leading to
larger values for the Bessel function indices. In this case one can apply
recurrence relations to reduce integrals containing higher powers of the
propagators to a set of master integrals~\cite{Groote:1999cx}. The master
integrals themselves show Laplace-type asymptotics (for a definition see e.g.\
Ref.~\cite{Gilkey:1975iq}). Finally we deal with nontrivial numerators in two
examples involving vacuum bubbles.

\subsection{Odd-dimensional case}

Compared to four-dimensional space-time, integrals containing products of
Bessel functions in odd dimensional space-time are analytically solvable
because the Bessel functions simplify significantly. It is interesting to note
that the evaluation of Eq.~(\ref{psun}) can be done in a closed form for any
number of internal lines in odd-dimensional space-time. One might suspect that
this special case is too simple to have any practical applications. However,
at the end of this subsection we will refer to many applications from various
fields of physics involving odd-dimensional space-time. As the simplest
example we take three-dimensional space-time $D\rightarrow D_0=3$. For $D_0=3$
(and $\lambda_0=(D_0-2)/2=1/2$) the  propagator in Eq.~(\ref{xprop}) reads
\begin{equation}
D(x,m)\rightarrow
  D_3(x,m)=\frac{\sqrt{mx}K_{1/2}(mx)}{(2\pi)^{3/2}x}=\frac{e^{-mx}}{4\pi x}.
\end{equation}
With $\lambda=\lambda_0=1/2$ the weight function becomes
\begin{equation}
\left(\frac{px}2\right)^{-1/2}J_{1/2}(px)
  =\frac2{\sqrt{\pi}}\frac{\sin(px)}{px}
\end{equation}
after angular integration. The explicit result for the $(n+1)$-line sunrise
diagram is then given by the integral
\begin{eqnarray}\label{corr3D}
\tilde\Pi(p)&=&4\pi\int_0^\infty\frac{\sin(px)}{px}
  \frac{e^{-Mx}}{(4\pi x)^{n-1}}(\mu x)^{2\epsilon}dx\nonumber\\
  &=&{\Gamma(1-n+2\epsilon)\over 2ip(4\pi)^n}
  \left[(M-ip)^{n-1-2\epsilon}-(M+ip)^{n-1-2\epsilon}\right]\mu^{2\epsilon}
\end{eqnarray}
where $\epsilon$ is used for regularization and $M=\sum m_i$. Note that we
have used an unorthodox regularization method by multiplying a factor
$(\mu x)^{2\eps}$. We will return to this point later on.

We consider some particular cases of Eq.~(\ref{corr3D}) for different values
of $n$. For $n=0$ we simply recover the (Euclidean) propagator function
$\tilde\Pi(p)=(M^2+p^2)^{-1}$ with the discontinuity 
\begin{equation}
\rho(s)=\frac{\mbox{Disc\,}\tilde\Pi(p)}{2\pi i}
  =\frac1{2\pi i}\left(\tilde\Pi(p)\Big|_{p^2=s\exp(-i\pi)}
  -\tilde\Pi(p)\Big|_{p^2=s\exp(i\pi)}\right)=\delta(s-m^2)
\end{equation}
where $s$ is the squared energy, $s=-p^2$. As remarked on earlier it is 
appropriate to call this expression the spectral density associated with the
diagram. For $n=1$ the answer for the polarization function $\tilde\Pi(p)$ is
still finite (no regularization is required) and is given by
\begin{equation}\label{2corr3D}
\tilde\Pi(p)=\frac1{8\pi ip}\ln\left(\frac{M+ip}{M-ip}\right).
\end{equation}
The spectral density, i.e.\ the discontinuity of Eq.~(\ref{2corr3D}) across
the cut in the complex $p^2$-plane is given by
\begin{equation}\label{2spec3D}
\rho(s)={1\over 8\pi\sqrt{s}}\theta(s-M^2),\qquad s=-p^2,\quad s>0 
\end{equation}
which is nothing but the three-dimensional two-particle phase space. This can 
be immediately checked by direct computation. The cases with $n>1$ have more
structure and therefore are more interesting. For the genuine sunrise diagram
with $n=2$, Eq.~(\ref{corr3D}) leads to
\begin{equation}\label{3corr3D}
\tilde\Pi(p)=\frac1{32\pi^2}\left(\frac1\epsilon-\frac{M}{ip}\ln\left(
  \frac{M+ip}{M-ip}\right)-\ln\left(\frac{M^2+p^2}{\mu^2}\right)\right).  
\end{equation}
While in Eq.~(\ref{3corr3D}) the arbitrary scale $\mu^2$ appears due to
regularization, the spectral density
\begin{equation}\label{3spec3D}
\rho(s)=\frac{\sqrt s-M}{32\pi^2\sqrt s}\,\theta(s-M^2)
\end{equation}
is independent of this scale. This is because the spectral density is again
finite and, therefore, independent of the regularization used. The general
formula for the spectral density for any $n>0$ in $D=3$ can be extracted from
Eq.~(\ref{corr3D}). It reads
\begin{equation}\label{spec3D}
\rho(s)=\frac{(\sqrt s-M)^{n-1}}{2(4\pi)^n(n-1)!\sqrt s}\,\theta(s-M^2).  
\end{equation}

We now want to comment on the relation between the momentum subtraction and
our unorthodox dimensional regularization. Taking Eq.~(\ref{corr3D}) for $n=2$
with momentum subtraction at the origin, one obtains
\begin{equation}\label{corr3Dsub}
\tilde\Pi(p)=\int_0^\infty \left(\frac{\sin(px)}{px}-1\right) 
  \frac{e^{-Mx}}{(4\pi)^2x}(\mu x)^{2\epsilon}dx
\end{equation}
which is UV-finite even for $\epsilon=0$ because there is no singularity at 
the origin. For practical computations it is convenient to keep the factor 
$(\mu x)^{2\epsilon}$ in the integrand since this factor gives a meaning to
each of the two terms in the round brackets in Eq.~(\ref{corr3Dsub}). Using
this factor we obtain
\begin{eqnarray}\label{3corr3Dsub}
\tilde\Pi(p)&=&\int_0^\infty\left(\frac{\sin(px)}{px}-1\right) 
  \frac{e^{-Mx}}{(4\pi)^2x}(\mu^2x^2)^\epsilon dx\nonumber\\
  &=&\frac{\Gamma(-1+2\epsilon)}{2ip(4\pi)^2}
  \left[(M-ip)^{1-2\epsilon}-(M+ip)^{1-2\epsilon}\right]\mu^{2\epsilon}
  -\frac{\Gamma(2\epsilon)}{(4\pi)^2}\left(\frac\mu{M}\right)^{2\epsilon}
  \qquad\nonumber\\
  &=&-\frac1{32\pi^2}\left\{\frac{M}{ip}\ln\left(\frac{M+ip}{M-ip}\right)
  +\ln\left(\frac{M^2+p^2}{M^2}\right)\right\}.
\end{eqnarray}
The poles cancel in this expression and the arbitrary scale $\mu$ changes to 
$M$. This corresponds to a transition from MS-type renormalization schemes to
the momentum subtraction scheme (in this particular case with a subtraction at
the origin). Since the spectral density $\rho(s)$ is finite it can be
computed using any regularization scheme as can be seen from
Eqs.~(\ref{3corr3D}) and~(\ref{3corr3Dsub}). 

We mention that in the three-dimensional case the spectral density $\rho(s)$
can also be found for general values of $n$ by traditional methods since the
three-dimensional case is sufficiently simple. By using the convolution for
the evaluation of spectral densities one stays in the same class of functions,
i.e.\ polynomials in the variable $\sqrt{s}$ divided by $\sqrt{s}$. The
general form of the convolution equation in $D$-dimensional space-time reads
\begin{equation}
\Phi_n(s)=\int\Phi_k(s_1)\Phi_p(s_2)\Phi_1(s,s_1,s_2)ds_1ds_2,\qquad k+p+1=n.
\end{equation}
For the particular case of three-dimensional space-time the kernel
$\Phi_1(p^2,m_1^2,m_2^2)$ is given by
\begin{equation}
(2\pi)^2\Phi_1(p^2,m_1^2,m_2^2)=\int\delta(k^2-m_1^2)
  \delta((p-k)^2-m_2^2)d^3k
\end{equation}
or explicitly by
\begin{equation}\label{convol}
\Phi_1(s,s_1,s_2)=\frac1{8\pi\sqrt s}\theta(s-(\sqrt{s_1}+\sqrt{s_2})^2).
\end{equation}
Eq.~(\ref{convol}) can be seen to be the two-particle phase space in three
dimensions (cf.\ Eq.~(\ref{2spec3D})). This is a rather simple example. 
However, our technique retains its efficiency for large $n$. 

We list some potential applications of the general results obtained in this 
paragraph for odd-dimensional space-time. In three space-time dimensions our
results can be used to compute phase space integrals for particles in jets
where the momentum along the direction of the jet is fixed~\cite{Mirkes}.
Another application can be found in phase transitions, for instance the
three-dimensional QCD which emerges as the high temperature limit of the
ordinary theory of strong interactions for the quark-gluon plasma (see
e.g.~\cite{Rajantie:1996cw,Gross:1980br,Polyakov:vu,Hatsuda:1991du}).
Three-dimensional models are also used to study the question of dynamical mass
generation and the infrared structure of the models of Quantum Field Theory in
general~\cite{Jackiw:1980kv,Mansfield:1985sj,Gusynin:1995bb} and some problems
of QCD~\cite{Bartels:2004jn}. 

Note that particular models of different space-time dimensions are very useful
because their properties may be simpler and may thus allow one to study
general features of the underlying field theory. For example, in
six-dimensional space-time the simplest model of quantum field theory $\phi^3$
is asymptotically free and can be used for simulations of some features of 
QCD~\cite{Mikhailov:1984cp}. 

As we have already mentioned before, closed form analytical results can be 
obtained for odd dimensional space-time $D=3,5,7,\ldots$. For example, the
propagator in five-dimensional space-time ($\lambda=3/2$) reads
\begin{equation}
D(x,m)\rightarrow D_5(x,m)=\frac{(mx)^{3/2}K_{3/2}(mx)}{(2\pi)^{5/2}x^3}
  =\frac{e^{-mx}}{8\pi^2 x^3}(1+mx)
\end{equation}
which assures that the basic integration can be performed in terms of
elementary functions (powers and logarithms) again.

Applications for odd space-time dimensions other than three can be found in
some models in unified field theories, or also in a general analysis of the
divergence structure in QFT. Five-dimensional models of QFT are rather popular
for general purposes~\cite{Slavnov:2000up,Dudas:2004ni,Cvetic:2004ny,
Kalinowski:1986pg}. They have useful applications for Yang-Mills theories
in five-dimensional space-time where the UV structure of the models can be 
analyzed~\cite{krasn5}.

Finally, non-standard space-time dimensions are useful in order to obtain
estimates for the standard case $D_0=4$. The feature that $K_\nu(x)>K_\mu(x)$ 
holds for modified Bessel functions of the second kind when $\nu>\mu$ and for
positive $x$-arguments leads to $K_{3/2}(x)>K_{1}(x)>K_{1/2}(x)$ which can be 
used for stringent numerical estimates for integrals containing these Bessel
functions.

We conclude this paragraph by noting that the $x$-space techniques allow 
one to compute sunrise-type diagrams in closed form in terms of elementary 
functions as long as one is dealing with odd-dimensional space-times. The 
resulting expressions are rather simple and can be directly used for 
applications. Having the complete formulas at hand, there is no need to 
expand in the parameters of the diagram such as masses or external momenta.

\subsection{Large powers of propagators}

\begin{figure}[t]\begin{center}
\epsfig{figure=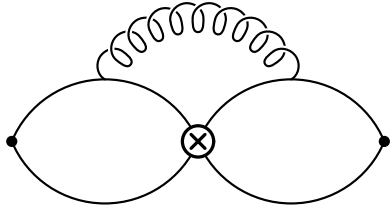, scale=0.5}
\caption{\label{bbmixing}A non-factorizable diagram at NLO for $B^0-\bar B^0$
  mixing}
\end{center}\end{figure}

\begin{figure}\begin{center}
\epsfig{figure=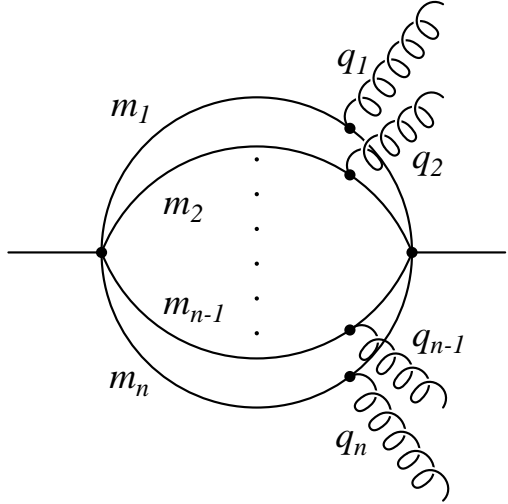, scale=0.5}
\caption{\label{mediag2}Sunrise-type diagram with additional external momenta}
\end{center}\end{figure}

In many physics applications one encounters large powers of propagators.
Recent examples are the calculation of the corrections to $B^0-\bar B^0$
mixing in perturbative QCD~\cite{Korner:2003zk,Pivovarov:2004jv} (see
Fig.~\ref{bbmixing}) or the large mass expansion for the contribution of
charged scalars to the muon anomalous magnetic moment~\cite{Kuhn:2003pu}.
When calculating moments of the spectral density of such a diagram by using
packages like MATAD for the automatic calculation of Feynman
diagrams~\cite{matad,Avdeev:db,Chetyrkin:1995ix}, high powers of propagators
are generated. Another example are sunrise-type diagrams with external lines
at small momenta $q_i$ (see Fig.~\ref{mediag2}). These diagrams appear in
calculations when there are weak external fields as, for example, in sum rule
applications~\cite{Krasulin:1997mi,Pivovarov:1992de} or in special cases of
high precision calculations in ChPT~\cite{Gasser:2002am}.

In momentum space larger powers of propagators are treated through recurrence
relations based on the integration-by-parts technique~\cite{Tkachov:wb,
Chetyrkin:qh,Baikov:1996rk}. In configuration space larger powers of the
propagators give rise to larger values of the indices for the corresponding
Bessel functions. We now describe a technique using recurrence relations which
allows one to reduce large indices of Bessel functions~\cite{Groote:1999cx}.

Note that the direct reduction of a sunrise-type diagram to a standard set of
master integrals with the help of algebraic computer systems is rather
time-consuming with present momentum space techniques. In practice the
computation proceeds through the use of a table of integrals with given powers
of the denominators. One would have to set up a three-dimensional table for a
given total power $N$. The number of entries (even when accounting for the
appropriate symmetries) still grows as fast as $N^3$ which is large for the
large values of $N$ needed in some present applications. Within our method one
first re-expresses the relevant integrals through a one-parameter set of
integrals which are then solved explicitly. For large $N$ the number of
entries increases as a first power of $N$ (the number of elements for the
$I_0(q)$ basis is given by $2[N/2]-5$ where $[z]$ is an integer part of $z$)
which considerably reduces the time consumption in a computer evaluation.

In order to show the applicability of our method we shall calculate three 
examples that were solved before with the help of momentum space 
techniques~\cite{Broadhurst:1991fi}. In Ref.~\cite{Broadhurst:1991fi}
Broadhurst considered general three-loop bubbles $B_N$. A subclass of these
are the sunrise-type three-loop bubbles. In momentum space they read
\begin{eqnarray}
\lefteqn{B_N(0,0,n_3,n_4,n_5,n_6)\ =\ \int\frac{d^Dk\ d^Dl\ d^Dp}{m^{3D}
  (\pi^{D/2}\Gamma(3-D/2))^3}\ \times}\nonumber\\&&\times\
  \frac{m^{2n_3}}{((p+k)^2+m^2)^{n_3}}\frac{m^{2n_4}}{((p+l)^2+m^2)^{n_4}}
  \frac{m^{2n_5}}{((p+k+l)^2+m^2)^{n_5}}\frac{m^{2n_6}}{(p^2+m^2)^{n_6}}
\end{eqnarray}
with two propagators absent ($n_1=n_2=0$). Actually we choose the remaining
indices $n_3$, $n_4$, $n_5$, and $n_6$ such that the results become finite.
We shall not only calculate the finite part but also the part porportional to
$\eps$ in order to be able to compare with~\cite{Broadhurst:1991fi}. Written
in configuration space, the particular subset of bubble diagrams $B_N$ is
given by
\begin{eqnarray}
\lefteqn{B_N(0,0,n_3,n_4,n_5,n_6)\ =\ 
  \frac{2(64\pi^4)^{2-\eps}}{(\Gamma(1+\eps))^3\Gamma(2-\eps)}
  m^{2(n_3+n_4+n_5+n_6)-12+6\eps}\ \times}\nonumber\\&&\times\
  \int_0^\infty D^{(n_3-1)}(x,m)D^{(n_4-1)}(x,m)D^{(n_5-1)}(x,m)
  D^{(n_6-1)}(x,m)x^{2\lambda+1}dx.
\end{eqnarray}
It is clear that one will end up with integrals of products of four Bessel
functions with non-integer indices and a non-integer power of $x$. We first
discuss three examples in the next three subsections which will be followed
by more general considerations on a general reduction procedure which allows
one to reduce all integrals containing products of Bessel functions to the
following set of two master integrals,
\begin{equation}\label{masterbasis}
L_4(r):=\int_0^\infty\left(K_0(\xi)\right)^4\xi^rd\xi\qquad\mbox{and}\qquad
L_4^l(r):=\int_0^\infty\left(K_0(x)\right)^4\xi^r\ln(e^{\gamma_E}\xi/2)\,dx
\end{equation}
where the index ``$l$'' in $L_4^l(r)$ reflects the logarithm appearing in the 
integrand. We will then add some considerations on the basic integrals
$L_4(r)$ and $L_4^l(r)$.

\begin{figure}[ht]\begin{center}
\epsfig{figure=diag2221.eps, scale=0.5}(a)\qquad
\epsfig{figure=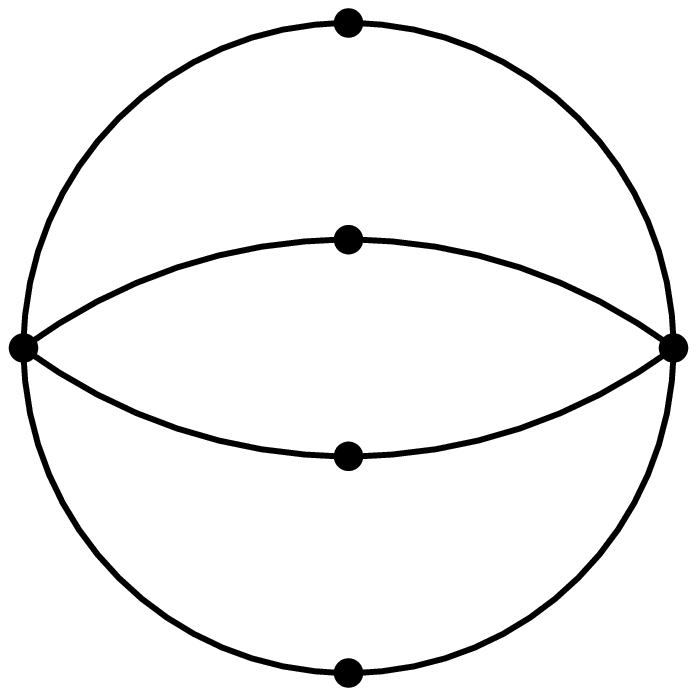, scale=0.5}(b)\qquad
\epsfig{figure=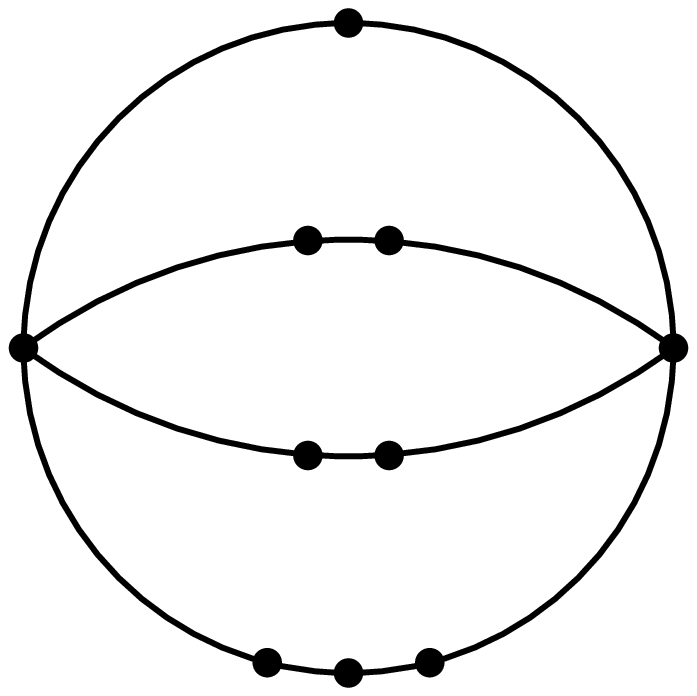, scale=0.5}(c)
\caption{\label{figBN}The diagrams for $B_N(0,0,2,2,2,1)$, $B_N(0,0,2,2,2,2)$,
  and $B_N(0,0,2,3,3,4)$.}
\end{center}\end{figure}

\subsection{The example $B_N(0,0,2,2,2,2)$}
We start with the example which is represented by the diagram in
Fig.~\ref{figBN}(b). Each of the lines is modified (indicated by the dots on
the lines) which means that instead of the propagators $D(x,m)$ we have to use
\begin{equation}
D^{(1)}(x,m)=\int\dDp\frac{e^{ip_\mu x^\mu}}{(p^2+m^2)^2}
  =\frac{(x/m)^{1-\lambda}}{2(2\pi)^{\lambda+1}}K_{\lambda-1}(mx)
  =\frac{(x/m)^\eps}{2(2\pi)^{2-\eps}}K_{-\eps}(mx).
\end{equation}
One obtains ($\xi=mx$)
\begin{equation}
B_N(0,0,2,2,2,2)=\frac{2^{1-2\eps}}{(\Gamma(1+\eps))^3\Gamma(2-\eps)}
  \int_0^\infty\left(K_{-\eps}(\xi)\right)^4\xi^{3+2\eps}d\xi.
\end{equation}
We now can use the general formula~\cite{Gradshteyn}
\begin{equation}\label{indder}
\left[\frac{\partial K_\nu(z)}{\partial\nu}\right]_{\nu=\pm n}
  =\pm\frac12n!\sum_{k=0}^{n-1}\pfrac z2^{k-n}\frac{K_k(z)}{k!(n-k)},\qquad
  n\in\{0,1,\ldots\}
\end{equation}
to expand the Bessel function in a series with respect to its index. In
case of $K_{-\eps}(z)$, however, the first derivative vanishes and we obtain
$K_{-\eps}(z)=K_0(z)+O(\eps^2)$. Therefore, in expanding
\begin{eqnarray}
\frac{2^{1-2\eps}\left(K_{-\eps}(\xi)\right)^4\xi^{3+2\eps}}{(\Gamma(1+\eps)
  )^3\Gamma(2-\eps)}
  &=&2\left(K_0(\xi)\right)^4\xi^3\left(1+(1+2\gamma_E-2\ln 2+2\ln\xi)\eps
  +O(\eps^2)\right)\ =\nonumber\\
  &=&2\left(K_0(\xi)\right)^4\xi^3\left(1+\left(1+2\ln(e^{\gamma_E}\xi/2)
  \right)\eps+O(\eps^2)\right)
\end{eqnarray}
we obtain
\begin{eqnarray}
B_N(0,0,2,2,2,2)&=&2(1+\eps)\int_0^\infty\left(K_0(\xi)\right)^4\xi^3d\xi
  +4\eps\int_0^\infty\left(K_0(\xi)\right)^4\xi^3\ln(e^{\gamma_E}\xi/2)d\xi
  \ =\nonumber\\&=&2(1+\eps)L_4(3)+4\eps L_4^l(3).
\end{eqnarray}
We have compared our numerical result with the analytical result
in~\cite{Broadhurst:1991fi} and have found agreement. In fact, from the
transcendality structure of the results in~\cite{Broadhurst:1991fi} one can
numerically obtain the appropriate coefficients multiplying the
transcendentals. One finds~\cite{Groote:1999cx}
\begin{eqnarray}
L_4(3)&=&-\frac3{16}+\frac7{32}\zeta(3),\nonumber\\
L_4^l(3)&=&\frac3{32}+\frac34\Li_4\pfrac12-\frac{17\pi^4}{1920}
  -\frac{\pi^2}{32}\left(\ln 2\right)^2+\frac1{32}\left(\ln 2\right)^4
  +\frac{49}{128}\zeta(3).
\end{eqnarray}

\subsection{The example $B_N(0,0,2,2,2,1)$\label{sec_diag2221}}

In the diagram of Fig.~\ref{figBN}(a) one of the lines is not modified.
Therefore, we have to deal with one regular propagator factor
\begin{equation}
D^{(0)}(x,m)=D(x,m)=\frac{(x/m)^{-\lambda}}{(2\pi)^{\lambda+1}}K_\lambda(mx)
  =\frac{(x/m)^{\eps-1}}{(2\pi)^{2-\eps}}K_{1-\eps}(mx).
\end{equation}
In this case we obtain
\begin{equation}
B_N(0,0,2,2,2,1)=\frac{2^{2-2\eps}}{(\Gamma(1+\eps))^3\Gamma(2-\eps)}
  \int_0^\infty\left(K_{-\eps}(\xi)\right)^3K_{1-\eps}(\xi)\xi^{2+2\eps}d\xi.
\end{equation}
Using Eq.~(\ref{indder}) one has
\begin{equation}
K_{1-\eps}(\xi)=K_1(\xi)-\frac\eps\xi K_0(\xi)+O(\eps^2)
\end{equation}
and
\begin{equation}
\frac{2^{2-2\eps}\left(K_{-\eps}(\xi)\right)^3K_{1-\eps}(\xi)
  \xi^{2+2\eps}}{(\Gamma(1+\eps))^3\Gamma(2-\eps)}
  =4\xi^2\Big(1+\Big(1-\frac1\xi+\ln(e^{\gamma_E}\xi/2)\Big)\eps
  +O(\eps^2)\Big).
\end{equation}
The result reads
\begin{eqnarray}
B_N(0,0,2,2,2,1)&=&4(1+\eps)\int_0^\infty\left(K_0(\xi)\right)^3
  K_1(\xi)\xi^2d\xi-4\eps\int_0^\infty\left(K_0(\xi)\right)^4\xi\,d\xi
  \nonumber\\&&\qquad
  +8\eps\int_0^\infty\left(K_0(\xi)\right)^3K_1(\xi)\xi^2
  \ln(e^{\gamma_E}\xi/2)d\xi.
\end{eqnarray}
This result is not yet written in terms of $L_4(r)$ and $L_4^l(r)$ and will be
dealt with after introduction of the reduction procedure (The result for
$\eps=0$ was given earlier in Sec.~\ref{finipart}).

\subsection{The example $B_N(0,0,2,3,3,4)$}

In order to demonstrate the power of the configuration space technique also in
a more complex setting we finally choose the diagram in Fig.~\ref{figBN}(c) as
an example. The modified propagators now read
\begin{equation}
D^{(2)}(x,m)=\frac{(x/m)^{1+\eps}}{8(2\pi)^{2-\eps}}K_{-1-\eps}(mx),\qquad
D^{(3)}(x,m)=\frac{(x/m)^{2+\eps}}{48(2\pi)^{2-\eps}}K_{-2-\eps}(mx)
\end{equation}
Using the expansions
\begin{equation}
K_{-1-\eps}(\xi)=K_1(\xi)+\frac\eps\xi K_0(\xi)+O(\eps^2),\qquad
K_{-2-\eps}(\xi)=K_2(\xi)+\frac{2\eps}\xi K_1(\xi)+\frac{2\eps}{\xi^2}K_0(\xi)
  +O(\eps^2)
\end{equation}
we obtain
\begin{eqnarray}\label{b2334}
\lefteqn{B_N(0,0,2,3,3,4)\ =\ \frac{2^{-6-2\eps}}{3(\Gamma(1+\eps))^2
  \Gamma(2-\eps)}\int_0^\infty K_{-\eps}(\xi)\left(K_{-1-\eps}(\xi)\right)^2
  K_{-2-\eps}(\xi)\xi^{7+2\eps}d\xi}\nonumber\\
  &=&\frac{1+\eps}{192}\int_0^\infty K_0(\xi)\left(K_1(\xi)\right)^2K_2(\xi)
  \xi^7d\xi+\frac\eps{96}\int_0^\infty K_0(\xi)\left(K_1(\xi)\right)^2K_2(\xi)
  \xi^6d\xi\qquad\qquad\nonumber\\&&
  +\frac\eps{96}\int_0^\infty K_0(\xi)\left(K_1(\xi)\right)^3\xi^6d\xi
  +\frac\eps{96}\int_0^\infty\left(K_0(\xi)\right)^2\left(K_1(\xi)\right)^2
  \xi^5d\xi\nonumber\\&&+\frac\eps{96}\int_0^\infty K_0(\xi)
  \left(K_1(\xi)\right)^2K_2(\xi)\xi^7\ln(e^{\gamma_E}\xi/2)d\xi.
\end{eqnarray}

\subsection{The reduction procedure\label{sec_reduct}}

Especially in the last expression given by Eq.~(\ref{b2334}) there are many
different integrals which differ from the basis $L_4(r)$ and $L_4^l(r)$.
However, the integrands can be reduced to integrands involving $K_0(\xi)$ and
$K_1(\xi)$ only by using the relation
\begin{equation}
K_n(\xi)=2\frac{n-1}\xi K_{n-1}(\xi)+K_{n-2}(\xi).
\end{equation}
After the first step, namely the expansion of Bessel functions for
non-integer indices, the above relation establishes the second step in our
reduction procedure. Finally, we use
\begin{eqnarray}
\frac{d}{d\xi}K_0(\xi)&=&-K_1(\xi)\qquad\mbox{and}\nonumber\\
\frac{d}{d\xi}K_1(\xi)&=&-\frac12\left(K_0(\xi)+K_2(\xi)\right)
  \ =\ -K_0(\xi)-\frac1\xi K_1(\xi)
\end{eqnarray}
to perform the third and last step. For instance one has
\begin{eqnarray}
L_4^{(1)}(r)&=&\int_0^\infty\left(K_0(\xi)\right)^3K_1(\xi)\xi^rd\xi
  \ =\ -\int_0^\infty\left(K_0(\xi)\right)^3\frac{dK_0(\xi)}{d\xi}\xi^rd\xi
  \ =\nonumber\\
  &=&-\Big[K_0(\xi)\left(K_0(\xi)\right)^3\xi^r\Big]_0^\infty
  +\int_0^\infty K_0(\xi)\frac{d}{d\xi}\left(K_0(\xi)\right)^3\xi^rd\xi
  \ =\nonumber\\
  &=&3\int_0^\infty K_0(\xi)\left(K_0(\xi)\right)^2\frac{dK_0(\xi)}{d\xi}
  \xi^rd\xi+r\int_0^\infty K_0(\xi)\left(K_0(\xi)\right)^3\xi^{r-1}d\xi
  \ =\nonumber\\
  &=&-3\int_0^\infty\left(K_0(\xi)\right)^3K_1(\xi)\xi^rd\xi
  +r\int_0^\infty\left(K_0(\xi)\right)^4\xi^{r-1}d\xi
\end{eqnarray}
and therefore
\begin{equation}
L_4^{(1)}(r)=\int_0^\infty\left(K_0(\xi)\right)^3K_1(\xi)\xi^rd\xi
  =\frac r4\int_0^\infty\left(K_0(\xi)\right)^4\xi^{r-1}d\xi
  =\frac r4L_4(r-1).
\end{equation}
The reduction formulas are given in general by
\begin{eqnarray}\label{genred}
L_n^{(m)}(r)&=&\frac1{n-m+1}\left((r-m+1)L_n^{(m-1)}(r-1)
  -(m-1)L_n^{(m-2)}(r)\right),\\
L_n^{l(m)}(r)&=&\frac1{n-m+1}\left((r-m+1)L_n^{l(m-1)}(r-1)
  +L_n^{(m-1)}(r-1)-(m-1)L_n^{l(m-2)}(r)\right)\nonumber
\end{eqnarray}
and are coded in MATHEMATICA in order to automatically reduce to the master
integrals~\cite{Groote:1999cx}. After executing the second and third step of
the recursion the results for the above examples read
\begin{eqnarray}
B_N(0,0,2,2,2,2)&=&2(1+\eps)L_4(3)+4\eps L_4^l(3)+O(\eps^2),\nonumber\\[7pt]
B_N(0,0,2,2,2,1)&=&2L_4(1)+4\eps L_4^l(1)+O(\eps^2),\nonumber\\[3pt]
B_N(0,0,2,3,3,4)&=&\frac1{36}L_4(3)-\frac1{144}L_4(5)
  -\frac1{576}L_4(7)+O(\eps).
\end{eqnarray}
Let us concentrate on the $O(\eps^0)$ terms. By matching our numerical results
to the results obtained by using the RECURSOR package~\cite{Broadhurst:1991fi},
\begin{eqnarray}
B_N(0,0,2,2,2,2)&=&-\frac38+\frac7{16}\zeta(3)+O(\eps),\nonumber\\
B_N(0,0,2,2,2,1)&=&\frac74\zeta(3)+O(\eps),\nonumber\\
B_N(0,0,2,3,3,4)&=&\frac1{576}+O(\eps)
\end{eqnarray}
we can determine the master integrals $L_4(r)$ for a few values of its
argument. One has
\begin{equation}
L_4(3)=-\frac3{16}+\frac7{32}\zeta(3),\qquad
L_4(1)=\frac78\zeta(3),\qquad
16L_4(3)-2L_4(5)-L_4(7)=1.
\end{equation}
In particular the last relation is interesting because it implies a sum rule
for the weighted product of three Bessel functions, namely
\begin{equation}
\int_0^\infty K_0(\xi)\left(K_1(\xi)\right)^2K_2(\xi)\xi^7d\xi=\frac13.
\end{equation}
This surprising identity has been cross-checked by numerical integration. Note
that for the $O(\eps^0)$ contribution only odd values of $r$ appear as
arguments in the master integral $L_4(r)$. 

\subsection{Laplace-type asymptotics}

The two master integrals Eq.~(\ref{masterbasis}) are the basic set for the
reduction procedure. A first estimate of the numerical magnitude of these
integrals can be easily inferred from the asymptotic expansion of the
integrals at large $q$,
\begin{eqnarray}\label{asymlog}
L_4(q)&=&\frac{\pi^2\Gamma(2q)}{4^{2q+1}}
  \left(1-\frac{1}{q-1/2}+O(1/q^2)\right),\nonumber\\
L_4^l(q)&=&\frac{\pi^2\Gamma(2q)}{4^{2q+1}}
  \left(\Psi(2q)+\gamma_E-3\ln2\right)\left(1-\frac{1}{q-1/2}+O(1/q^2)\right)
\end{eqnarray}
where $\Psi(x)=\Gamma'(x)/\Gamma(x)$ is the logarithmic derivative of the
$\Gamma$-function. The direct configuration space representation is the most
convenient for numerical evaluation. To see this, we modify the last term of
the integrals $L_4(q)$ and $L_4^l(q)$ by introducing parameters $\kappa$ and
$\kappa_l$, respectively. In doing this, we find that the relations
\begin{equation}
\label{kapasymlog}
L_4(q)=\frac{\pi^2\Gamma(2q)}{4^{2q+1}}
  \left(1-\frac{1}{q+\kappa}\right),
\end{equation}
\begin{equation}
L_4^l(q)=\frac{\pi^2\Gamma(2q)}{4^{2q+1}}\left(\Psi(2q)+\gamma_E-3\ln2\right)
  \left(1-\frac1{q+\kappa_l}\right)
\end{equation}
with $\kappa=0.97$ ($\kappa_l=1.17$) gives good results for the basis rational
(resp.\ log-type) integrals with an accuracy better than $1\%$ for all values
of the arguments $q>1$ ($q>3$). For $q>3$ ($q>5$) the relative accuracy is
better than $10^{-3}$. The leading order asymptotic formulas for the basis
log-type integrals are less precise for small $q$ than its analogue in the
rational case because the integrand in Eq.~(\ref{masterbasis}) is not positive
for log-type integrals. Obviously, an exact solution via recurrence relations
in momentum space will be much more complicated than these simple asymptotic
formulae. Because the cancellation of significant figures in the numerical
evaluation of terms with different transcendental structure is a dominating
feature of the momentum space recurrence procedure for large $q$, the use of
asymptotic formulas is worth considering.

\subsection{Irreducible numerator: three-loop vacuum bubble case}

Loop integrals may have numerator factors which involve the loop momenta and
cannot be expanded in terms of the denominator pole factors. In such a case
one speaks of irreducible numerator factors. Momentum space techniques can run
into problems when non-trivial numerator factors appear. Take for example the
numerator factor $(k_1\cdot k_2)$ for a $n$-loop bubble with $n+1$ massive
lines. For $n<3$ the momentum space integral can be reduced to scalar
integrals. For $n=3$, however, the numerator factorintegral is no longer
reducible~\cite{Laporta:2002pg}. The problem of irreducible numerator factors
has a straightforward solution in configuration space by the
integration-by-parts technique~\cite{Groote:2004qq}. The starting expression
involving the non-trivial numerator factor $(k_1\cdot k_2)$ (or any other
scalar product of linear independent inner moments) corresponds to
\begin{equation}
\tilde\Pi_3^*(0)=\int D(x,m)^2
  \left(\partial_\mu D(x,m)\right)\left(\partial^\mu(D(x,m))\right)d^Dx
\end{equation}
where the asterix denotes the fact that we are dealing with a non-scalar
master integral. Careful use of the integration-by-parts
identities~\cite{Chetyrkin:nn,Tkachov:1999nk}
\begin{equation}
\int\partial_\mu\left(D(x,m)\cdots D(x,m)\right)d^Dx=0,\qquad
\int\dalembertian\left(D(x,m)\cdots D(x,m)\right)d^Dx=0
\end{equation}
leads to the result
\begin{equation}
\int D(x,m)^2\left(\partial_\mu D(x,m)\right)\left(\partial^\mu(D(x,m))\right)
  d^Dx=-\frac13\int D(x,m)^3\dalembertian D(x,m)d^Dx.
\end{equation}
For the last integral we have used $(-\dalembertian+m^2)D(x,m)=\delta(x)$ and 
end up with
\begin{equation}
\int D(x,m)^3\dalembertian D(x,m)d^Dx=m^2\int D(x,m)^4d^Dx-D(0,m)^3.
\end{equation}
The value of $D(0,m)$ can be found for instance by integration in momentum
space,
\begin{equation}
D(0,m)=\frac1{(2\pi)^D}\int\frac{d^Dp}{m^2+p^2}
  =\frac{2\pi^{D/2}}{(2\pi)^D\Gamma(D/2)}\int_0^\infty
  \frac{p^{D-1}dp}{p^2+m^2}=\frac{m^{D-2}}{(4\pi)^{D/2}}\Gamma(1-D/2).
\end{equation}
In the end we obtain
\begin{equation}
\tilde\Pi_3^*(0)=-\frac{m^2}3\tilde\Pi_3(0)+\frac13D(0,m)^3
\end{equation}
where $\tilde\Pi_3(0)$ is the scalar three-loop sunrise-type bubble diagram 
without a numerator factor.

\subsection{Irreducible numerator: four-loop vacuum bubble case}

Next we consider a four-loop diagram which appears as a second independent
master integral $V_2$ of the sunrise topology in the classification of
Ref.~\cite{Laporta:2002pg}. In momentum space the second master integral $V_2$
has the additional numerator factor $(k_1\cdot k_4)^2$ as compared to the
first scalar master integral $V_1$. The second master integral reads
\begin{equation}
\tilde\Pi_4^*(0)=\int\frac{(k_1\cdot k_4)^2(2\pi)^{-4D}
  d^Dk_1d^Dk_2d^Dk_3d^Dk_4}{(m_1^2+k_1^2)
  (m_2^2+(k_2-k_1)^2)(m_3^2+(k_3-k_2)^2)(m_4^2+(k_4-k_3)^2)(m_5^2+k_4^2)}.
\end{equation}
Turning again to the equal mass case and using the configuration space
representation, this integral can be written as
\begin{equation}
\tilde\Pi_4^*(0)=\int D(x,m)^3
  (\partial_\mu\partial_\nu D(x,m))(\partial_\mu\partial_\nu D(x,m))d^Dx.
\end{equation}
It is apparent that by using integration-by-parts techniques this integral
cannot be reduced to scalar integrals and/or integrals containing
d'Alembertians. The easiest way to evaluate such an integral is to compute the
derivatives directly. This is done with the help of the relation
\begin{equation}\label{iterK}
\frac1z\frac{d}{dz}\left(z^{-\nu}K_\nu(z)\right)
  =-\left(z^{-\nu-1}K_{\nu+1}(z)\right)
\end{equation}
(cf.\ Eq.~(\ref{diffK})). Eq.(\ref{iterK}) can be iterated and gives results
for arbitrary high order derivatives of Bessel functions $K_\lambda(z)$ in
terms of the same class of Bessel functions with shifted indices and powers in
$z$. For the first derivative we obtain
\begin{equation}
\partial_\mu D(x,m)=-x_\mu\frac{m^{2\lambda+2}}{(2\pi)^{\lambda+1}}
\frac{K_{\lambda+1}(mx)}{(mx)^{\lambda+1}}.
\end{equation}
Since the resulting analytical expression for a given line of the diagram
lies in the same class as the original line, the procedure of evaluating the
integral is similar to the usual one. However, we cannot use the second
derivative
\begin{equation}\label{secondDer}
\partial_\mu\partial_\nu D(x,m)=\frac{m^{2\lambda+2}}{(2\pi)^{\lambda+1}
  (mx)^{\lambda+1}}\left(g_{\mu\nu}K_{\lambda+2}(mx)-\frac{x_\mu x_\nu}{x^2}
  K_{\lambda+1}(mx)\right)
\end{equation}
directly under the integration sign. The reason is that the propagator has to
be regarded as a distribution. There is a $\delta$-function singularity at the
origin which is not taken into account in Eq.~(\ref{secondDer}). Indeed,
contracting the indices $\mu$ an $\nu$ in Eq.~(\ref{secondDer}) one obtains
\begin{equation}
\partial_\mu\partial^\mu D(x,m)=m^2D(x,m)
\end{equation}
while the correct equation for the propagator reads
$(-\partial^2+m^2)D(x,m)=\delta(x)$. Thus, a straightforward evaluation of
derivatives is valid only for $x\neq 0$. The behaviour at the origin ($x=0$)
requires special consideration. In practice, to treat this case one should not
use higher order derivatives but stay at the level of the first derivative.

In order to deal with this case we introduce another master integral
\begin{equation}
\tilde\Pi_4^{**}(0)=\int D(x,m)\partial_\mu D(x,m)\partial_\nu
  D(x,m)\partial^\mu D(x,m)\partial^\nu D(x,m)d^Dx.
\end{equation}
The relation between the two master integrals $\Pi_4^*(0)$
and $\Pi_4^{**}(0)$ is found to be
\begin{equation}\label{v2v2bar}
\tilde\Pi_4^*(0)=3\tilde\Pi_4^{**}(0)-\frac18m^4\tilde\Pi_4(0)
  -\frac78m^2D(0,m)^4.
\end{equation}
The quantity $\tilde\Pi_4^{**}(0)$ can be calculated with the explicit use of
first order derivatives within our technique. The analytical result for the
pole part reads
\begin{equation}
{\cal N}_4\tilde\Pi_4^{**}(0)=m^{10}\left(-\frac3{8\eps^4}
  -\frac{277}{144\eps^3}-\frac{37837}{6912\eps^2}-\frac{4936643}{414720\eps}
  +O(\eps^0)\right)
\end{equation}
(${\cal N}_4=((4\pi)^{2-\eps}/\Gamma(1=\eps))^4$, cf.\ Eq.(\ref{normfac})) and
the $\eps$-expansion in the form
\begin{eqnarray}
{\cal N}_4\tilde\Pi_4^{**}(0)
  &=&m^{10}\Big(-0.375\eps^{-4}-1.923611\eps^{-3}-5.474103\eps^{-2}
  -11.90356\eps^{-1}\nonumber\\&&
  -27.99303-104.5384\eps-663.6123\eps^2-3703.241\eps^3+O(\eps^4)\Big).
\end{eqnarray}
Since the results for $\tilde\Pi_4(0)$, $\tilde\Pi_4^{**}(0)$, and $D(0,m)^4$
are known, Eq.~(\ref{v2v2bar}) can be used to obtain the final result for the
original integral,
\begin{eqnarray}
{\cal N}_4\tilde\Pi_4^*(0)&=&m^{10}\Big(-1.6875\eps^{-4}-7.8125\eps^{-3}
  -21.20964\eps^{-2}-44.76955\eps^{-1}\nonumber\\&&
  -97.07652-290.9234\eps-1719.809\eps^2-8934.731\eps^3
  +O(\eps^4)\Big)\qquad
\end{eqnarray}
which again verifies the result given in Ref.~\cite{Laporta:2002pg}.

Differentiation of the massive propagator leads to expressions of a similar
functional form which makes the configuration space technique a universal tool
for calculating any master integral of the sunrise topology. This technique is
also useful for finding master integrals. Indeed, new master integrals appear
when there is a possibility to add new derivatives into the integrands which
cannot be eventually removed by using the equations of motion or
integration-by-parts recurrence relations. But once again: without explicit
inclusion of the $\delta$-function only one derivative is allowed. Otherwise
one misses tadpole parts of the result. Therefore, the new master integral
should contain just one derivative for each line except for one line. This
allows one to enumerate the number of non-scalar master integrals, i.e.\
master integrals including non-trivial numerator factors. For instance, in the
five-loop case (six propagators) there will be only two non-scalar master
integrals.

\section{Generalization: Still within reach}

Apart from the diagrams with sunrise-type topology treated in the previous 
sections there are more involved topologies close to the sunset-type topology
that still allow for quite a simple numerical evaluation. We conclude this
report by discussing several such cases which are still in reach of the
methods presented here.

\subsection{Generalization to the spectacle topology}

\begin{figure}[t]\begin{center}
\epsfig{figure=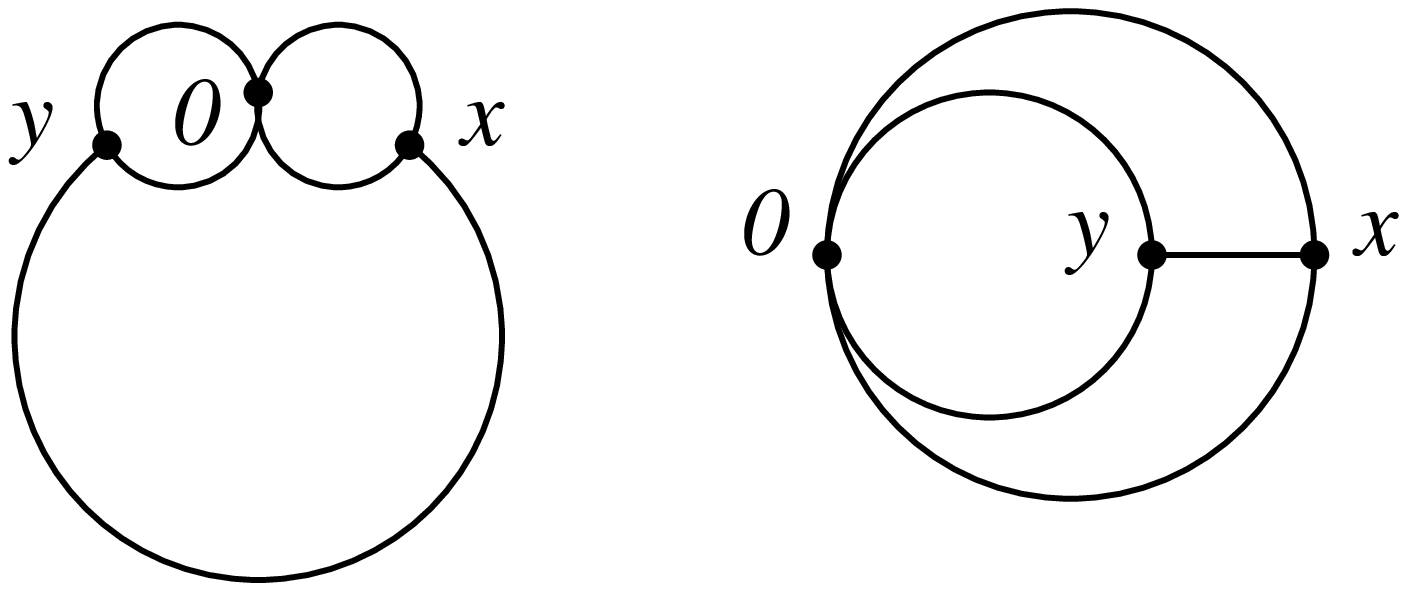, scale=0.5}\kern-4truecm(a)\kern4truecm(b)
\caption{\label{masmel2}spectacle topology diagram in two different
representations (a) (left hand side) and (b) (right hand side). The
configurations space points $0$, $x$ and $y$ are indicated.}
\end{center}\end{figure}

In this subsection we give a formula for a more general topology which occurs
when one propagator is removed from an initial three-loop, tetrahedron-shaped
bubble diagram. In the original classification of Ref.~\cite{Avdeev:1995eu}
these diagrams are class $E$ diagrams belonging to the spectacle topology (see
Fig.~\ref{masmel2}). The formulas obtained in this subsection are efficient
for the numerical integration of diagrams of the spectacle topology. We have,
however, not been able to find an analytical solution to the problem. The main
obstacle of generalizing the configuration space technique to a general
multi-loop diagram is the angular integration. The configuration space
technique proved to be rather successful for general diagrams in the massless
case~\cite{Chetyrkin:pr} but it brings no essential simplification in the
general massive case (see e.g.\ Ref.~\cite{Mendels:wc}). However, for special
configurations the angular integration can be explicitly performed with a
reasonably simple integrand left for the radial integration. Diagrams of the
spectacle topology such as the ones shown in Fig.~\ref{masmel2} are examples
of such a configuration.

The configuration space expression of a spectacle topology diagram can be
written in a form suitable for our purpose (see Fig.~\ref{masmel2}(a)),
\begin{equation}
\int D(x-y,m)D(x,m)^2d^DxD(y,m)^2d^Dy.
\label{spectxexp}
\end{equation}
The key relation which leads to the simplification of the configuration space
integral with spectacle topology is an addition theorem for Bessel functions
in Eq.~(\ref{addth}), allowing to perform the integration over the relative
angle in the propagator $D(x-y,m)$. Using Eq.~(\ref{addth}) for $Z=H^+$ and
substituting $m=e^{i\pi/2}$, we can perform an analytic continuation in order
to obtain a relation for the modified Bessel functions $K$ and $I$,
\begin{eqnarray}
\frac{K_\lambda(R)}{R^\lambda}
  &=&2^\lambda\Gamma(\lambda)\sum_{k=0}^\infty(\lambda+k)
  \frac{I_{\lambda+k}(\rho)}{\rho^\lambda}
  \frac{K_{\lambda+k}(r)}{r^\lambda}C_k^\lambda(\cos \varphi)
\label{eqn1}
\end{eqnarray}
where we have $R=\sqrt{r^2+\rho^2-2r\rho\cos\varphi}$,
\begin{eqnarray}
K_\lambda(z)&=&\frac{i\pi}2e^{\pi\lambda i/2}H_\lambda^+(iz),\nonumber\\
I_\lambda(z)&=&e^{-\pi\lambda i/2}J_\lambda(e^{\pi i/2}z)\qquad
  \mbox{for\ }-\pi<\mbox{arg}\,z\le\frac\pi2,\nonumber\\
I_\lambda(z)&=&e^{3\pi\lambda i/2}J_\lambda(e^{-3\pi i/2}z)\qquad
  \mbox{for\ }\frac\pi2<\mbox{arg}\,z\le\pi.
\end{eqnarray}
Using the orthogonality relations for Gegenbauer polynomials in
Eq.~(\ref{gegenort}), the sum disappears after integration over the
relative angle and only one term contributes. We obtain
\begin{eqnarray}
\int\frac{K_\lambda(R)}{R^\lambda}d\Omega_\rho
  &=&2^\lambda\Gamma(\lambda)\sum_{k=0}^\infty(\lambda+k)
  \frac{I_{\lambda+k}(\rho)}{\rho^\lambda}\frac{K_{\lambda+k}(r)}{r^\lambda}
  \int C_k^\lambda(\cos\varphi)d\Omega_\varphi\nonumber\\
  &=&2^\lambda\Gamma(\lambda)\lambda
  \frac{I_{\lambda}(\rho)}{\rho^\lambda}\frac{K_{\lambda}(r)}{r^\lambda}
  \frac{2\pi^{\lambda+1}}{\Gamma(\lambda+1)}C_0^\lambda(1)\nonumber\\
  &=& (2\pi)^{\lambda+1}\frac{I_\lambda(\rho)}{\rho^\lambda}
  \frac{K_\lambda(r)}{r^\lambda}, \qquad {r>\rho},
\end{eqnarray}
where the first equality is a consequence of the orthogonality relation
with the trivial factor $C_0^\lambda(1)=1$. This result allows one to write
down an expression for any spectacle-type diagram in the form of a two-fold
integral with a simple integration measure,
\begin{eqnarray}\label{spect}
&&\int_0^\infty D(x,m)^2x^{2\lambda+1}dx
\int_0^\infty D(y,m)^2y^{2\lambda+1}dy\ \times\nonumber\\&&\qquad\left(
  \frac{K_\lambda(x)}{x^\lambda}\frac{I_\lambda(y)}{y^\lambda}\theta(x-y)
  +\frac{K_\lambda(y)}{y^\lambda}\frac{I_\lambda(x)}{x^\lambda}\theta(y-x)
  \right)\nonumber\\
  &=&\int_0^\infty D(x,m)^2I_\lambda(x)x^{\lambda+1}dx
  \int_x^\infty D(y,m)^2K_\lambda(y)y^{\lambda+1}dy\qquad
\end{eqnarray}
where $\theta(x)$ is the standard step-function
distribution~\cite{Groote:1999cx}.

Note that the integration measure $D(x,m)^2x^{2\lambda+1}dx$ allows one to
perform the integration by using efficient integration routines for numerical
evaluation. The form of the weight function is close to $e^{-ax}x^{\alpha}$
which suggests the use of Laguerre polynomials
\begin{equation}
L_n(x)=\frac{e^x}{n!}\frac{d^n}{dx^n}\left(x^ne^{-x}\right)
\end{equation}
as a convenient choice within the Gaussian numerical integration method. In
fact, any modified propagator (with any power of the denominator) can be used
as a factor in the integration measure $D(x,m)^2x^{2\lambda+1}dx$. This fact
makes the representation universal and useful for the case of higher powers of
denominators of the lines associated with pairs $(x,0)$ and $(y,0)$ of
space-time points. If the angular structure of the diagram is preserved, the
generalization to higher loops in the expressions for the radial measures is
straightforward.

\subsection{Example for the spectacle diagram}

To show how this technique works, we give an example of the evaluation of a
spectacle diagram. Consider an integer dimension space-time which, without
loss of generality, can be chosen to be two-dimensional (an odd number of
dimensions is trivial because, as we have seen earlier, the propagators
degenerate to simple exponentials). The spectacle-type three-loop diagram can
be obtained in a closed form. Indeed, in momentum space representation we have
\begin{equation}\label{sp1}
S(M)=\int\frac{{\tilde\Pi(p)}^2}{p^2+M^2}d^2p
\end{equation}
for the basic spectacle diagram $S$ with $\tilde\Pi(p)$ and the mass $M$ of
the connecting propagator kept different. After the substitutions
$p=2m\sinh(\xi/2)$ and $t=e^{-\xi}$ we obtain
\begin{equation}\label{spcha}
S(M)=\frac1{2\pi m^4}\int_0^1\frac{t\,\ln^2t\,dt}{(1-t^2)(1-2\gamma t+t^2)}
\end{equation}
where $\gamma=1-M^2/2m^2$. Performing the integration, we finally obtain
\begin{equation}\label{manint}
S(M)=\frac{f(t_1)-f(t_2)}{t_1-t_2}
\end{equation}
with $t_{1,2}=\gamma\pm\sqrt{\gamma^2-1}$ and
\begin{equation}
f(t)=\frac{8t{\rm Li}_3(1/t)-(t+7)\zeta(3)}{8\pi m^4(t^2-1)}.
\end{equation}
$\Li_3(z)$ is the trilogarithm function
\begin{equation}
\Li_3(z)=\sum_{k=1}^\infty\frac{z^k}{k^3}, \quad |z|<1.
\end{equation}
For $M=2m$ (i.e.\ $\gamma=-1$) the integral in Eq.~(\ref{spcha}) simplifies
and one finds a simple answer in terms of the (for the present context)
standard transcendental numbers $\ln 2$ and $\zeta(3)$,
\begin{equation}
S(2m)=\frac1{4\pi m^4}\left(\frac78\zeta(3)-\ln 2\right).
\end{equation}
For the equal mass case $M=m$ we obtain a result in terms of Clausen's
trilogarithm $\Cl_3(2\pi/3)$. As one can conclude from these results, the
conjugate pair of the sixth order roots of unity, $\exp(\pm 2\pi i/3)$ play
an important role in this case again in accordance with the general analysis
in~\cite{Broadhurst:1998rz}. The origin of the appearance of the sixth order
roots of unity as the parameters of the analytical expressions of the diagrams
lies in the mismatch of masses along the lines of the diagrams. However, the
exceptional case $M=2m$, where one line has twice the mass of the other lines
(which results in the drastic simplification), also keeps us within the set of
the sixth order roots of unity. The key parameter in this case is simply the
natural number $1$ which is definitely one of the sixth order roots of unity.

Turning to the configuration space representation in Eq.~(\ref{spect}) we find
\begin{eqnarray}\label{xMc00}
S(M)&=&\int_0^\infty xK_0(mx)^2dx\int_0^\infty yK_0(my)^2dy\ \times\nonumber\\
  &&\Bigg(K_0(Mx)I_0(My)\theta(x-y)+K_0(My)I_0(Mx)\theta(y-x)\Bigg).
\end{eqnarray}
for the basic spectacle diagram. An explicit numerical integration of
Eq.~(\ref{xMc00}) shows coincidence with the analytical result in
Eq.~(\ref{manint}) which we have checked numerically for arbitrary values of 
$M$. In this example the analytical result has a rather simple form. This is
not true if high powers of the denominators enter. Then the corresponding
one-loop insertions are rather cumbersome and an explicit integration in
configuration space is more convenient.

\subsection{About the occurrence of Clausen's dilogarithms}

We want to have a closer look at how the square of Clausen's dilogarithm
$\Cl_2(\pi/3)^2$ can emerge at the level of spectacle topology diagrams. As
transcendental number, Clausen's dilogarithm $\Cl_2(\pi/3)$ characterizes the
analytical results for three-loop bubbles. Its presence was discovered in the
impressive treatise of David Broadhurst on the role of the sixth order roots
of unity for the transcendental structure of results for Feynman diagrams in
quantum field theory~\cite{Broadhurst:1998rz}.

We consider the spectacle diagram in the form shown in Fig.~\ref{masmel2}(b).
Here the expression for the generalized middle line is a product of the
one-loop propagator and the standard particle propagator. We express the
one-loop propagator using a dispersion representation with the spectral
density $\rho(s)$ and obtain
\begin{equation}\label{midline}
\frac1{p^2+m^2}\tilde\Pi(p)
  =\frac1{p^2+m^2}\int_{4m^2}^\infty\frac{\rho(s)ds}{s+p^2}
  =\int_{4m^2}^\infty\frac{\rho(s)ds}{s-m^2}
  \left(\frac1{p^2+m^2}-\frac1{s+p^2}\right).
\end{equation}
We take only the first term which is sufficient for obtaining the result we
are aiming for. In this case the integral becomes independent of $p^2$ and can
be considered separately. One has
\begin{equation}\label{ii}
I=\frac1{p^2+m^2}\int_{4m^2}^\infty\frac{\rho(s)ds}{s-m^2}
\end{equation}
which leads to the sunrise diagram after the two other line shown in
Fig.~\ref{masmel2}(b) have been added with a normalization factor given by the
integral. One factor $\Cl_2(\pi/3)$ results from integrating the overall
sunrise diagram which is composed of the propagator $(p^2+m^2)^{-1}$ from
Eq.~(\ref{ii}). The two other lines of the diagram are shown in
Fig.~\ref{masmel2}(b). The second factor $\Cl_2(\pi/3)$ has to be found in the
normalization factor given by the integral in Eq.~(\ref{ii}). Note that
the very structure of this contribution -- the square of a number which first
appeared at the lower loop level -- suggests a hint for its search. It should
emerge as an iteration of a lower order contribution in accordance with the
iterative structure of the R-operation which provides a general framework for
the analysis of multi-loop diagrams. The following consideration confirms this
conjecture. Consider the quantity
\begin{equation}
N=\int_{4m^2}^\infty\frac{\rho(s)ds}{s-m^2}
\end{equation}
and take $\rho(s)$ to be the spectral density in $D$-dimensional
space-time~\cite{Groote:1998wy},
\begin{equation}
\rho(s)=\frac{(s-4m^2)^{\lambda-1/2}}
  {2^{4\lambda+1}\pi^{\lambda+1/2}\Gamma(\lambda+1/2)\sqrt s},\qquad\sqrt s>2m.
\end{equation}
Next consider the first order contribution of the expansion in $\eps$ near the
space-time dimension $D_0=2$. The expansion in $\lambda=\eps$ results in
\begin{equation}
\frac{(s-4m^2)^{\eps-1/2}}{\mu^{2\eps}\sqrt s}=\frac1{\sqrt{s(s-4m^2)}}
  \left(1+\eps\ln\left(\frac{s-4m^2}{\mu^2}\right)+O(\eps^2)\right).
\end{equation}
Therefore, the relevant first order term in $\eps$ reads
\begin{equation}
\Delta_\eps\rho(s)=\frac{\ln((s-4m^2)/m^2)}{2\pi\sqrt{s(s-4m^2)}}
\end{equation}
where $\mu=m$ has been chosen for convenience. Next we change variables
according to
\begin{equation}
\sqrt s=2m\cosh(\xi/2),\qquad t=e^{-\xi}
\end{equation}
to obtain
\begin{equation}
\Delta_\eps\rho(4m^2\cosh^2(\xi/2))
  =\frac{(-\ln t+2\ln(1-t))t}{2\pi m^2(1-t^2)}.
\end{equation}
For the quantity in question we find
\begin{equation}\label{delnorm}
\Delta_\eps N=\int_{4m^2}^\infty\frac{\Delta_\eps\rho(s)ds}{s-m^2}
  =\frac1{2\pi m^2}\int_0^1\frac{(-\ln t+2\ln(1-t))dt}{1+t+t^2}.
\end{equation}
The roots of the denominator of the integrand in Eq.~(\ref{delnorm}) are now
given by $t_{3,4}=\exp({\pm 2\pi i/3})$ which again is a conjugate pair of the
sixth order roots of unity. After integrating this equation we readily find
\begin{eqnarray}
\Delta_\eps N&=&-\frac1{2\pi m^2\sqrt3}\left(-\frac\pi3\ln3
  +{\rm Im}\left(\Li_2\left(e^{2i\pi/3}\right)
  -\Li_2\left(e^{-2i\pi/3}\right)\right)\right)\nonumber\\
  &=&-\frac1{\pi m^2\sqrt3}\left(\Cl_2\left(\frac{2\pi}3\right)
  -\frac\pi6\ln3\right).
\end{eqnarray}
Using the relation
\begin{equation}
\Cl_2\left(\frac{2\pi}3\right)=\frac23\Cl_2\left(\frac\pi3\right)
\end{equation}
we finally obtain
\begin{equation}
\Delta_\eps N
  =-\frac2{3\pi m^2\sqrt3}\left(\Cl_2\left(\frac\pi3\right)
  -\frac\pi4\ln3\right).
\end{equation}
Therefore, in the first order of the $\eps$ expansion of the spectacle diagram
we indeed found the remarkable contribution proportional to $\Cl_2(\pi/3)^2$.
In our calculation it emerges naturally as the iteration of the lower order
term~\cite{Groote:1999cn}. Originally, the presence of this contribution has 
been conjectured and confirmed in~\cite{Broadhurst:1998rz} by direct numerical
computation of the finite part of the general three-loop bubble in
four-dimensional space-time. 

\subsection{The insertion of massless irreducible loops}

\begin{figure}[t]\begin{center}
\epsfig{figure=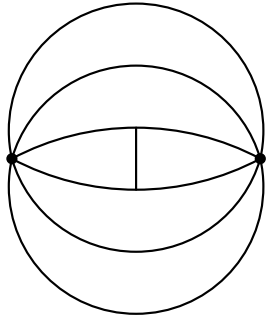, scale=0.5}
\caption{\label{irrloop} Irreducible loop replacing one of the propagators}
\end{center}\end{figure}

One can consider a sunrise-type diagram with one or more irreducible loops by
which we mean a generalized line more complicated than an ordinary standard
propagator or a (large) power of them. Consider, for instance, the replacement 
of a line by a subdiagram of fish-type topology as shown in Fig.~\ref{irrloop}.
The configuration space technique leads to a numerical solution in this case
because we can replace one of the propagator factors $D(x,m)$ by the two-point
propagator in the irreducible subdiagram, using the fact that
\begin{equation}
\Pi(x)=\int D(x,\sqrt s)\rho(s)ds
\end{equation}
where $\rho(s)$ is the spectral density of the subdiagram. If the subdiagram
is a massless fish-type diagram, the spectral density is given by
\begin{equation}
\rho(s)=6\zeta(3)\delta(s).
\end{equation}
The known result in momentum space can be restored by using the dispersion
relation~(\ref{disprel}). On the other hand, we obtain
\begin{equation}
D(x)=6\zeta(3)\int D(x,\sqrt s)\delta(s)ds=6\zeta(3)D(x,0)
\end{equation}
which up to a factor is the usual massless propagator.

\subsection{The insertion of massive irreducible loops}

Insertions of massive irreducible loops were used analytically for the
calculation of corrections to baryon correlators~\cite{Ovchinnikov:1991mu,
Groote:1999zp,Groote:2000py} and numerically for the mixing of heavy neutral
mesons~\cite{Narison:1994zt}.
The massive insertion can be incorporated mainly because the two-loop diagrams
are well analyzed with any mass configurations~\cite{Broadhurst:1987ei,
Generalis:1990id,Kreimer:1991jv,Brucher:1997bv}. For the semi-massive
fish-type one-loop diagram with one massive and one massless line we can use
\begin{equation}
\rho(s)=\frac4{(4\pi)^4s}\left(\Li_2\pfrac{m^2}s
  +\frac12\ln\left(1-\frac{m^2}s\right)\ln\pfrac{m^2}s\right).
\end{equation}
But even in the case with two different masses a result for the spectral
density is available,
\begin{equation}
\rho(s)=\frac{-4}{(4\pi)^4s}\left(\Li_2(x_1)+\Li_2(x_2)
  -\frac12\ln(-x_1)\ln(1-x_1)-\frac12\ln(-x_2)\ln(1-x_2)\right)
\end{equation}
where
\begin{eqnarray}
x_1&=&\frac{2m_1^2}{m_1^2+m_2^2-s+\sqrt{(m_1^2+m_2^2-s)^2-4m_1^2m_2^2}},
  \nonumber\\
x_2&=&\frac{2m_2^2}{m_1^2+m_2^2-s+\sqrt{(m_1^2+m_2^2-s)^2-4m_1^2m_2^2}}
\end{eqnarray}

\newpage

\section{Summary and conclusion}

To conclude, we have presented a review of configuration space techniques for
the calculation of sunrise-type diagrams. We have shown that the singular or
pole parts of any sunrise-type diagram are calculable analytically in the
simplest possible manner. For the finite parts we obtained one-dimensional
integrals of well-known functions which is very convenient for numerical
evaluation. We have presented many sample calculations of multi-loop
sunrise-type diagrams and have compared them with momentum space results in
the literature when available. We have found agreement in every case. The
agreement provides for a mutual check of the results which have been derived
using very different methods. In the asymptotic analysis have dealed with
different expansions. We can conclude that any kind of expansions is given as
an expansion in parameters of the propagators and thus leads to the expansion
coefficients in a straightforward way. Finally, extensions to non-standard
propagators and other exotic settings figured out to be feasible as well as
some considerations for slightly different topologies. All in all, we can
stress again that the benefit of configuration space techiques is the fact
that there is 'almost' no integration in sunrise-type diagrams, as
Eq.~(\ref{xsun}) shows.

\subsection*{Acknowledgements}

We thank Kostja Chetyrkin for discussion, Robert Delbourgo for kind attention
and enthusiasm in advertising $x$-space, Andrey Davydychev for communication
and help in finding references,  Andrey Grozin for providing us with the
source code of RECURSOR, with which a part of the calculation was
cross-checked, David Broadhurst for criticism and friendly remarks, and
Giampiero Passarino for communication. SG thanks the Mainz xloops-GiNaC-group
for permanent interest, valuable comments and numerical cross-checks. AAP
thanks V.A.~Matveev for encouragement, attention and support, and P.~Baikov
for illuminating discussions of the present status of the optimization of
recurrence relations under study by him. This work was supported by the INTAS
grant No.~03-51-4007. SG acknowledges support by the DFG as a guest scientist
in Mainz, by the Estonian target financed project No.~0182647s04, and by the
grant No.~6216 given by the Estonian Science Foundation. The work of AAP was
supported in part by the Russian Fund for Basic Research under contract
No.~03-02-17177 and by the grant NS-2184.2003.2.

\newpage

\begin{appendix}

\section{Properties of Bessel functions\label{sec_bessel}}
\setcounter{equation}{0}\def\theequation{A\arabic{equation}}

Since Bessel functions play a crucial role in our calculations we collect a
number of definitions and properties of Bessel functions, mostly taken from
the handbooks of Watson (1944)~\cite{Watson}, Prudnikov, Brychkov and
Marichev (1990)~\cite{Prudnikov}, and Gradshteyn and Ryshik
(1994)~\cite{Gradshteyn}.

\subsection*{Bessel functions and Hankel functions}
Ordinary Bessel functions are solutions of the {\em Bessel differential
equation}
\begin{equation}\label{bessdiff}
z^2\frac{d^2Z_\nu(z)}{dz^2}+z\frac{dZ_\nu(z)}{dz}
  +(z^2-\nu^2)Z_\nu(z)=0
\end{equation}
($\nu$ need not be an integer). There are two classes of solutions $Z_\nu(z)$.
{\em Bessel functions of the first kind $J_\nu(z)$} are nonsingular at the
origin $z=0$ while {\em Bessel functions of the second kind $Y_\nu(z)$} are
singular at $z=0$. Combining Bessel functions of both classes, we end up with
{\em Hankel functions}
\begin{equation}\label{defH}
H_\nu^\pm(z)=J_\nu(z)\pm iY_\nu(z)
\end{equation}
(instead of $H_\nu^\pm(z)$ the notation $H_\nu^{(1)}=H_\nu^+$ and
$H_\nu^{(2)}=H_\nu^-$ are quite common).

Bessel functions of the first kind have the series expansion 
\begin{equation}\label{serJ}
J_\nu(z)=\pfrac z2^\nu\sum_{k=0}^\infty
  \frac{(-z^2/4)^k}{k!\Gamma(\nu+k+1)}
  =\frac{(z/2)^\nu}{\Gamma(1+\nu)}\left(1-\frac{(z/2)^2}{1+\nu}
  +\ldots\ \right).
\end{equation}
Including a factor occuring in practical applications, we can also write
\begin{equation}\label{serJfac}
\pfrac z2^{-\nu}J_\nu(z)
  =\sum_{k=0}^\infty\frac{(-z^2/4)^k}{k!\Gamma(\nu+k+1)}
  =\frac1{\Gamma(1+\nu)}\left(1-\frac{(z/2)^2}{1+\nu}+\ldots\ \right).
\end{equation}
$\Gamma(z)$ is Euler's Gamma function. Functional equations are of help in
order to relate Bessel functions of different degree. They are given by
\begin{equation}\label{funceq}
zZ_{\nu-1}(z)+zZ_{\nu+1}(z)=2\nu Z_\nu(z),\qquad
Z_{\nu-1}(z)-Z_{\nu+1}(z)=2\frac{d}{dz}Z_\nu(z)
\end{equation}
where $Z$ is any of the Bessel functions $J$, $Y$, or $H^{\pm}$.
As a consequence we obtain
\begin{equation}\label{diffJ}
\frac{d^k}{dz^k}\left(z^{-\nu}J_\nu(z)\right)=-z^{-\nu}J_{\nu+k}(z).
\end{equation}

\subsection*{Modified Bessel functions}

We can continue Bessel functions into the complex plane, ending up with
{\em modified Bessel functions of the first kind $I_\nu(z)$} and {\em modified
Bessel functions of the second kind $K_\nu(z)$}, sometimes also known as
McDonald functions. $I_\nu(z)$ and $K_\nu(z)$ are solutions of the
{\em modified Bessel differential equation}
\begin{equation}\label{mbessdiff}
z^2\frac{d^2Z_\nu(z)}{dz^2}+z\frac{dZ_\nu(z)}{dz}-(z^2-\nu^2)Z_\nu(z)=0
\end{equation}
The modified Bessel function of the second kind
\begin{equation}\label{defK}
K_\nu(z)=\frac\pi2\frac{I_{-\nu}(z)-I_\nu(z)}{\sin(\pi\nu)},
  \qquad\Gamma(\nu)\Gamma(1-\nu)=\frac\pi{\sin(\pi\nu)}
\end{equation}
can be expressed by the modified Bessel function of the first kind $I_\nu(z)$
with series expansion
\begin{equation}\label{serI}
I_\nu(z)=\pfrac z2^\nu
  \sum_{k=0}^\infty\frac{(z^2/4)^k}{k!\Gamma(\nu+k+1)}
\end{equation}
Therefore, one has 
\begin{equation}\label{serKfac}
\left(\frac z2\right)^\nu K_\nu(z)
  =\frac{\Gamma(\nu)}2\left[1+\frac1{1-\nu}\left(\frac z2\right)^2
  -\frac{\Gamma(1-\nu)}{\Gamma(1+\nu)}
  \left(\frac z2\right)^{2\nu}\right]+O(z^4,z^{2+2\nu}).
\end{equation}
For Eq.~(\ref{funceq}) one can we derive functional equations for the 
modified Bessel functions,
\begin{eqnarray}\label{mfunceq}
zI_{\nu-1}(z)-zI_{\nu+1}(z)=2\nu I_\nu(z),&&
I_{\nu-1}(z)+I_{\nu+1}(z)=2\frac{d}{dz}I_\nu(z),\nonumber\\
zK_{\nu-1}(z)-zK_{\nu+1}(z)=-2\nu K_\nu(z),&&
K_{\nu-1}(z)+K_{\nu+1}(z)=-2\frac{d}{dz}K_\nu(z),\qquad\qquad
\end{eqnarray}
For the differentiation of the modified Bessel function including an
appropriate factor we derive
\begin{equation}\label{diffK}
\frac{d}{dz}\left(z^{-\nu}K_\nu(z)\right)=-z^{-\nu}K_{\nu+1}(z).
\end{equation}
Bessel functions can also be differentiated with respect to their index, if
this index is given by an integer. For $K_\nu(x)$ we for instance obtain
\begin{equation}\label{serKind}
\left[\frac{\partial K_\nu(z)}{\partial\nu}\right]_{\nu=\pm n}
  =\pm\frac12n!\sum_{k=0}^{n-1}\left(\frac z2\right)^{k-n}
  \frac{K_k(z)}{k!(n-k)},\qquad n\in\{0,1,\ldots\,\}.
\end{equation}
Finally, the modified Bessel function $K_\nu$ and the Hankel function
$H_\nu^+$ are related by
\begin{equation}\label{KtoH}
K_\nu(z)=\frac{\pi i}2e^{i\nu\pi/2}H_\nu^+(iz).
\end{equation}

\subsection*{Asymptotic behaviour}

Asymptotic expansions determine the behaviour of the modified Bessel functions 
of the second kind and the Hankel functions $H_\nu^-$ when the arguments
become large. For the two functions just related to each other we obtain
\begin{equation}\label{asyK}
K_{\nu,N}^{as}(z)=\pfrac\pi{2z}^{1/2}e^{-z}\Bigg[\sum_{n=0}^{N-1}
  \frac{(\nu,n)}{(2z)^n}+\theta\frac{(\nu,N)}{(2z)^N}\Bigg],\quad
  (\nu,n):=\frac{\Gamma(\nu+n-1/2)}{n!\Gamma(\nu-n-1/2)}
\end{equation}
and
\begin{equation}\label{asyH}
H_{\nu,N}^{-{as}}(iz)=\pfrac2{\pi z}^{1/2}e^{z+i\nu\pi/2}
  \left[\sum_{n=0}^{N-1}\frac{(-1)^n(\nu,n)}{(2z)^n}
  +\theta\frac{(-1)^N(\nu,N)}{(2z)^N}\right]
\end{equation}
where $\theta\in[0,1]$ is chosen appropriately. For the asymptotic behaviour
of $I_\nu(z)$ see Eq.~(\ref{asymptotI}).

\subsection*{Addition theorem}

The addition theorem for Bessel functions is given by
\begin{equation}\label{addth}
\frac{Z_\nu(mR)}{R^\nu}=2^\nu m^{-\nu}\Gamma(\nu)
  \sum_{k=0}^\infty(\nu+k)\frac{J_{\nu+k}(m\rho)}{\rho^\nu}
  \frac{Z_{\nu+k}(mr)}{r^\nu}C^\nu_k(\cos\varphi)
\end{equation}
where $C^\nu_k$ are the Gegenbauer polynomials (cf.\
Appendix~\ref{sec_gegenbauer}), $Z$ is any of the Bessel functions $J$, $Y$,
or $H^{\pm}$, $R=\sqrt{r^2+\rho^2-2r\rho\cos\varphi}$, and $r>\rho$. For
$r<\rho$ the arguments of the Bessel functions on the right hand side of
Eq.~(\ref{addth}) should be interchanged.

\section{Properties of Gegenbauer polynomials\label{sec_gegenbauer}}
\setcounter{equation}{0}\def\theequation{B\arabic{equation}}

Gegenbauer polynomials can be generated using the characteristic polynomial
\begin{equation}\label{gegengen}
(t^2-2tx+1)^{-\nu}=\sum_{n=0}^\infty t^nC_n^\nu(x).
\end{equation}
In particular, we have $C_0^\nu(x)=1$, $C_1^\nu(x)=2\nu x$ and
\begin{equation}\label{gegenrec}
(n+1)C_{n+1}^\nu(x)=2(n+\nu)xC_n^\nu(x)
  -(n+2\nu-1)C_{n-1}^\nu(x).
\end{equation}
Gegenbauer polynomials obey the orthogonality relations~\cite{Chetyrkin:pr}
\begin{equation}\label{gegenort}
\int C_m^\nu(\hat x_1\cdot\hat x_2)C_n^\nu(\hat x_2\cdot\hat x_3)
  d\Omega_2=\frac{2\pi^{\nu+1}}{\Gamma(\nu+1)}\frac{\nu\delta_{mn}}{n+\nu}
  C_n^\nu(\hat x_1\cdot\hat x_3),\quad
  \int d\Omega_2=\frac{2\pi^{\nu+1}}{\Gamma(\nu+1)}
\end{equation}
where $\hat x_i$ are unit four-vectors and $d\Omega_i$ is the angular part of
the integration measure. Finally, we have
\begin{equation}\label{gegenone}
C_n^\nu(1)=\frac{\Gamma(n+2\nu)}{n!\Gamma(2\nu)}.
\end{equation}

\section{Cuts and discontinuities\label{sec_disc}}
\setcounter{equation}{0}\def\theequation{C\arabic{equation}}

Functions which are continuous at least on parts of the real axis can be
continued to the whole complex plane. However, if singularities occur on the
real axis or elsewhere in the complex plane, this continuation need no longer
be unique. Examples which occur in the context of our calculations are
logarithms and polylogarithms but also powers with non-integer exponent and
Bessel functions. We will deal with these special cases in this appendix.

\subsection*{The case of logarithms}

For $x=0$ the function $\ln(x)$ has a non-removable singularity while for
negative real values it is not defined. However, we can continue to the
negative real axis by following a path which circumvents the origin. It makes 
a difference whether we turn around the origin in the mathematically positive
or negative sense. If we for instance start at the real value $x$ and turn
around in the positive sense to continue to a value at $-x$, we can
parametrize this by writing $z=xe^{i\varphi}$ where $\varphi$ starts at
$\varphi=0$ and increases until it reaches the negative real axis for
$\varphi=\pi$. By doing so, we obtain
\begin{equation}
\ln(xe^{+i\pi})=\ln x+i\pi.
\end{equation}
However, we can reach the negative real axis as well by starting with
$\varphi$ again but going to negative values, reaching the negative real axis
for $\varphi=-\pi$. In this case we obtain
\begin{equation}
\ln(xe^{-i\pi})=\ln x-i\pi.
\end{equation}
Obviously, the value ``$\ln(-x)$'' is not unique. Indeed, we can even think
of the continuation to the complex plane as a sheet which winds up and rises
each time we turn around the origin, the so-called Riemannian sheet.
Nevertheless, for only one turn we can calculate the difference between the
values. This quantity is known as discontinuity, and it can be calculated
actually for every complex value $z$ except for singular points. The
definition is given by
\begin{equation}
\Disc f(z)=f(ze^{+i0})-f(ze^{-i0}),\qquad
  e^{\pm i0}=\lim_{\epsilon\to 0}e^{\pm i\epsilon},\ \epsilon>0
\end{equation}
where $z=0$ is the singular point of the function $f(z)$. If we apply this
definition to the case of a logarithm with negative argument, we obtain
\begin{equation}
\Disc\ln(-x)=\ln(-xe^{+i0})-\ln(-xe^{-i0})=\ln(xe^{-i\pi})-\ln(xe^{+i\pi})
  =-2\pi i\theta(x)
\end{equation}
where $\theta(x)$ is the step function ($\theta(x)=1$ for $x>0$, $\theta(x)=0$
for $x<0$). Note that we have been careful in replacing $-xe^{\pm i0}$ by
$xe^{\mp i\pi}$. The discontinuity of powers of logarithms can be calculated
as well, for instance
\begin{eqnarray}
\Disc\ln^2(-x)&=&\ln^2(xe^{-i\pi})-\ln^2(xe^{+i\pi})\nonumber\\[7pt]
  &=&\left((\ln x-i\pi)^2-(\ln x+i\pi)^2\right)\theta(x)
  \ =\ -4\pi i\ln x\ \theta(x).
\end{eqnarray}
For later convenience we divide the discontinuity by the factor $2\pi i$ to
obtain
\begin{equation}
\frac1{2\pi i}\Disc\ln(-x)=-\theta(x),\qquad
\frac1{2\pi i}\Disc\ln^2(-x)=-2\ln x\ \theta(x),\qquad\ldots
\end{equation}

\subsection*{The case of polylogarithms}

Polylogarithms can be defined iteratively by
\begin{equation}\label{polylogit}
\Li_n(x)=\int_0^x\frac{\Li_{n-1}(x')}{x'}dx',\qquad
\Li_1(x)=-\ln(1-x).
\end{equation}
For all the polylogarithms a discontinuity occurs for $x>1$. We obtain
\begin{equation}
\Li_1(xe^{\pm i0})=-\ln(1-xe^{\pm i0})=-\ln(1+xe^{\mp i\pi})
  =\left(-\ln(x-1)\pm i\pi\right)\theta(x-1)
\end{equation}
where we have extracted a factor $e^{\mp i\pi}$ from the argument, and
therefore
\begin{equation}
\frac1{2\pi i}\Disc\Li_1(x)=\theta(x-1).
\end{equation}
In order to calculate the discontinuity of $\Li_2(x)$, for $x>1$ we split the
integral into two parts,
\begin{eqnarray}
\Li_2(xe^{\pm i0})&=&-\int_0^1\frac{\ln(1-x')}{x'}dx'
  -\int_1^x\frac{\ln(1-x')}{x'}dx'\nonumber\\
  &=&\Li_2(1)-\int_1^x\frac{\ln(x'-1)}{x'}dx\pm\pi i\int_1^x\frac{dx'}{x'}.
\end{eqnarray}
The real parts cancel out when taking the difference needed for the
determination of the discontinuity. Contrary to this the imaginary parts add
up and one obtains
\begin{eqnarray}
\frac1{2\pi i}\Disc\Li_2(x)&=&\frac1{2\pi i}\int\frac{dx'}{x'}\Disc\Li_1(x')
  =\int_1^x\frac{dx'}{x'}=\ln x\ \theta(x-1),\quad
  \mbox{in the same way}\nonumber\\
\frac1{2\pi i}\Disc\Li_3(x)&=&\frac1{2\pi i}\int\frac{dx'}{x'}\Disc\Li_2(x')
  =\int_1^x\frac{dx'}{x'}\ln x'=\frac12\ln^2x\ \theta(x-1),\quad
  \mbox{in general}\nonumber\\
\frac1{2\pi i}\Disc\Li_n(x)&=&\frac1{n-1}\ln^{n-1}x\ \theta(x-1).
\end{eqnarray}

\subsection*{The case of non-integer powers}

For powers with exponents close to integer values such as $(b-ax)^{-n}$
the calculation of the discontinuity is closely related to the calculation of
the discontinuity of the logarithm because
\begin{equation}
x^\eps=1+\eps\ln x+O(\eps^2).
\end{equation}
Similarly, the discontinuity of $(b-ax)^{-n}$ can be obtained in the more 
general case in closed form. In order to calculate the discontinuity of a
power $(b-ax)^{-n}$ with non-integer exponent $-n$ (we include the minus sign
for later convenience) we first have to ponder where the discontinuities can
appear. Obviously, this is the case for $b-ax<0$. Taking $a>0$ and $b\ge 0$
which is the case for all applications in this report, we obtain $x>b/a$.
Therefore, we obtain
\begin{eqnarray}
\lefteqn{\Disc(b-ax)^{-n}\ =\ (b-ax^{+i0})^{-n}-(b-ax^{-i0})^{-n}
  \ =\ (b+axe^{-i\pi})^{-n}-(b+axe^{+i\pi})^{-n}}\nonumber\\[7pt]
  &=&\left(e^{+i\pi n}(ax-b)^{-n}-e^{-i\pi n}(ax-b)^{-n}\right)\theta(ax-b)
  \ =\ 2i\sin(\pi n)(ax-b)^{-n}\theta(ax-b).\qquad\quad
\end{eqnarray}
Using
\begin{equation}
\sin(\pi n)=\frac\pi{\Gamma(n)\Gamma(1-n)}=-\sin(-\pi n)
\end{equation}
where $\Gamma(x)$ is Euler's Gamma function, we finally obtain
\begin{equation}\label{discpow}
\frac1{2\pi i}\Disc(b-ax)^{-n}
  =\frac{(ax-b)^{-n}\theta(ax-b)}{\Gamma(n)\Gamma(1-n)}
\end{equation}
Note that for negative integer values of $n$ the discontinuity vanishes.

\subsection*{The case of Bessel functions}

The final item about ambiguities found in the continuation into the complex
plane we want to mention here is the so-called Stokes' phenomenon which occurs
for the asymptotic expansion of Bessel functions and is mentioned already in
the handbook of Watson~\cite{Watson}. The asymptotic expansion of the modified
Bessel function of the first kind reads
\begin{eqnarray}\label{asymptotI}
I_\nu(z)&=&\frac{e^z}{\sqrt{2\pi z}}\sum_{k=0}^\infty
  \frac{(-1)^k}{(2z)^k}\,\frac{\Gamma(\nu+k+1/2)}
  {\Gamma(\nu-k+1/2)k!}\nonumber\\&&\qquad
  +\frac{e^{-z\pm(\nu+1/2)\pi i}}{\sqrt{2\pi z}}\sum_{k=0}^\infty
  \frac1{(2z)^k}\,\frac{\Gamma(\nu+k+1/2)}{\Gamma(\nu-k+1/2)k!},
\end{eqnarray}
where the plus sign is valid for $-\pi/2<{\rm arg\,}z<3\pi/2$ while the 
minus sign has to be taken for $-3\pi/2<{\rm arg\,}z<\pi/2$. The surprising
fact is that there are two different asymptotic expansions available at every
point of the complex plane, except for the arguments $\pm i\pi/2$. Therefore,
in continuing $I_\nu(\pm ix)=I_\nu(xe^{\pm i\pi/2})$ to the complex plane, the
choice is still unique.

\section{Singular contributions up to four loops\label{sec_sing}}
\setcounter{equation}{0}\def\theequation{D\arabic{equation}}

In the following we present complete results for the singular parts of
sunrise-type diagrams with arbitrary masses up to four-loop order. The results
are given in the $\overline{\rm MS}$-scheme in the Euclidean domain.
\begin{eqnarray}
\lefteqn{\tilde\Pi_1^s(p,m_1,m_2)\ =\ \frac1\eps,}\nonumber\\[7pt]
\lefteqn{\tilde\Pi_2^s(p,m_1,m_2,m_3)\ =\ -\frac1{2\eps^2}\sum_im_i^2
  -\frac1{4\eps}\left(p^2+2\sum_im_i^2(3-2\ell_i)\right),}\nonumber\\[7pt]
\lefteqn{\tilde\Pi_3^s(p,m_1,m_2,m_3,m_4)\ =\ \frac1{6\eps^3}\sum_{i\neq j}
  m_i^2m_j^2
  +\frac1{12\eps^2}\left(p^2\sum_im_i^2-\sum_im_i^4+2\sum_{i\neq j}m_i^2m_j^2
  \left(4-3\ell_i\right)\right)\,+}\nonumber\\&&\kern-8pt
  +\frac1{72\eps}\left(2p^4+9p^2\sum_im_i^2(3-2\ell_i)-9\sum_im_i^4(5-2\ell_i)
  +6\sum_{i\neq j}m_i^2m_j^2(20-24\ell_i+3\ell_i^2+6\ell_i\ell_j)\right),
  \nonumber\\[7pt]
\lefteqn{\tilde\Pi_4^s(p,m_1,m_2,m_3,m_4,m_5)\ =\ -\frac1{24\eps^4}
  \sum_{i\neq j\neq k}m_i^2m_j^2m_k^2\,+}\nonumber\\&&
  -\frac1{48\eps^3}\left(p^2\sum_{i\neq j}m_i^2m_j^2
  -\sum_{i\neq j}(m_i^4m_j^2+m_i^2m_j^4)
  +2\sum_{i\neq j\neq k}m_i^2m_j^2m_k^2(5-4\ell_i)\right)\nonumber\\&&
  -\frac1{288\eps^2}\Bigg(2p^4\sum_im_i^2-6p^2\sum_im_i^4+2\sum_im_i^6
  +3p^2\sum_{i\neq j}m_i^2m_j^2(11-8\ell_i)\nonumber\\&&\qquad
  -6\sum_{i\neq j}(m_i^4m_j^2+m_i^2m_j^4)(7-4\ell_i)
  +12\sum_{i\neq j\neq k}m_i^2m_j^2m_k^2(15-20\ell_i+2\ell_i^2+6\ell_i\ell_j)
  \Bigg)\nonumber\\&&
  -\frac1{1728\eps}\Bigg(3p^6+2p^4\sum_im_i^2(35-24\ell_i)
  -18p^2\sum_im_i^4(21-8\ell_i)+2\sum_im_i^6(77-24\ell_i)\nonumber\\&&\qquad
  +9p^2\sum_{i\neq j}m_i^2m_j^2(71-88\ell_i+8\ell_i^2+24\ell_i\ell_j)
  -216\sum_{i \neq j}(m_i^4m_j^2-m_i^2m_j^4)\ell_i\nonumber\\&&\qquad
  -18\sum_{i\neq j}(m_i^4m_j^2+m_i^2m_j^4)
  (49-56\ell_i+4\ell_i^2+12\ell_i\ell_j)\nonumber\\&&\qquad
  +24\sum_{i\neq j\neq k}m_i^2m_j^2m_k^2(105-180\ell_i+30\ell_i^2
  +90\ell_i\ell_j-2\ell_i^3-18\ell_i^2\ell_j-12\ell_i\ell_j\ell_k)\Bigg)
\end{eqnarray}
where $\ell_i=\ln(m_i^2/\mu^2)$. The indices $i$, $j$, and $k$ run over all
mass indices. One can check that the general results listed in this Appendix
reproduce the equal mass results in the main text.

\section{Subtraction terms for the small $x$ singularities\label{sec_smallx}}
\setcounter{equation}{0}\def\theequation{E\arabic{equation}}

The leading singularity at small $x$ is given by the massless propagator of
the form
\begin{equation}
D(x,0)=\frac{\Gamma(\lambda)}{4\pi^{\lambda+1}x^{2\lambda}}.
\end{equation}
The next order of the small $x$-expansion for the propagator $D(x,m)$ 
is explicitly given by
\begin{equation}
D_1(x,0)=\frac1{4\pi^{\lambda+1}x^{2\lambda}}\left(\pfrac x2^2
  \frac{\Gamma(\lambda)}{1-\lambda}-\pfrac x2^{2\lambda}
  \frac{\Gamma(1-\lambda)}\lambda\right).
\end{equation}
This term is suppressed relative to the first term by one power of $x^2$ at
small $x$ in four-dimensional space-time (however, this
is not the case for two-dimensional space-time with $\lambda=0$). The term 
\begin{equation}
D_2(x,0)=\frac1{4\pi^{\lambda+1}x^{2\lambda}}\pfrac x2^2
\left(\pfrac x2^2\frac{\Gamma(\lambda)}{2(1-\lambda)(2-\lambda)}
  -\pfrac x2^{2\lambda}\frac{\Gamma(1-\lambda)}{\lambda(\lambda+1)}\right)
\end{equation}
is further suppressed by one power of $x^2$ at small $x$. Therefore, the full
subtraction of the three terms gives a rather smooth behaviour at small $x$
which is sufficient to obtain a regular integrand for the numerical
integration.

\section{Analytical results for the four-loop sunrise
  diagram\label{sec_fourloop}}
\setcounter{equation}{0}\def\theequation{F\arabic{equation}}

In this appendix we present some more details of our calculations for the
four-loop sunrise diagram. For the analytical evaluation we take the last two
terms of the integrand from Eq.~(\ref{V4main}). One has to integrate a product
of two Bessel functions multiplied with powers of $x$ which can be done 
analytically. The explicit expression for the $\eps$-expansion of that part of
the integral which is evaluated analytically reads
\begin{eqnarray}
\tilde\Pi_4^{\rm ana}(0)&=&m^6\Bigg\{-\frac5{2\eps^4}-\frac{35}{3\eps^3}
  -\frac{4565}{144\eps^2}-\frac{58345}{864\eps}\nonumber\\&&
  -\frac{1456940638037}{7779240000}-\frac{17099\pi^2}{24}
  -\frac{3857\pi^4}{10}+\frac{2525968\zeta(3)}{105}\nonumber\\&&
  +\Bigg(-\frac{55171475321621447}{1633640400000}+\frac{2457509\pi^2}{144}
  -\frac{1292537\pi^4}{175}\nonumber\\&&\qquad
  +\frac{6752474831\zeta(3)}{44100}+16530\pi^2\zeta(3)+59508\zeta(5)\Bigg)\eps
  \nonumber\\&&
  +\Bigg(-\frac{10610679621089130529}{68612896800000}
  +\frac{92781949\pi^2}{864}-\frac{4290113759\pi^4}{110250}
  -\frac{22591\pi^6}{14}\nonumber\\&&\qquad
  +\frac{952412727629\zeta(3)}{9261000}+244476\pi^2\zeta(3)-168606\zeta(3)^2
  +\frac{32210272\zeta(5)}{35}\Bigg)\eps^2\nonumber\\&&
  +\Bigg(-\frac{5963907632629558995931}{14408708328000000}
  +\frac{1325204033\pi^2}{5184}\nonumber\\&&\qquad
  -\frac{464379085699\pi^4}{6615000}-\frac{48529231\pi^6}{2205}
  \nonumber\\&&\qquad
  -\frac{312138383154103\zeta(3)}{277830000}+\frac{7285043\pi^2\zeta(3)}{6}
  +43529\pi^4\zeta(3)-\frac{238229084\zeta(3)^2}{105}\nonumber\\&&\qquad
  +\frac{13583011297\zeta(5)}{2940}+247950\pi^2\zeta(5)+1190160\zeta(7)\Bigg)
  \eps^3+O(\eps)^4\Bigg\}.
\end{eqnarray}
This expression shows the real complexity of the calculation which reveals
itself in the structure of the results. The main feature is that the terms
cannot be simultaneously simplified to all orders in $\eps$. By a special
choice of the normalization factor one can make the leading term and, in fact,
even all pole terms simple, but then the higher order terms contain rather lengthy
combinations of transcendental numbers that are not reducible in terms of
standard quantities such as the Riemann $\zeta$-functions. Note also that the
rational coefficients of transcendental numbers are very big and there is a
huge numerical cancellation between the rational and transcendental parts 
of the answer~(see also the discussion in Ref.~\cite{Groote:1999cx}).

\section{Integrand for the numerical integration\label{sec_numint}}
\setcounter{equation}{0}\def\theequation{G\arabic{equation}}

For the numerical evaluation we take the first three terms of the integrand
from Eq.~(\ref{V4main}). One has to integrate them numerically as there is a
product of three or more Bessel functions which is too complicated to be done
analytically. To find the $\eps$-expansion of the integral one has to first
expand the integrand in $\eps$. The expression for the $\eps$-expansion is
quite lengthy. We therefore give explicit results only for $\eps=0$. For this
part the integrand for the numerical integration over $z=mx$ reads
\begin{eqnarray}\label{integrand}
\Pi_4^{\rm num}(x)&=&m^6\Bigg(66-108l-144l^2+192l^3+\frac{384}{z^6}
  -\frac{384}{z^4}+\frac{768l}{z^4}+\frac{24}{z^2}-\frac{480l}{z^2}
  +\frac{576l^2}{z^2}\nonumber\\&&\qquad
  -\frac{111z^2}{16}+\frac{147lz^2}{2}-117l^2z^2+24l^3z^2+24l^4z^2+
  -\frac{165z^4}{32}\nonumber\\&&\qquad
  +\frac{201lz^4}{16}+\frac{9l^2z^4}{2}-24l^3z^4+12l^4z^4+\frac{75z^6}{512}
  -\frac{405lz^6}{128}+\frac{531l^2z^6}{64}\nonumber\\&&\qquad
  -\frac{15l^3z^6}{2}+\frac{9l^4z^6}{4}+\frac{375z^8}{2048}
  -\frac{825lz^8}{1024}+\frac{315l^2z^8}{256}-\frac{51l^3z^8}{64}
  +\frac{3l^4z^8}{16}\nonumber\\&&\qquad
  +\frac{1875z^{10}}{131072}-\frac{375lz^{10}}{8192}+\frac{225l^2z^{10}}{4096}
  -\frac{15l^3z^{10}}{512}+\frac{3l^4z^{10}}{512}\Bigg)K_1(z)\nonumber\\&&
  +m^6\Bigg(\frac{-512}{z^5}+\frac{384}{z^3}-\frac{768l}{z^3}+\frac{24}{z}
  +\frac{288l}{z}-\frac{384l^2}{z}-52z+120lz-64l^3z\nonumber\\&&\qquad
  -\frac{15z^3}{8}-21lz^3+48l^2z^3-24l^3z^3+\frac{75z^5}{32}
  -\frac{135lz^5}{16}+9l^2z^5-3l^3z^5\nonumber\\&&\qquad
  +\frac{125z^7}{512}-\frac{75lz^7}{128}+\frac{15l^2z^7}{32}
  -\frac{l^3z^7}{8}\Bigg)K_1(z)^2+m^6\frac{128K_1(z)^5}{z^2}.
\end{eqnarray}
Here $l=\ln(e^\gamma_Ez/2)$, $z=mx$, and $\gamma_E=-\Gamma'(1)=0.577\ldots$ is
Euler's constant. As shown in Fig.~\ref{numplot5}, the plot of this function
as well as the shapes of the corresponding functions in higher orders of
$\eps$ are very smooth and quite similar. The analytical expressions for
higher orders in $\eps$, however, become much longer. Note that the new
functions $f_n(z)$ first appear at order $\eps^2$. 

The smoothness of the zeroth order integrand as shown in Eq.~(\ref{integrand})
implies that the numerical integration is quite easy to execute. Because
the integrand vanishes exponentially for large values of $z$ and has no
singularities of the kind $z\ln z$ for small values of $z$, the integration
can in principle range from $0$ to $\infty$. However, for practical reasons
we had to instruct Wolfram's Mathematica system for symbolic manipulations
(which we used for all of our calculations presented here) that the integrand
vanishes for $z=0$. On the other hand, the asymptotic expansion of the
integrand together with the integration measure is dominated by the part
\begin{equation}
\frac{6\pi^2m^2}{512}z^{10}\ln^4(e^\gamma z/2)K_1(z)z^3dz\qquad
(K_\lambda(z)\to\sqrt{\frac\pi{2z}}e^{-z}\mbox{\ for\ }z\to\infty).
\end{equation}
Integrated from $\Lambda$ to $\infty$, this part gives a contribution
\begin{equation}
\frac{6\pi^2m^2}{512}\Lambda^{25/2}\ln^4(e^\gamma\Lambda/2)e^{-\Lambda}
\end{equation}
and terms which are of subleading order. Therefore, the result can be
well estimated by
\begin{equation}
\tilde\Pi_4^{\rm num}(0)=2\pi^2\int_0^\infty\Pi_4^{\rm num}(x)x^3dx
  \approx 2\pi^2\int_0^\Lambda\Pi_4^{\rm num}(x)x^3dx
  +\frac{6\pi^2m^2}{512}\Lambda^{25/2}\ln^4(e^\gamma\Lambda/2)e^{-\Lambda}
\end{equation}
and $\Lambda$ can be adjusted to any desired precision.

A possibility to avoid any cutoff is to change the integration variable such
that the interval $[0,\infty]$ is mapped onto $[0,1]$. Then the integration
can be done numerically with the additional information that the integrand
vanishes identically at both end points. Possible transformations of this kind
are for instance $z=\ln(1/t)$ or $z=(1-t)/t$ for $t\in[0,1]$.

\end{appendix}

\end{document}